%% file: main.tex
\documentclass{egpubl}
\usepackage{eg2023}
 
\STAR                   
\CGFStandardLicense

\usepackage[T1]{fontenc}
\usepackage{dfadobe}  

\usepackage{cite}  
\BibtexOrBiblatex
\electronicVersion
\PrintedOrElectronic
\ifpdf \usepackage[pdftex]{graphicx} \pdfcompresslevel=9
\else \usepackage[dvips]{graphicx} \fi

\usepackage{egweblnk} 

\usepackage{colortbl}
\usepackage{booktabs}
\usepackage[dvipsnames]{xcolor}
\usepackage{adjustbox}
\usepackage{inconsolata}   
\input{macros}
\input{icons}

\title[Neurosymbolic Models for Computer Graphics]%
      {Neurosymbolic Models for Computer Graphics}
      
\author[D. Ritchie, P. Guerrero, R. Jones, N. Mitra, A. Schulz, K. Willis, \& J. Wu]
{\parbox{\textwidth}{\centering 
        Daniel Ritchie$^{1}$ \quad
        Paul Guerrero$^{2}$ \quad
        R. Kenny Jones$^{1}$ \quad
        Niloy J. Mitra$^{2,3}$ \quad
        Adriana Schulz$^{4}$ \quad 
        Karl D. D. Willis$^{5}$ \quad
        Jiajun Wu$^{6}$
        }
        \\  
{\parbox{\textwidth}{\centering $^1$Brown University\quad
         $^2$ Adobe Research \quad
         $^3$University College London \quad
         $^4$University of Washington\quad
         $^5$Autodesk Research\quad
         $^6$Stanford University
       }
}
}

\begin{document}

\teaser{
 \includegraphics[width=\linewidth]{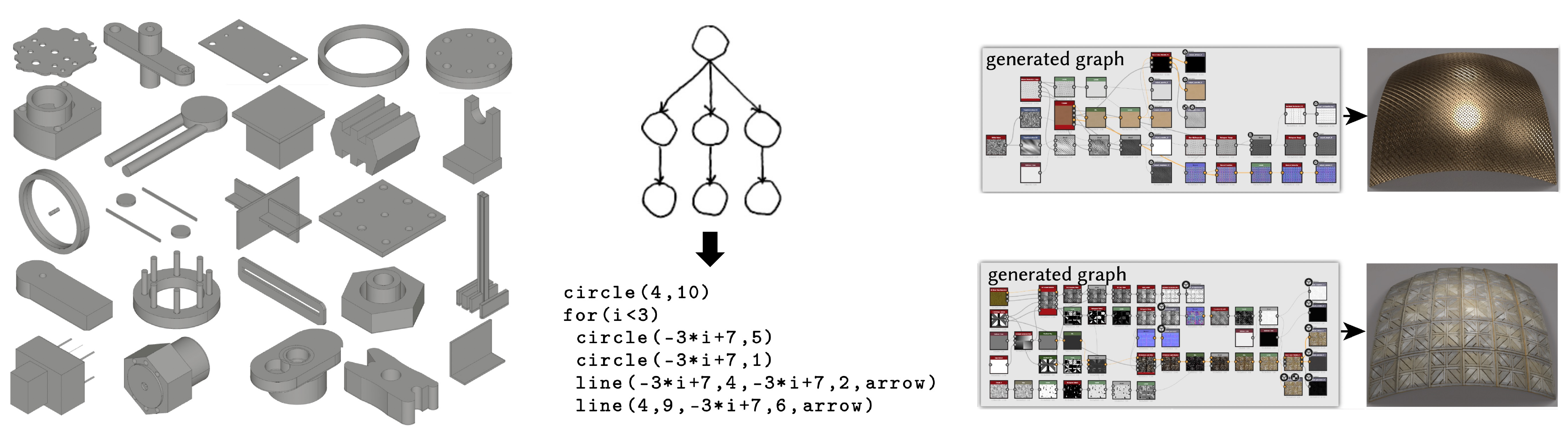}
 \centering
  \caption{\emph{Neurosymbolic models} produce visual data via a combination of symbolic programs and machine learning.
  From left to right:
  outputs of CAD programs written by a neural network~\cite{xu2022skexgen};
  inferring a 2D drawing program that reproduces an input hand-drawn diagram~\cite{ellis2018learning}; 
  procedural material programs (i.e. node graphs) generated by a neural network~\cite{guerrero2022matformer}.
  }
\label{fig:teaser}
}

\maketitle

\begin{abstract}
\input{00-abstract}


\end{abstract}

\input{01-intro}
\input{02-background}

\input{03-designspace}
\input{04-datasets}

\input{05-applications}
\input{06-conclusion}

\input{07-bios}
\input{08-acks}

\bibliographystyle{eg-alpha-doi}  
\bibliography{main}        

\end{document}

%% file: macros.tex
\newcommand{\rev}[1]{#1}

%% file: icons.tex
\newcommand{\icon}[1]{\raisebox{-0.25em}{\includegraphics[width=1.2em]{#1}}}

\newcommand{\neural}{\icon{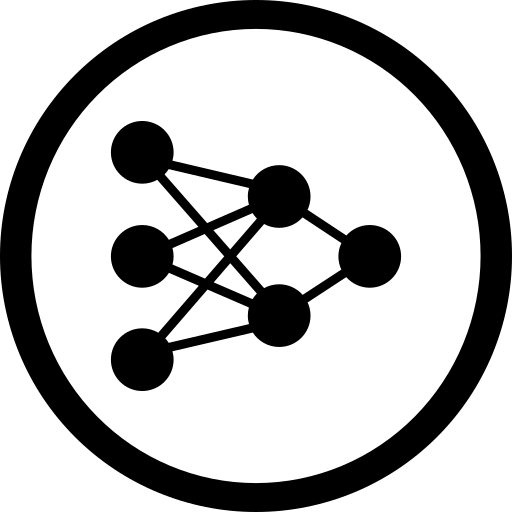}}

\newcommand{\visualtarget}{\icon{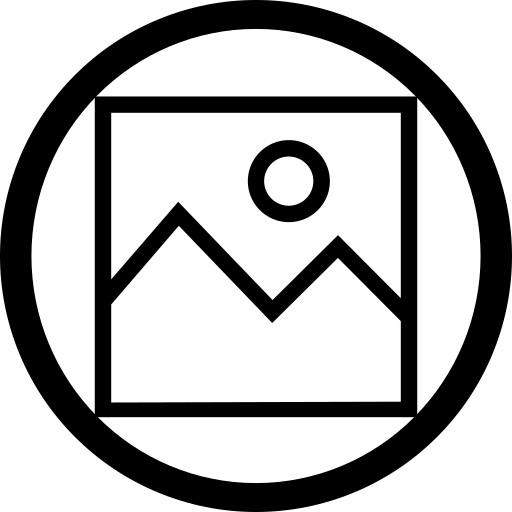}}
\newcommand{\visualexamples}{\icon{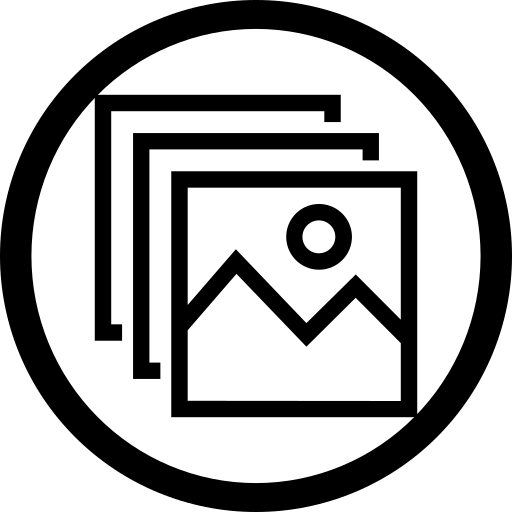}}
\newcommand{\blackbox}{\icon{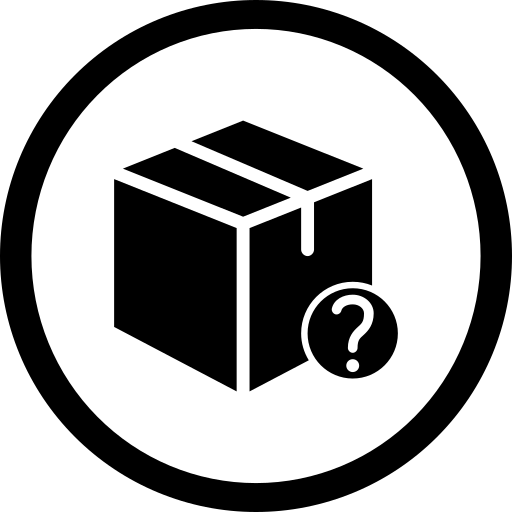}}
\newcommand{\textprompt}{\icon{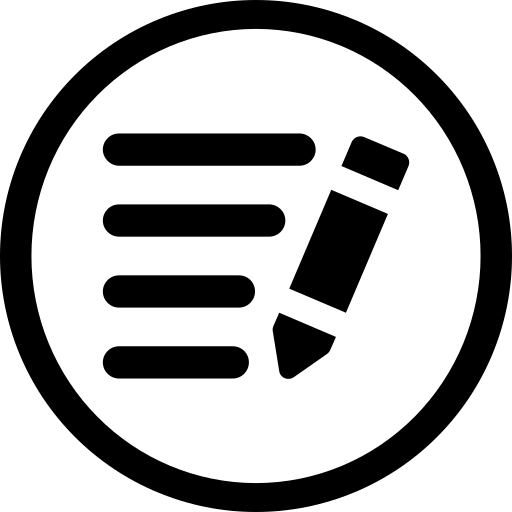}}
\newcommand{\objectivefns}{\icon{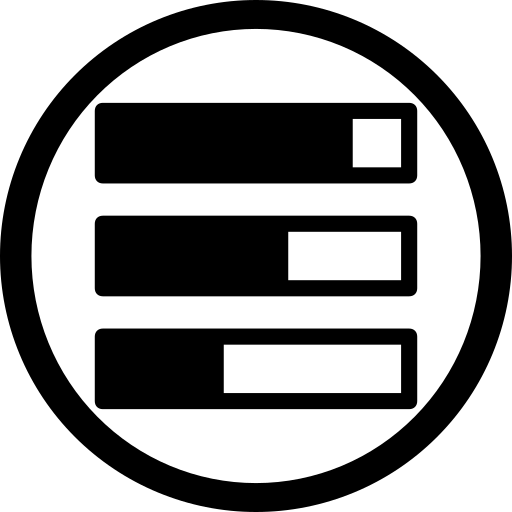}}

\newcommand{\detprog}{\icon{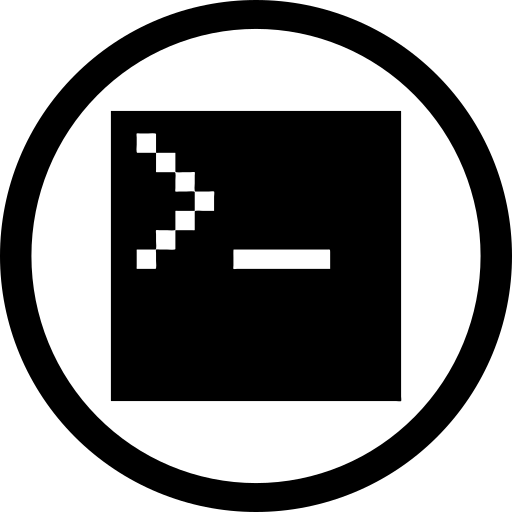}}
\newcommand{\probprog}{\icon{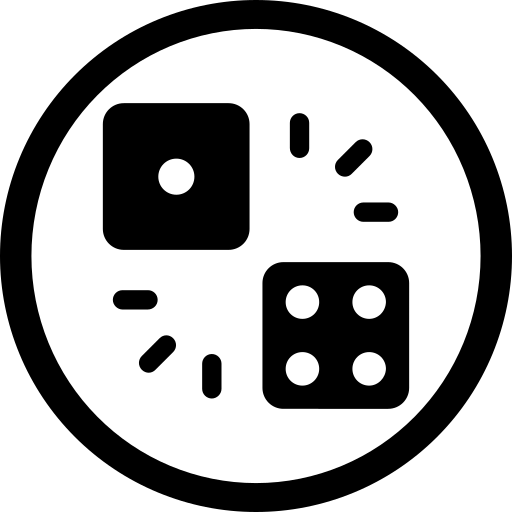}}
\newcommand{\progdist}{\icon{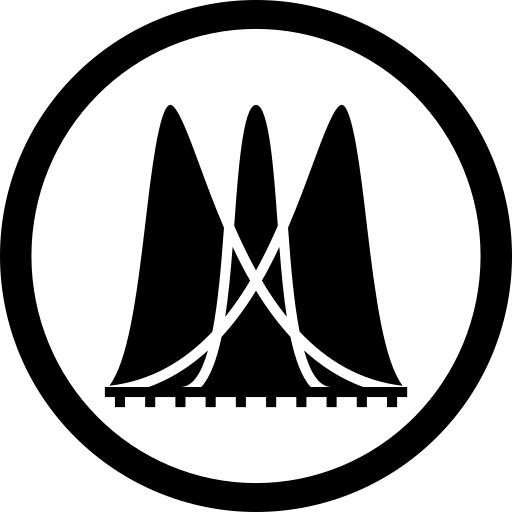}}

\newcommand{\functional}{\icon{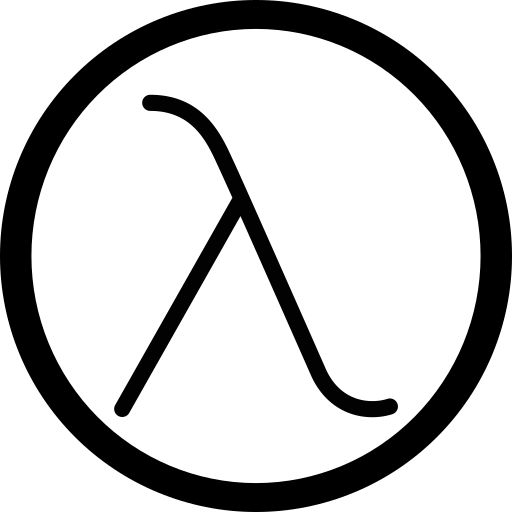}}
\newcommand{\imperative}{\icon{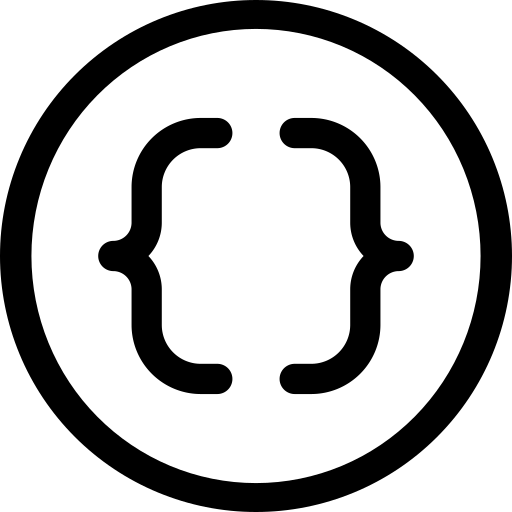}}
\newcommand{\constraint}{\icon{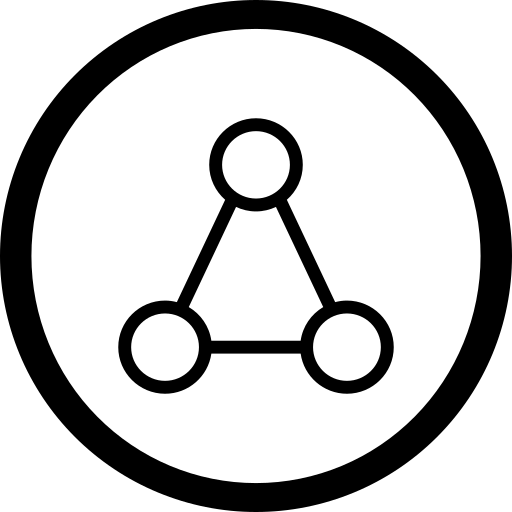}}

\newcommand{\neuralprims}{\neural}
\newcommand{\hardcodedprims}{\icon{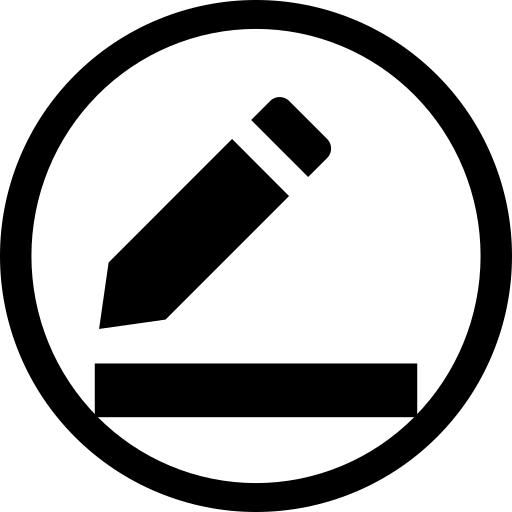}}

\newcommand{\fixed}{\icon{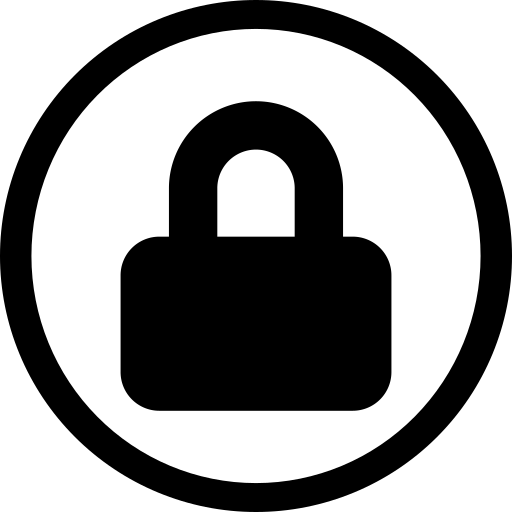}}
\newcommand{\preprocessed}{\icon{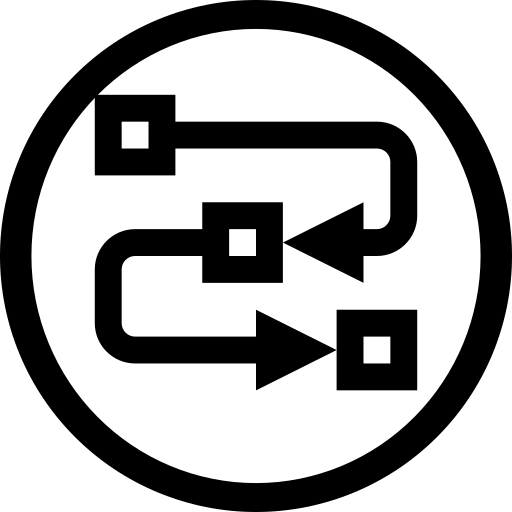}}
\newcommand{\evolving}{\icon{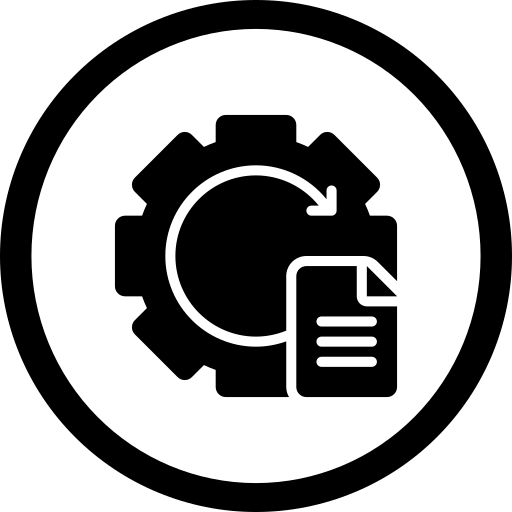}}
\newcommand{\invented}{\icon{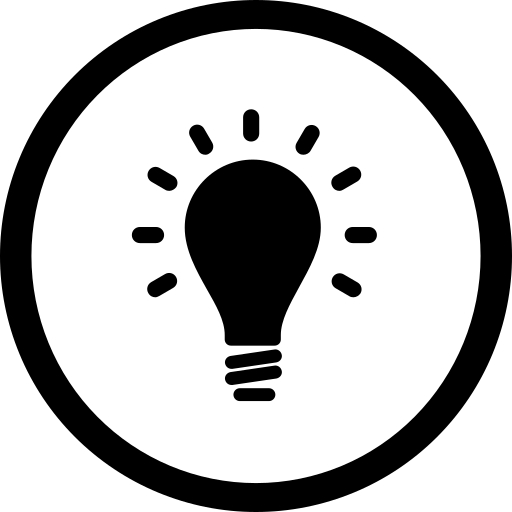}}

\newcommand{\retrieval}{\icon{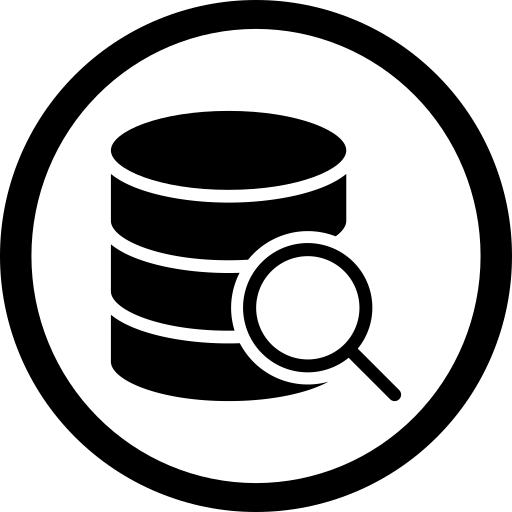}}
\newcommand{\enumeration}{\icon{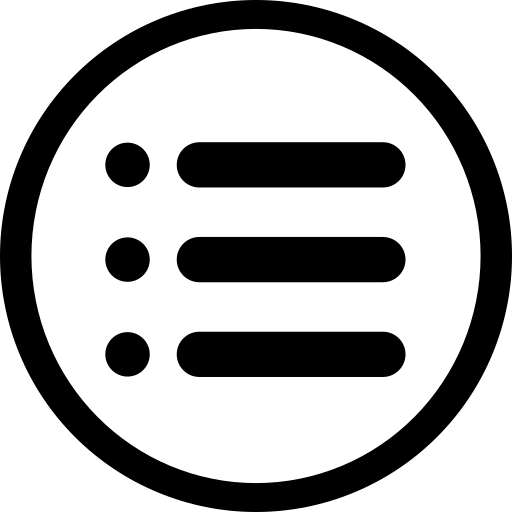}}
\newcommand{\stochastic}{\icon{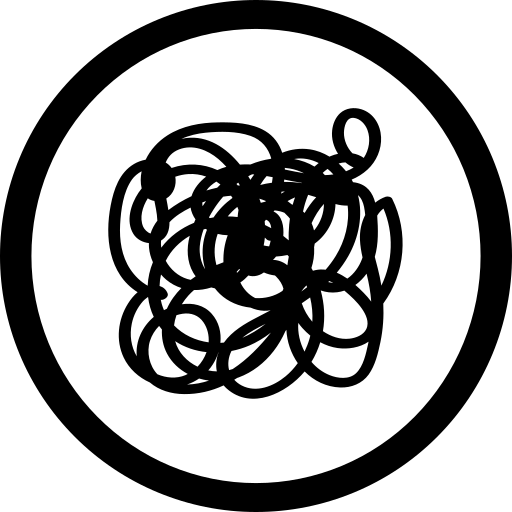}}
\newcommand{\constraintsat}{\icon{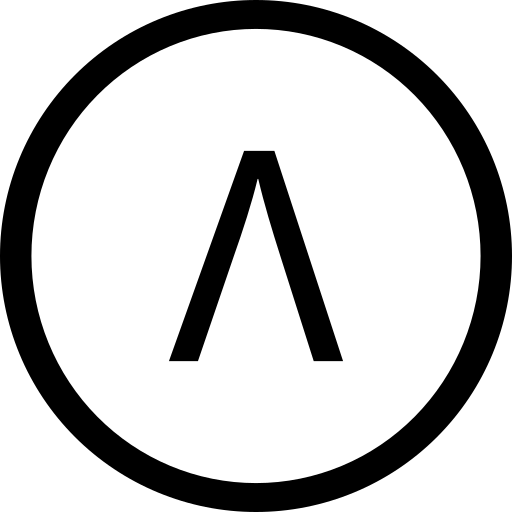}}
\newcommand{\userintheloop}{\icon{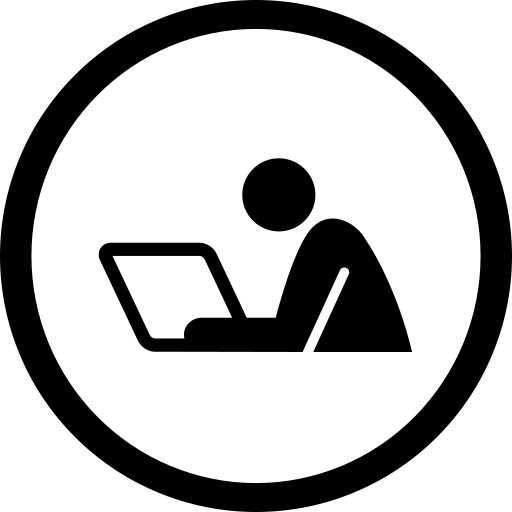}}

\newcommand{\synthguide}{\neural}

\newcommand{\direct}{\icon{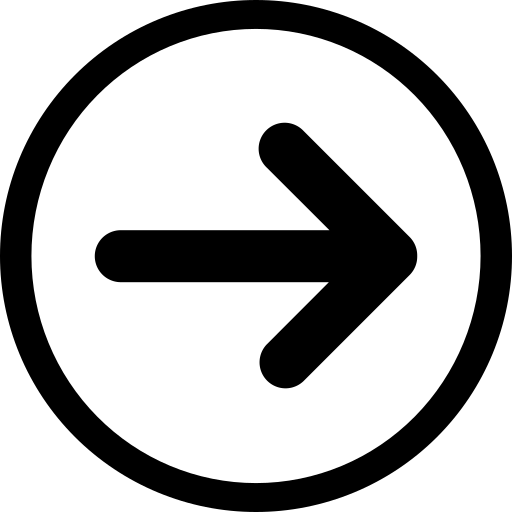}}
\newcommand{\solver}{\icon{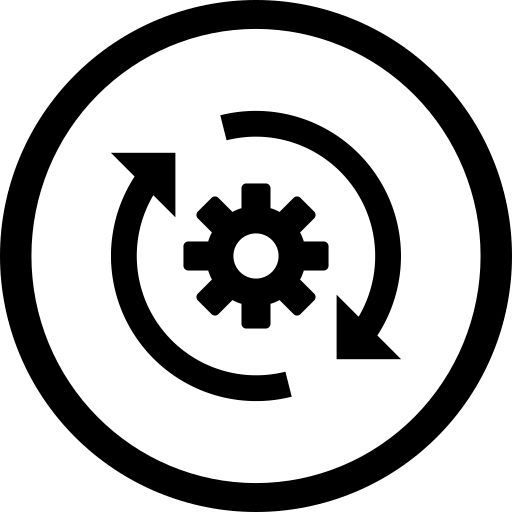}}
\newcommand{\solverguide}{\neural}
\newcommand{\proxy}{\neural}

\newcommand{\neuralpostproc}{\neural}

\newcommand{\smoothrelax}{\icon{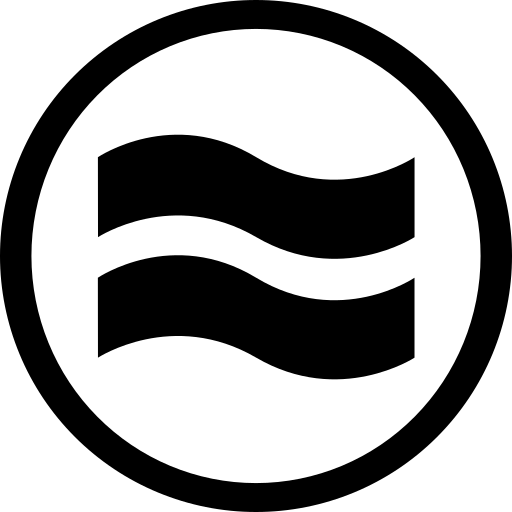}}
\newcommand{\rl}{\icon{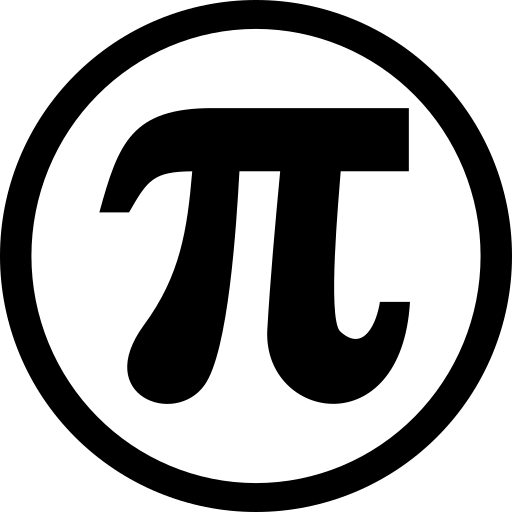}}
\newcommand{\people}{\icon{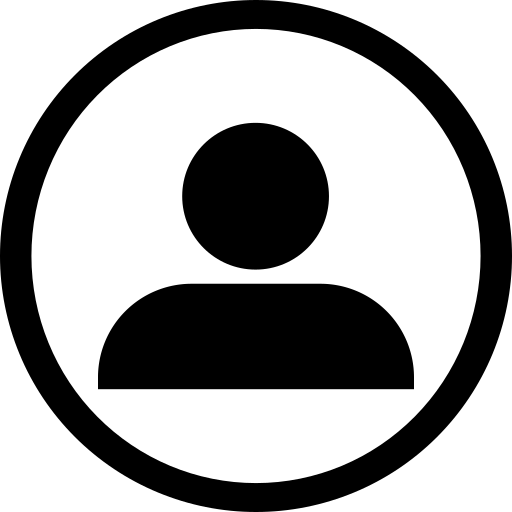}}
\newcommand{\synthetic}{\icon{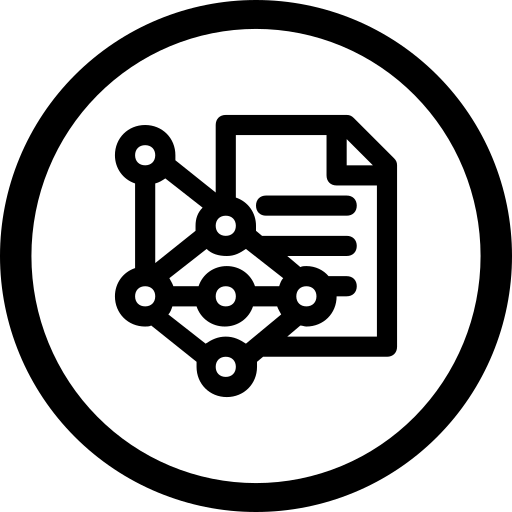}}
\newcommand{\bootstrap}{\icon{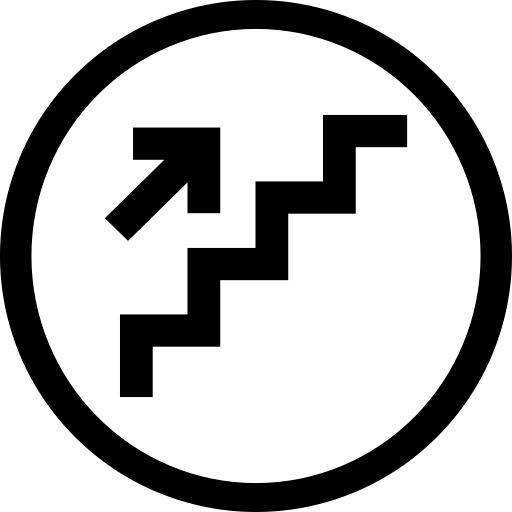}}

\newcommand{\iconlegend}{
\setlength{\tabcolsep}{3pt}
\begin{tabular}{lllllll}
    \neural{} Neural &
    \visualtarget{} Visual Target &
    \visualexamples{} Visual Examples &
    \detprog{} Deterministic Program &
    \probprog{} Probabilistic Program &
    \progdist{} Program Distribution &
    \functional{} Functional \\
    \imperative{} Imperative &
    \constraint{} Constraint &
    \hardcodedprims{} Hard-coded Primitives &
    \fixed{} Fixed &
    \preprocessed{} Preprocessed &
    \evolving{} Evolving &
    \retrieval{} Database Retrieval \\
    \enumeration{} Explicit Enumeration &
    \stochastic{} Stochastic Search &
    \constraintsat{} Constraint Satisfaction &
    \userintheloop{} User-in-the-loop &
    \direct{} Direct &
    \solver{} Solver &
    \proxy{} Learned Proxy \\
    \smoothrelax{} Smooth Relaxation &
    \rl{} Reinforcement Learning &
    \people{} Programs from People &
    \synthetic{} Synthetic Programs &
    \bootstrap{} Boostrapping
\end{tabular}
}

%% file: 00-abstract.tex

Procedural models (i.e. \rev{symbolic} programs that output visual data) are a historically-popular method for representing graphics content: vegetation, buildings, textures, etc.
They offer many advantages: interpretable design parameters, stochastic variations, high-quality outputs, compact representation, and more.
But they also have some limitations, such as the difficulty of authoring a procedural model from scratch.
More recently, AI-based methods, and especially neural networks, have become popular for creating graphic content.
These techniques allow users to directly specify desired properties of the artifact they want to create (via examples, constraints, or objectives), while a search, optimization, or learning algorithm takes care of the details.
However, this ease of use comes at a cost, as it's often hard to interpret or manipulate these representations.
In this state-of-the-art report, we summarize research on \textbf{neurosymbolic} models in computer graphics: methods that combine the strengths of both AI and symbolic programs to represent, generate, and manipulate visual data.
We survey recent work applying these techniques to represent 2D shapes, 3D shapes, and materials \& textures.
Along the way, we situate each prior work in a unified design space for neurosymbolic models, which helps reveal underexplored areas and opportunities for future research.

%% file: 01-intro.tex
\section{Introduction}
\label{sec:intro}

Throughout the history of computer graphics, progress in the field has often been driven by advancements in \emph{representations} of visual data.
New representations of visual data have enabled new creative capabilities, or have facilitated existing capabilities more efficiently.
For example, texture maps as a representation of surface detail have enabled virtual objects to efficiently mimic the appearance of real-world objects~\cite{CatmulThesis}. 
Level-of-detail systems for polygonal meshes have enabled the rendering of densely-populated virtual scenes at interactive rates~\cite{JimClarkLOD}.
Various types of splines have enabled higher-fidelity modeling of smooth, curved surfaces~\cite{SplineBook}.
Implicit surface representations have enabled beautiful scenery to be encoded in compact algebraic expressions~\cite{Shadertoy} and have proven useful for reconstructing 3D geometry from image observations~\cite{niessner2013hashing,mildenhall2020nerf}.

One broad class of representation that has been historically popular throughout computer graphics is \emph{procedural models}: \rev{symbolic} programs that output some visual datum when executed.
Procedural models are in some sense as old as computer graphics itself, dating back to the constraint programs used to produce engineering sketches in Ivan Sutherland's SketchPad system~\cite{SketchPad}.
Nowadays, procedural approaches are widely used for modeling certain classes of 3D shapes, including trees and other vegetation, buildings, and whole cityscapes~\cite{SpeedTree,CityEngine,Houdini}.
They are also used for modeling object surface appearance, i.e. materials and textures~\cite{SubstanceDesigner}, and even the behaviors of virtual characters in a crowd~\cite{Massive}.
Procedural models have achieved such longevity and popularity because they have many desirable properties.
They offer interpretable parameters that can be manipulated to change attributes of the visual data they generate (e.g. the height of a virtual building).
Furthermore, randomization of these parameters allows a single procedural model to generate a wide variety of different visual outputs---this is useful for rapidly exploring their design space or for populating large virtual worlds with non-repetitive content.
Procedural models do have some limitations, though.
First and foremost, creating a procedural model is challenging: doing so typically requires both programming expertise and artistic/design acumen, a combination that only certain practitioners possess.
Also, the types of variations achievable by a single procedural model are usually limited to parametric variations; more complex structural variations usually require non-trivial modifications to the structure of the procedural model itself (e.g. changing a model of sports cars to produce trucks).

Recently, machine learning models have become popular for producing visual data, with deep neural networks being especially prevalent.
Use of such models has rapidly become widespread for applications in image synthesis, processing, and manipulation~\cite{pix2pix2017,CycleGAN2017,GatysStyleTransfer,StyleGAN,DALLE2,Imagen,StableDiffusion}, 3D shape modeling~\cite{3DGAN,GRASS,StructureNet,jones2020SA,IMNet,li2021spgan,3DWaveletDiffusion,zeng2022lion}, material and texture modeling~\cite{TileGAN,SingleImageSVBRDF,henzler2021neuralmaterial}, 2D drawing and sketching~\cite{SketchRNN,Sketchformer2020,vinker2022clipasso}, and more.
In principle, these models are easy to create: just provide examples/training data, and a learning algorithm takes care of the rest.
What's more, they are quite general: the same model architecture (and sometimes even the same trained model) can represent a huge variety of different kinds of visual data (e.g. the space of all human faces).
They are not without their own limitations, however.
The representations these models learn are usually opaque and uninterpretable, making them hard to edit or manipulate (though some researchers have made progress in this space~\cite{GanDissection,bau2020rewriting}).
Additionally, as machine learning methods produce statistical \emph{approximations} of the true function implied by their training data, such models may generate outputs that exhibit artifacts: failing to generalize beyond their training set, producing e.g. blurry images, blobby geometry, etc.

\begin{table}[t!]
    \centering
    \begin{tabular}{rcc}
        \toprule
        \textbf{Property} & \textbf{Procedural/Symbolic} & \textbf{Learned/Neural} \\
        \midrule
        Ease of authoring & \color{Red}{Hard} & \color{Green}{Easy}\\
        Interpretability & \color{Green}{High} & \color{Red}{Low} \\
        Output artifacts & \color{Green}{No} & \color{Red}{Yes}\\
        Output variability & \color{Orange}{Medium} & \color{Green}{High}\\
        \bottomrule
    \end{tabular}
    \caption{Procedural/symbolic models and learned/neural models have complementary strengths and weaknesses for generating and manipulating visual data. This report discusses research that combines them to get the best of both worlds.
    }
    \label{tab:neural_vs_symbolic}
\end{table}

As summarized in Table~\ref{tab:neural_vs_symbolic}, procedural/symbolic models and learned/neural models have complementary strengths and weaknesses.
One might naturally ask: can one get the best of both worlds by somehow combining these two types of representations?
In this report, we summarize the state-of-the-art for research that focuses on such \textbf{neurosymbolic} models which generate visual data using \rev{symbolic programs augmented with AI/ML techniques}.
We first define a design space for such neurosymbolic models, taxonomizing the ways in which neural and symbolic representations can be hybridized to represent visual data (Section~\ref{sec:designspace}).
We then survey recent work which applies these representations to several different computer graphics domains: 2D shapes, 3D shapes, and materials/textures (Sections~\ref{sec:app_layout}-\ref{sec:app_materials}; see Figure~\ref{fig:teaser} for examples).
In the process, we situate each prior work within our design space.
Finally, we conclude with a look at open problems and future research opportunities in the field, including pointing out potentially fruitful regions of the neurosymbolic design space which have yet to be explored (Section~\ref{sec:conclusion}).

%% file: 02-background.tex
\section{Background \& Scope}
\label{sec:background}

The subject matter of this report may be of interest to many potential readers; we have written it with new graduate students in visual computing (e.g. computer graphics, computer vision) in mind.
We assume the reader has a solid background in the fundamentals of computer graphics (including its mathematical prerequisites, e.g. linear algebra).
We also assume some familiarity with basic machine learning concepts (e.g. training vs. test datasets, overfitting) and basic neural network concepts (network weights, stochastic gradient descent, etc.)
Some familiarity with basic programming language concepts is also helpful (e.g. different programming paradigms, abstract syntax trees).

To keep this report clear and concise, it is intentionally limited in scope.
Specifically, we focus on the task of generating visual data using symbolic programs augmented with AI/ML techniques.
In this report, when we say that a neurosymbolic model ``uses symbolic programs'', we mean that it explicitly constructs an intermediate symbolic representation of the visual data it generates: the visual data is represented by discrete symbols composed into a structure using one or more symbolic operators (i.e. functions with well-defined symbolic implementations).
By contrast, visual data may be represented with non-symbolic functions (e.g. a neural network which generates the data) or data structures (e.g. a raster image or a signed distance field).

Research on neurosymbolic modeling is interdisciplinary, drawing on ideas from deep learning, procedural modeling, and program synthesis.
Each of these areas would itself merit a full report; in this section, we discuss which material from each is relevant to our discussion, and which is out of scope.

\subsection{Deep Learning}


Deep learning and neural networks have seen widespread adoption across the field of computer graphics.
Neural networks are often used as learned function approximators: given a source domain $X$ and a target domain $Y$, a neural network can find a mapping from $X$ to $Y$ using gradient-based optimization techniques \cite{lecun2015deep}. 
In deep learning, complex neural architectures are developed by composing many simple neural layers and non-linear operations together.
These approaches have found success across many computer graphics tasks that can be framed as mapping problems (e.g. feature detection, denoising, rendering, animation, etc.) \cite{creativeAI}. The questions of how deep learning should be applied, and when it will be successful, are complex and dependent on many problem-specific factors outside the scope of this report. 

Neural approaches have proven useful for reconstruction tasks.
With enough data and compute, learning-based methods are able to produce visual outputs that correspond to partially observed or under-specified inputs.
Typically, such systems learn to directly predict visual outputs in an end-to-end, differentiable fashion (e.g. methods for image-based 3D reconstruction \cite{han2019image}).
\rev{
However, we instead focus on approaches that create (either explicitly or implicitly), a non-trivial symbolic intermediary representation capable of producing visual data.
This symbolic representation must be more than a trivial combination of primitives, and as such, we consider neural methods that directly predict low-level representations entities (e.g. triangle meshes \cite{polygen}) or even complex geometric representations (e.g. hierarchies of bounding boxes or boundary representations \cite{StructureNet,sharmaParseNet,GuoComplexGen2022}) to be outside the scope of this report.
}
Relatedly, neural approaches have also been investigated that learn how to represent visual data in a structure-aware fashion \cite{3d_struct_star}.
While neurosymbolic representations are always structured, not all structure-aware representations fall under our definition of a ``symbolic program''.
As such, our framework does not capture some approaches that represent visual data as sequences of discrete codes \cite{yan2022shapeformer}, or those that simply combine primitives or \cite{abstractionTulsiani17} learned parts \cite{Paschalidou2021CVPR}. 
In later sections, we discuss methods for visual reconstruction tasks that use learning methods to produce symbolic representations that, when executed, generate visual outputs that reconstruct the input \cite{tian2018learning,ellis2019repl,ellis2018learning,jones2022PLAD}. 

Beyond reconstruction, neural approaches have also proven useful as generative models of computer graphics content. 
Deep generative models learn to represent the probability distribution over an input domain $X$ (e.g. a collection of visual data). 
This probability distribution can then be sampled to synthesize novel instances from $X$ (e.g. new visual data).  
There are a multitude of deep generative modeling paradigms \cite{deep_gen_survey, croitoru2022diffusion}, all with different pros and cons: generative adversarial networks (GANs), variational autoencoders (VAEs), normalizing flows, auto-regressive models, and diffusion models.
These deep generative models have been applied across many visual domains such as natural images~\cite{StyleGAN}, materials~\cite{Guo:2020:MaterialGAN}, sketches~\cite{vinker2022clipasso}, scenes~\cite{deep_synth}, voxels\cite{3DGAN}, meshes \cite{polygen}, implicit shapes \cite{IMNet}, and character motion \cite{ling2020character}.
As in the reconstruction context, many of these approaches have been designed to produce visual outputs directly, without any intermediate symbolic representation, and thus are outside the scope of what our survey covers.
Generative neurosymbolic models will often use one of the aforementioned learning paradigms, but will instead train this network to synthesize novel symbolic representations that can be executed to produce new visual outputs.

Finally, there has been growing interest centered around approaches that \emph{represent} visual data neurally \cite{neural_fields}.
In some applications, a neural representation will be specialized to a particular datum. 
For instance, in Neural Radiance Fields (NeRFs), a scene is represented by a neural network that learns a mapping from spatial location and viewing direction input pairs to volume density and emitted radiance output pairs \cite{mildenhall2020nerf}.
Neural representations can also generalize across a distribution of data. 
For instance, a neural network can learn to map a point and a latent code to an occupancy output prediction. 
This network can then train on a large collection of shapes and compress each complex geometry into a small latent code
\cite{overfit_implicits}.
Most approaches that represent visual data neurally fall outside the scope of our report, as they lack any form of symbolic representation.
However, as we discuss in Section \ref{sec:dsl}, some neurosymbolic models augment their symbolic languages with learned neural primitives.

\subsection{Procedural Modeling}

Procedural modeling, or the use of symbolic programs to generate visual data, is as old as computer graphics itself.
Procedural approaches have been proposed for data at virtually every stage of the graphics pipeline, but they have seen the most widespread use in 3D geometry and texture modeling.
As we discuss applications of neurosymbolic models to these domains in Sections~\ref{sec:app_shapes} and \ref{sec:app_materials}, respectively, we review some relevant background material here.
A full treatment of the history of these fields is outside the scope of this report.

\rev{Note that throughout this report, we use ``procedural model'' to refer to a program in \emph{any} symbolic programming paradigm which generates visual data.
In the programming languages literature, the term ``procedural programming'' is sometimes used to denote a programming paradigm in which a program is interpreted as a sequence of instructions to be executed; this is contrasted with other programming paradigms, such as functional and declarative programming.
In this report, we refer to programs which provide sequences of instructions as ``imperative programs,'' to avoid overloading the word ``procedural.''
}

\begin{figure*}[t!]
    \centering
    \includegraphics[width=\linewidth]{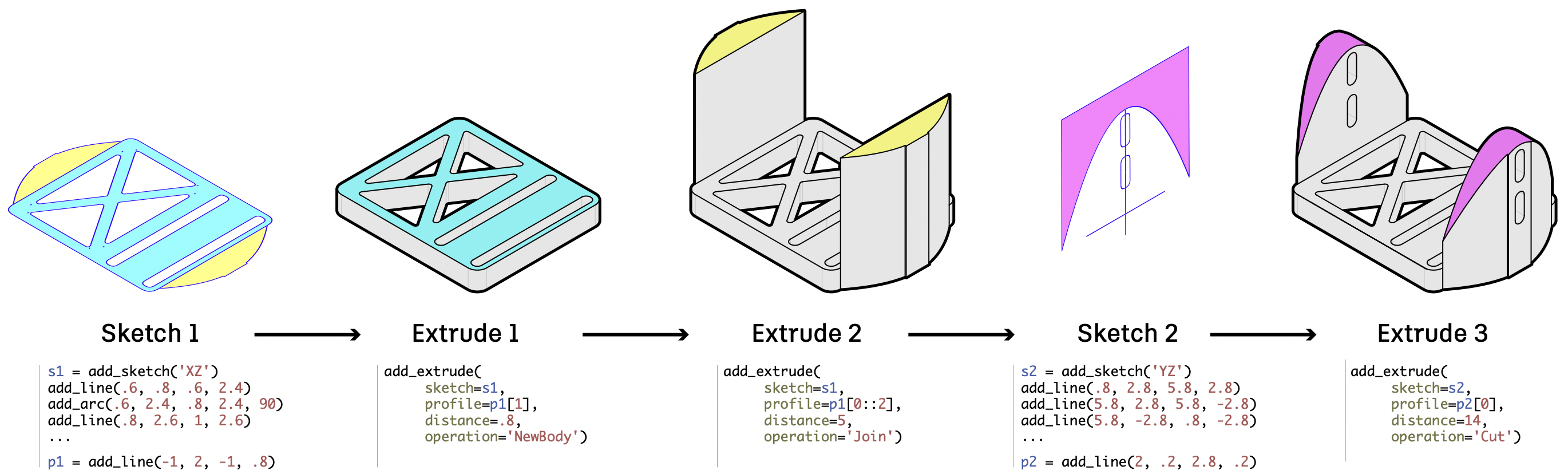}
    \caption{Procedural modeling as applied to CAD: using \emph{feature-based modeling} to produce solid geometry as a composition of operations on 2D sketches extruded into boundary representations~\cite{Fusion360Gallery}.}
    \label{fig:sketchextrude}
\end{figure*}

\paragraph*{Procedural geometry:}
The earliest application of procedural geometric modeling is to computer-aided design (CAD), dating back to Ivan Sutherland's SketchPad system~\cite{SketchPad}.
A CAD model consists of a set of parameterized operations (possibly including constraints); these parameters can be adjusted and the modeling program re-executed to produce updated geometry.
\emph{Constructive solid geometry} (CSG) is one such programming model for 3D geometry generation: sets of parametric primitives which are combined via Boolean set operations (intersection, union, difference) to create more complex shapes~\cite{CSGBookChapter}.
By construction, CSG programs produce watertight surface geometry that bounds a solid object, making it well-suited for modeling objects for manufacturing.
\emph{Boundary representations} (or B-reps) are an alternative to CSG for representing solid geometry~\cite{BrepBook}.
A B-rep consists of a set of connected surface patches (of any manifold shape) which collectively form the boundary of a solid region.
One can perform Boolean operations on B-reps (as with CSG), but they also support additional useful modeling operations, such as filleting or chamfering edges.
They are also often authored by lifting 2D engineering sketches into 3D via extrusion, revolution, or other operations (Figure~\ref{fig:sketchextrude}).
Taken together, this paradigm of solid modeling is typically referred to as \emph{feature-based modeling}~\cite{hoffmann1989geometric}; it is now the dominant form of solid modeling adopted by industry-standard CAD software~\cite{Solidworks,Fusion360}.
We survey neurosymbolic CAD modeling work in Section~\ref{sec:app_shapes}. 
\rev{Some prior work seeks to infer B-reps from ``raw'' input geometry such as point clouds~\cite{GuoComplexGen2022}; such inference systems do not consider the higher-level CAD programs which could produce such B-reps and thus fall outside the scope of this report.}

\emph{Context-free grammars} are another popular paradigm for procedural geometry generation.
\emph{L-systems} are a particularly prominent example: context-free string rewriting systems whose output strings are interpreted to produce geometry~\cite{LSystems}.
Their recursive nature makes them well-suited for modeling naturally-occurring fractal structures such as trees and other types of vegetation~\cite{ABOP,SyntheticTopiary}.
Other popular instantiations of context-free grammars include \emph{shape grammars}: rewriting systems that operate directly on geometry, typically by subdividing regions of space~\cite{ShapeGrammars}.
They are widely used for regular, repeating structures, particularly those occurring in architecture: buildings~\cite{CGAShape}, facades~\cite{FacadeParsing}, and cities~\cite{CityEngineBookChapter}.
The neurosymbolic modeling techniques we discuss later in this report are applicable to such grammars, because they apply to programs in any language (of which context-free grammars are a subset).

\rev{Since trees are common use case for procedural models:
there exists prior work on inferring tree models from unstructured inputs (e.g. images or point clouds), much of which uses non-procedural tree representations~\cite{UrbanTreeGenerator,TreePartNet21,SingleImageTrees}.
Such work is outside the scope of this report.
}

\paragraph*{Procedural textures:}
Procedural techniques have also been popular for creating textures, as they can create intricate details with infinite spatial resolution.
The foundation of most procedural texturing systems is a library of primitive patterns, especially \emph{noise} functions such as Perlin noise~\cite{Perlin:1985:ImageSynthesizer} and Worley noise~\cite{Worley:1996:Cellular}.
More complex textures can then be created by systematically composing these texture primitives via various mixing and blending functions.
This paradigm, first elucidated in the Shade Trees paper~\cite{ShadeTrees}, has become widespread in commercial modeling and rendering software, such as Maya's Hypershade~\cite{MayaHypershade} and Substance Designer~\cite{SubstanceDesigner}.
We survey neurosymbolic modeling work applied to material and texture authoring in Section~\ref{sec:app_materials}.

\subsection{Program Synthesis}

Program synthesis is an important area of research in programming languages that investigates methods for automatically generating programs that satisfy some high-level specifications. While the dream of automating code generation can be traced back to the birth of computer science, and the first paper that proposed a synthesis algorithm dates from 1957~\cite{backus1957fortran}, it was not until the last couple of decades that the field saw great advances in what is called \emph{inductive synthesis} \rev{~\cite{solar2006combinatorial,alur2013syntax}}. In inductive synthesis, users give a (potentially partial) specification to describe the desired \emph{intent}, and search methods are used to explore the space of possible programs generating one that satisfies the specifications. This involves (i)~developing methods for specifying user intent, (ii)~defining a search space by restricting an existent language or designing novel domain-specific languages, and (iii)~developing search algorithms to efficiently explore this space.   

The field of program synthesis has advanced by proposing different interfaces for specifying intent, ranging from input-output examples, demonstrations, natural languages, partial programs, and assertions. Search algorithms include enumeration, constraint solving, probabilistic search, and combinations thereof. The first commercial application of program synthesis was FlashFill in Excel 2013, which derives small programs from data manipulation examples~\cite{gulwani2011automating}. Today, program synthesis is used in many applications, and frameworks such as Sketch~\cite{solar2006combinatorial} and Rosette~\cite{torlak2013growing} make it easy to develop synthesis tools for new languages. We refer the reader to~\cite{gulwani2017program} for an overview of traditional synthesis techniques; we will discuss search algorithms relevant to neural symbolic reasoning in further detail in Section~\ref{sec:progsynth}.

In recent years, program synthesis techniques have also been applied in novel and exciting ways to solve problems in computer graphics. 
These applications are inspired by procedural representations of shapes which transform modeling into a code generation task. 
Recent work includes reverse-engineering CAD programs from 3D shapes~\cite{du2018inversecsg,nandi2018functional} and shape program manipulation ~\cite{hempel2019sketch}. 

\rev{More recently, work in program rewrites has been used to generate 3D CAD code that is more compact and easy to manipulate~\cite{nandi2020synthesizing}. They leverage a data structure called E-graphs~\cite{nelson1980fast}. E-graphs are popular for code optimization, which requires searching over a large number of programs that are syntactically different but semantically equivalent. The key insight is that programs are typically viewed as treelike structures containing smaller sub-programs, many of which are shared across the different semantically equivalent variations. The E-graph data structure is capable of representing many equivalent programs efficiently by sharing sub-programs whenever possible, and further supports operations for extracting programs that have minimal cost. 
Please refer to \cite{2021-egg} for a more comprehensive overview.
}
While such program optimization methods live at the boundary of program synthesis and compilers, they are becoming increasingly popular within computer graphics, for example, to optimize design and fabrication plans for carpentry~\cite{zhao2021co,wu2019carpentry}.  

Despite these advances, there are still fundamental challenges in expanding the reasoning capabilities of synthesis techniques to complex domains. This is because (i)~the search space grows exponentially with the size of the synthesized code and (ii)~because inferring intent from natural forms of human interaction is challenging~\cite{gottschlich2018three}. 
Inspired by deep learning's successes over search and inference tasks, the programming languages community has looked for ways to apply these ideas to enable automatic code generation; for instance, synthesizing complex code fragments from natural language specifications ~\cite{li2022competition}.
We refer the reader to a recent survey on neurosymbolic programming~\cite{chaudhuri2021neurosymbolic} for more examples outside the domain of computer graphics.


We also note that some research has explored using neurosymbolic methods for computer graphics applications, but does not \emph{represent} visual data with a neurosymbolic model.
For example, recent work uses neurosymbolic representations and search algorithms to discover demosaic algorithms that trade-off performance and output quality~\cite{ma2022searching}. This line of work, however, is concerned with finding \emph{transformative} programs (that take input data to output data), instead of \emph{generative} programs (that generate some visual datum) and therefore falls outside the scope of our report.

%% file: 03-designspace.tex
\begin{figure*}[t!]
    \centering
    \includegraphics[width=\linewidth]{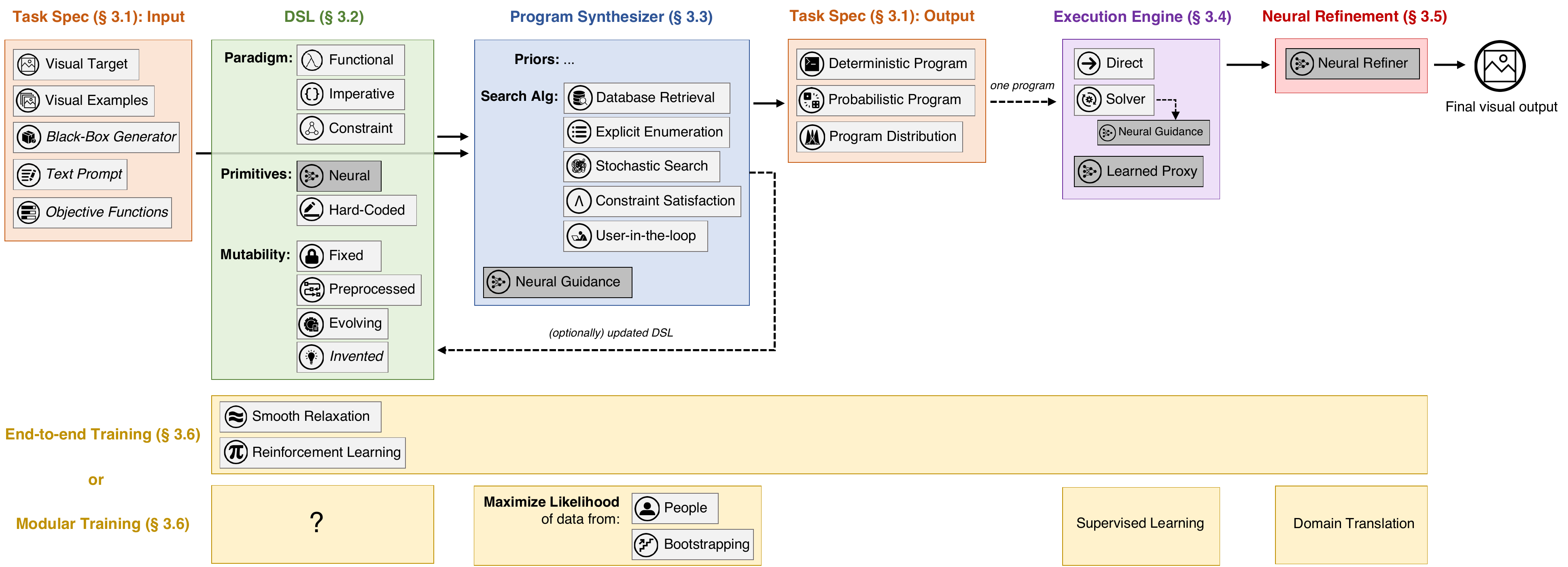}
    \caption{
    The design space of neurosymbolic models.
    A neurosymbolic model takes as input a specification of user intent and a domain-specific language (DSL), and it synthesizes programs that output visual data that satisfy the user intent.
    The visual output may then be further refined in a non-symbolic, neural postprocess.
    Several approaches have been proposed for learning the various neural components of such a model.
    \emph{Note: the icons and color scheme introduced in this figure are used throughout the report to identify where in this design space different prior works fall.}
    }
    \label{fig:designspace}
\end{figure*}

\section{A Design Space for Neurosymbolic Models}
\label{sec:designspace}

Thus far, we have described neurosymbolic models in broad terms: symbolic programs augmented with machine learning.
There are many possible realizations of this core idea; in this section, we attempt to organize these different instantiations into a formal \emph{design space}.
For our purposes, a design space is a set of \emph{design axes} along which the design of a system may vary, coupled with a set of \emph{design choices} for each axis.
The design space we present in this section was created to span a large set of neurosymbolic models presented in prior research, which we will discuss later in this report.
Given its ability to fit such a broad range of work under its umbrella, we believe that this design space should remain a useful intellectual framework for the field of neurosymbolic modeling as it progresses.
In Section~\ref{sec:conclusion}, we highlight some new, unexplored points in this design space that may merit further investigation.


Figure~\ref{fig:designspace} gives a diagrammatic overview of our design space, organized left to right according to the pipeline of stages followed by typical neurosymbolic models.
A neurosymbolic model takes as input some specification of user intent (Section~\ref{sec:taskspec}) and produces programs (Section~\ref{sec:progsynth}) which can then be executed (Section~\ref{sec:execengine}) to produce visual data which satisfies the user intent.
This output visual data might then be further refined by a non-symbolic (e.g. neural) postprocessing step (Section~\ref{sec:neuralrefine}).
To perform this task, the model must also take as input a specification of the language in which its output program should be expressed (Section~\ref{sec:dsl}); some neurosymbolic models will make changes to this language as part of their operation.
Various stages of the model contain neural components which much be learned in some way (Section~\ref{sec:learningalgs}).

Figure~\ref{fig:designspace} uses specific colors to refer to different neurosymbolic model stages, and it uses specific icons to refer to different design choices.
These colors and icons will be used throughout this report to identify where in the design space different prior works fall.
Finally, we note that some of the design choices identified here are speculative, in that they are possible design choices but have not yet been executed by any prior work of which we are aware.
We identify these design choices with an \emph{italic} font.

\subsection{Task Specification}
\label{sec:taskspec}

The task that a neurosymbolic model should perform can be specified by its given inputs and the outputs it should produce.
In general, a neurosymbolic model will take some user input about the kind of visual data the user would like the model to generate; it will then provide as output one or more programs, which can be executed to generate visual data which satisfies these inputs.
The model is also typically given a formal specification for the language in which these output programs should be expressed (see Section~\ref{sec:dsl}).

There are many ways for a user to specify their intent to a neurosymbolic model.
Potential user inputs include (but are not limited to):
\begin{itemize} 
    \item \visualtarget{} \textbf{Visual Target}: A single visual datum (e.g. an image or a 3D shape) to be reconstructed.
    \item \visualexamples{} \textbf{Visual Examples:} a collection of visual data whose distribution should be emulated.
    \item \blackbox{} \textbf{\emph{Black-Box Generator}:} an existing visual data generator whose behavior should be emulated (e.g. a pretrained \rev{deep generative model of 3D shapes~\cite{li2021spgan,3DGAN,3DWaveletDiffusion}}). \rev{We are not aware of any existing work which implements this particular design choice.}
    \item \textprompt{} \textbf{\emph{Text Prompt}:} a natural-language description of the type of visual data to generate. \rev{We are not aware of any existing work which implements this particular design choice.}
    \item \objectivefns{} \textbf{\emph{Objective Functions}:} functions which map a program's visual output to a desirability score \rev{(e.g. the stability of a generated 3D shape under gravity~\cite{Mezghanni_2022_CVPR}). We are not aware of any existing work which implements this particular design choice.}
\end{itemize}
As output, the neurosymbolic model could produce:
\begin{itemize}
    \item \detprog{} \textbf{Deterministic Program:} a single program that returns the same visual output each time it is executed. Such a program could include continuous parameters that can be varied to produce a range of possible visual outputs. \rev{Many CAD programs fall into this category~\cite{Fusion360Gallery,DeepCAD}.}
    \item \probprog{} \textbf{Probabilistic Program:} a single program that makes random choices, producing a different visual output each time it is executed \rev{(e.g. the type and position of wings on a spaceship~\cite{ExampleBasedProcMod})}.
    As these random choices can affect both continuous and structural elements, these programs can capture a wider variety of visual inputs compared with Deterministic Programs.
    \item \progdist{} \textbf{Program Distribution:} a sampler for a distribution of different programs. Sampled programs could have a range of different structures, allowing this representation to capture an even wider range of visual outputs than either of the two previous types.
    \rev{In the works we will survey in this report, such samplers are usually represented generative neural networks that output programs}.
\end{itemize}
Additionally, the model may produce a new language in which its output programs are expressed (if no such language was provided as input, or if the input language was modified by the neurosymbolic model).

\subsection{Domain-Specific Language (DSL)}
\label{sec:dsl}

The heart of a neurosymbolic model is the programs it produces; in order to specify a program, one needs a language in which those programs are to be expressed.
Although it is in principle possible to use general-purpose programming languages such as C++ or Python, the standard practice is instead to use a \emph{domain-specific language}, or DSL: a language whose data types and operators are specialized for a particular domain of interest (e.g. generating 3D shapes)~\cite{fowler2010domain}.
Such languages allow useful programs in that domain to be expressed more concisely, which makes them both easier to use and modify as well as easier for the neurosymbolic model to produce (as the space of possible output programs is smaller).

For the purposes of neurosymbolic modeling, there are several important design axes to consider when specifying a DSL:

\paragraph*{Programming paradigm:}
A DSL can adopt any of the many programming language paradigms.
The work we survey in this report includes examples of \functional{} \textbf{Functional} languages \rev{(e.g. node-based dataflow programs representing procedural materials~\cite{guerrero2022matformer})}, \imperative{} \textbf{Imperative} programs \rev{(e.g. scalable vector graphics, SVG, files describe sequence of drawing commands~\cite{DeepSVG,reddy2021im2vec})}, and \constraint{} \textbf{Constraint} languages \rev{(e.g. CAD engineering sketch languages which specify constrained relationships between the geometric attributes of different sketch elements~\cite{seff2021vitruvion,Ganin2021ComputerAidedDA,para2021sketchgen})}.

\paragraph*{Primitive types:}
One of the ways that a symbolic program can be augmented with machine learning is to have learned \neuralprims{} \textbf{Neural Primitives} within the language---either primitive data types or primitive functions.
\rev{For example, a language for 3D shapes could include a library of 3D parts, each of which is represented by a learned neural field~\cite{deng2022unsupervised}).}
If a DSL does not have such primitives, we say that it uses \hardcodedprims{} \textbf{Hard-coded Primitives} only.

\paragraph*{Mutability:}
The language given as input (if any) to a neurosymbolic model isn't necessarily fixed; the model could choose to modify the language, if this allows the model to better explain the visual data it is trying to produce.
We classify a language's mutability into one of the following categories:
\begin{itemize}
    \item \fixed{} \textbf{Fixed:} the language given as input remains unchanged.
    \item \preprocessed{} \textbf{Preprocessed:} the language given as input is modified by the neurosymbolic model in a preprocessing step before the model learns to produce programs \rev{(e.g. by finding common patterns in some example programs and factoring them out into new procedures~\cite{jones2021shapeMOD})}.  
    \item \evolving{} \textbf{Evolving:} the language given as input is continually modified by the neurosymbolic model as it learns to produce programs \rev{(e.g. by discovering increasingly complex subroutines which make it easier to produce more complex programs~\cite{DreamCoder})}.
    \item \invented{} \textbf{\emph{Invented}:} no language is given as input; the neurosymbolic model invents the entire language itself.
    \rev{The process of ``inventing'' a language would involve proposing a set of primitive data types and functions which combine those data types to parsimoniously represent visual data from a particular domain. We are not aware of any existing work which implements this particular design choice. We briefly discuss potential future work ideas along this direction in Section~\ref{sec:conclusion}.}
\end{itemize}

\subsection{Program Synthesizer}
\label{sec:progsynth}

Given a task specification and a domain-specific language (DSL), the neurosymbolic model must produce generative programs that satisfy the specification in that DSL.
This problem is fundamentally a search program: searching through the space of programs expressible in the DSL to find satisfying ones.
This is a challenging search problem, and there are many possible strategies for solving it.
We can organize these strategies along the following design axes:

\paragraph*{Program priors:}
In any nontrivially complex DSL, it is likely there will exist multiple programs that satisfy the input specification equally well.
To disambiguate between them, we must ask ourselves the question: what kind of programs would we prefer \emph{a priori}?
To the extent that this question has been considered by prior work (in both neurosymbolic models for graphics and in the field of program synthesis more generally), the answer has been to prefer shorter/simpler programs (i.e. following Occam's razor).
This design axis could admit different/additional priors, however; we will come back to this point in Section~\ref{sec:conclusion}.

\paragraph*{Search algorithm:}
The heart of a program synthesizer is the algorithm it uses to search through the (typically vast) space of possible programs.
Prior work has used methods of the following types:
\begin{itemize}
    \item \retrieval{} \textbf{Database Retrieval:} the simplest form of program ``synthesis'': rather than generate a new program, retrieve one from a database of existing programs.
    Some methods will then modify the retrieved program, e.g. by tweaking its parameters to better satisfy user goals.
    \item \enumeration{} \textbf{Explicit Enumeration:} deterministic exploration of possible programs in the language.
    Enumeration can proceed either in \emph{top-down} or \emph{bottom-up} fashion, i.e. whether the search starts from the root or the leaves of the program's abstract syntax tree.
    This approach is typically only viable for finding short programs in relatively simple languages; additional techniques are often required to make the search tractable in more complex settings.
    The most common such approach is to restrict the search space using some heuristic(s).
    Greedy search is the most extreme form of restriction. Variants of \emph{beam search} are also widely used~\cite{BeamSearch}. In top-down search, the \emph{branch and bound} strategy can be applied~\cite{Clausen2003BranchAB}.
    A more recent class of techniques involves compressing the search space by compactly representing large classes of equivalent sub-programs; these representations include version spaces~\cite{Mitchell1977VersionSA} and e-graphs~\cite{eqsat-lmcs}.
    \item \stochastic{} \textbf{Stochastic Search:} for languages with very large program spaces where enumeration is not tractable, a randomized form of search can be a viable alternative.
    These algorithms start with some initial program(s) (which could be random programs from the language or initialized using some heuristic) and then iteratively make random changes to the program(s) in the hopes of improving them.
    Some of these algorithms maintain a single candidate program: these include Markov Chain Monte Carlo (MCMC)~\cite{MCMCIntro} and simulated annealing~\cite{Laarhoven1987SimulatedAT}.
    Other algorithms maintain a set, or \emph{population}, of programs; the random changes the algorithm uses may involve reasoning about multiple programs at once.
    Examples of such algorithms include Sequential Monte Carlo (SMC)~\cite{SMCBook} and genetic programming~\cite{GeneticProgrammingBook}.
    \item \constraintsat{} \textbf{Constraint Satisfaction:} in some cases, it is possible to convert the task specification into a set of constraints and treat the problem as one of constraint satisfaction.
    One common approach is to convert the spec into a Boolean satisfiability problem.
    If this can be done without intractable blowup in the size of the problem, then efficient SAT solvers can be applied~\cite{SAT}.
    If such a conversion is not possible, it may be possible to convert the spec into a first-order logic satisfiability problem, which permits the use of SMT solvers~\cite{SMT}.
    \item \userintheloop{} \textbf{User-in-the-loop:} rather than synthesize a program automatically, involve a person to help the synthesizer make decisions about what different parts of the output program should look like \rev{(e.g. sparse user scribbles to guide the decomposition of a texture image into multiple procedural components~\cite{hu2022inverse})}.
\end{itemize}
Finally, it is worth discussing the search algorithm's termination conditions.
The enumeration-based and constraint-based algorithms will terminate once they have explored all possible programs; in practice, reaching this termination condition can require an intractable amount of computing time.
Furthermore, the stochastic search algorithms have no natural termination condition: they can, in principle, be run forever and may continue to produce better programs.
With the above in mind, most program synthesizers treat their search algorithms as \emph{anytime algorithms}, terminating and returning the best program(s) found so far once some fixed compute budget is exceeded.

\noindent
\newline
\solverguide{} \textbf{Neural Guidance:}
Many of the search algorithms discussed above have steps that can benefit from some form of heuristic guidance: beam search needs to rank the set of possible next search states to explore; branch-and-bound needs to bound the possible score of part of the search tree before deciding whether to expand it; stochastic search methods can use non-uniform distributions for their random changes.
In prior work thus far, this is the most common place for machine learning to come into play: \emph{learning} a search guidance heuristic.
This heuristic often takes the form of ``suggestions for what to add to the program next.''
As such, autoregressive language models are the most widely-used neural guidance architectures, particularly transformers~\cite{GPT3Paper}.
Some past work also uses pointer networks~\cite{PointerNetworks}, so that the neural guide can suggest repetitions of prior parts of the program (e.g. variable re-use).
Various forms of graph convolution networks~\cite{NeuralMessagePassing,RelationalInductiveBiases} have also been applied, though these have largely fallen out of favor since the advent of transformers and other attention-based models.
For systems that ``synthesize'' programs by retrieving them from a database, 
a learned deep feature space (where similarity search is performed) 
acts as a form of neural guidance.
Section~\ref{sec:learningalgs} discusses how to train these guidance networks.

We note that some models in prior work perform program synthesis via a single forward pass of such a guide network, i.e. there ``is no search'' being performed.
In our categorization of prior work, we label such models as either using \enumeration{} \textbf{Explicit Enumeration} with neural guidance if the network takes the argmax over discrete program choices (i.e. beam search with a beam width of one) or \stochastic{} \textbf{Stochastic Search} with neural guidance if it randomly samples them (i.e. Sequential Monte Carlo with a single particle).

\subsection{Execution Engine}
\label{sec:execengine}

When the neurosymbolic model has produced a program, its job is not yet finished---the program must then be \emph{executed} to produce a visual output.
The process of executing a program can take different forms:
\begin{itemize}
    \item \direct{} \textbf{Direct:} the program is linearized and its operations executed in sequential order. This type of execution is most common; it corresponds to the execution model used by most (non-parallel) general-purpose programming languages (whether interpreted or compiled).
    \item \solver{} \textbf{Solver:} in some languages, executing the program requires solving a search, optimization, or constraint-satisfaction problem.
    We think it worthwhile to assign such execution engines a distinct point in the neurosymbolic model design space for two reasons.
    First, using them can make the neural components of the model harder to learn, if learning requires computing gradients w.r.t. the execution engine (see Section~\ref{sec:learningalgs}).  
    Second, just as search during program \emph{synthesis} presents opportunities for neural guidance, so too does search during program \emph{execution}.
    \item \proxy{} \textbf{Learned Proxy:} instead of a hardcoded execution engine (either direct or solver-based), one may instead opt to learn a mapping between program text and program output.
    As with any learned function, such an executor can only be approximately correct; however, it can have advantages for learning the other neural components of a neurosymbolic model (see Section~\ref{sec:learningalgs}).
\end{itemize}

\subsection{\neuralpostproc{} Optional Neural Postprocessing}
\label{sec:neuralrefine}

There is one final place in a neurosymbolic model where learning can come into play: applying learned postprocessing to the visual data output by a synthesized program.
Such a step can help close the ``reality gap'' between procedurally-generated data and data which is acquired from the world via sensors (as the former is almost always ``cleaner'' than the latter).

\subsection{Learning Algorithm}
\label{sec:learningalgs}

At this point, we have completed our tour through the pipeline of a neurosymbolic model: from input specification to final visual outputs.
However, there is still one critical design axis to discuss; one that touches multiple parts of the neurosymbolic model: how are its learnable components trained?

\paragraph*{End-to-end \rev{(Unsupervised)} Training:}
The most obvious approach is to train all the neural components of the model at once by backpropagating the final task loss function through the entire model.
\rev{This is desirable because it is conceptually simple and does not require any ground-truth supervision for intermediate steps in the pipeline---most notably, no pairs of (input space, ground-truth program) need be provided.}
As part of this process, loss gradients must be backpropagated through every choice the synthesizer guide network makes in constructing the program.
If those choices represent continuous values in the program, then this backpropagation is well-defined.
However, programs also typically exhibit complex structure dictated by discrete choices (e.g. branching decisions, types of primitives to use).
Gradients through such discrete choices are not well-defined.
There are at least two approaches to work around this conundrum:
\begin{itemize}
    \item \smoothrelax{} \textbf{Smooth Relaxation:} defining a continuous relaxation of discrete structural program choices, so that a single program can smoothly blend between different discrete structures. This strategy also requires the design of a program executor which can produce a correspondingly (and semantically meaningful) smoothed output in the program's output domain. Designing such relaxations (and their executors) is challenging, time-consuming, and domain-specific: the creation of a new relaxation typically warrants a research publication\rev{~\cite{kania2020ucsgnet,ren2021csg}}.
    Some methods use end-to-end differentiable learning to train only the continuous parts of their model; for simplicity, we also group such methods under this category.
    \item \rl{} \textbf{Reinforcement Learning:} specifically, policy gradient RL with a score function gradient estimator (e.g. REINFORCE~\cite{REINFORCE} or PPO~\cite{PPO}).
    These algorithms stochastically estimate gradients and do not require any part of the model to be differentiable (aside from the neural networks themselves).
    While unbiased, these gradient estimates can have extremely high variance, resulting in training that converges slowly, to a poor local optimum, or not at all.
\end{itemize}
It is worth noting that using a solver-based executor in end-to-end training requires running an iterative solver loop within the outer iterative optimization loop of neurosymbolic model training, which will be slow, not to mention likely not (easily) differentiable.
Thus, it is usually best to use a learned proxy executor instead, whenever the approximation error from doing so is acceptable.
On a related note: in the end-to-end learning paradigm, a learned proxy executor may be incompatible with having neural primitives in the program (as gradients may never propagate through the actual executor in order to train the neural primitives).

\paragraph*{Modular Training:}
The alternative to end-to-end training is to train different stages of the neurosymbolic model separately.
\rev{This often, but not always, necessitates supervision in the form of ground-truth outputs for the stage of the model being trained. For some stages, this supervision is easier to come by than others (e.g. a ``ground-truth program'' for a given input spec may be difficult to obtain).
In the remaining paragraphs in this section, we describe how modular training can be applied to different stages of the neurosymbolic modeling pipeline from Figure~\ref{fig:designspace}:
}

\emph{\rev{Modular training for a program synthesizer guidance network}:}
If one can produce paired (task spec, program) data, then this network can be trained to maximize the likelihood of that data.
There are several ways such data could be obtained/constructed:
\begin{itemize}
    \item \people{} \textbf{People:} in the best-case scenario, one has access to a curated set of programs written by people to satisfy a particular task specification. Such data is rare, but examples do exist; see Section~\ref{sec:datasets}.
    \item \synthetic{} \textbf{Synthetic Programs:} in the absence of ``ground-truth'' programs created by people, one can instead write a procedure to create a variety of synthetic programs along with task specs that they satisfy. 
    Such a procedure could be as simple as sampling random derivations from the DSL grammar, or it could make use of domain knowledge.
    Models trained on such data typically do not generalize well to actual task specs of interest, so they are usually only used to provide an initialization for another learning method.
    \item \bootstrap{} \textbf{Bootstrapping:} starting from a model pretrained on synthetic data, some algorithms can improve the model by iteratively training on variations of the model's own predictions~\cite{jones2022PLAD}.
\end{itemize}
Alternatively, the \smoothrelax{} \textbf{Smooth Relaxation} or \rl{} \textbf{Reinforcement Learning} approaches can be used here (typically also initialized with maximum likelihood pretraining on synthetic data).

\emph{\rev{Modular training for a learned proxy execution engine or a solver guidance network}:}
Both types of networks can be trained using supervised learning on (program, program output) pairs.
\rev{As with training a synthesizer guidance network, such data can be obtained by by creating and executing synthetic programs; however, networks trained on such data may not generalize to ``real'' data presented to the model at inference time.}

\emph{\rev{Modular training for a neural refinement network}:}
Refining the output of a procedural modeling program can be phrased as a type of \emph{domain translation} problem: transforming the output of the program from the ``looks procedurally-generated'' appearance domain to the ``looks realistic'' appearance domain.
Since one typically does not have access to paired (procedural, real) data, unpaired domain translation approaches are useful here~\cite{CycleGAN2017,UNIT-DDPM}.

\emph{\rev{Modular training for neural language primitives}:}
\rev{We are unaware of any existing work which learns neural language primitives using a modular training paradigm.}
If one could obtain (input, output) example pairs for the part of the program's execution that a neural primitive should perform, then such a primitive could be trained with supervised learning.
\rev{It is less clear} how to learn neural primitives in the absence of such data (i.e. when only the \emph{entire} program's desired output is available as a training signal) within a modular training regime; \rev{this remains an open problem for future work.}

%% file: 04-datasets.tex
\section{Datasets for Neurosymbolic Modeling}
\label{sec:datasets}

\begin{figure*}
    \centering
    \includegraphics[width=\linewidth]{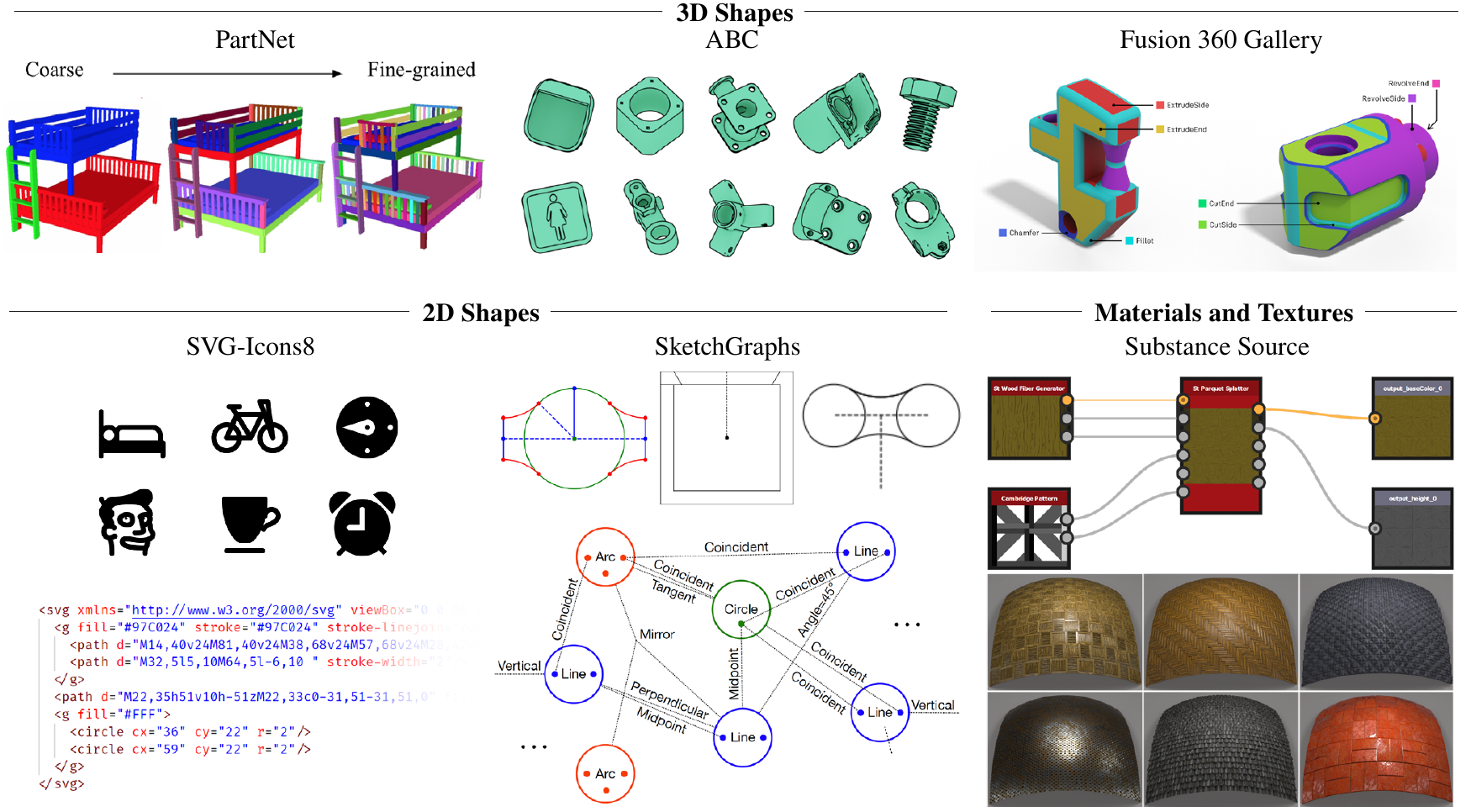}
    \caption{Examples of datasets that provide structured data that is more amenable to neurosymbolic models. We show examples of datasets for 2D shapes, 3D shapes, and materials. All datasets provide some form of structural information, such as part decompositions, construction sequences, or geometric relationships between parts.}
    \label{fig:datasets}
\end{figure*}


Several neurosymbolic methods that we discuss in this paper require some form of supervision. For some of the methods, a traditional dataset such as a set of 3D shapes or 2D images is sufficient. Other methods, however, require more structured data as supervision that is more amenable to neurosymbolic modeling. Here, we give an overview of datasets that provide a more symbolic representation of the data, for example in the form of a program or a high-level structure. Examples of several of these datasets are shown in Figure~\ref{fig:datasets}.

\paragraph*{2D Shapes:} Several datasets provide higher-level descriptions of 2D shapes, such as their composition from parametric primitives, or the set of strokes that were used to create a sketch. CAD programs are a possible source for this type of data. They typically provide an interface for authoring 2D engineering sketches. These sketches are represented by parametric primitives, such as circles, lines, and arcs, and spatial relationships between primitives, such as adjacency or tangency. Two datasets provide 2D sketches from CAD programs. \emph{SketchGraphs}~\cite{Ari2020} provides $15$ million 2D sketches from OnShape~\cite{OnShape}; however, it is known to contain a significant amount of duplication in the data samples. The \emph{CAD as Language} dataset~\cite{Ganin2021ComputerAidedDA} contains roughly $4.5$ million sketches, also scraped from OnShape, and tries to address some of these issues.
%
Wong et al.~\cite{wong2022identifying} introduce two procedurally generated datasets of 2D drawings. The first contains $1$k drawings from categories such as furniture, vehicles, and gadgets, each consisting of simple primitives like lines and curves. The second dataset contains $1$k sketches of simple 2D buildings such as towers, bridges, and houses composed of blue and red blocks.
Several datasets provide high-level information about strokes in sketches, in the form of vectorized strokes and the order in which they were drawn. The \emph{OmniGlot} dataset~\cite{Brenden:2015:Omniglot, lake2019omniglot} contains $1.6$k handwritten characters from $50$ different alphabets, each drawn by $20$ different people. The \emph{Quick, Draw!}~\cite{SketchRNN, QuickDraw} dataset contains $50$ million drawings across 345 categories.
Scalable Vector Graphics (SVG) is a popular format for vector graphic content and can be seen as a simple programming language to generate parametric shapes. Several datasets provide 2D shapes in this format. DeepSVG~\cite{DeepSVG} introduced the \emph{SVG-Icons8} dataset, which consists of $100$k icons obtained from the Icons8~\cite{Icons8} website. The icons depict various categories of objects, such as beds, bicycles, and cups. \emph{SVG-Fonts}~\cite{lopes2019learned} provides $14$ million vectorized characters in different fonts.

\paragraph*{3D Shapes:} Three datasets provide higher-level structural information for 3D shapes, in addition to meshes. 
\emph{PartNet}~\cite{Mo_2019_CVPR} is a dataset that provides $26$k synthetic 3D shapes across several categories of manufactured objects typically found in interior scenes, such as chairs, tables, and lamps. These shapes originate from the ShapeNet~\cite{shapenet2015} dataset, and PartNet additionally provides a hierarchical decomposition of each shape into its constituent parts. For example, a chair could be first decomposed into backrest, seat, and base parts, which could be further decomposed into smaller constituent parts. This decomposition is consistent across all shapes of a category.
The \emph{ABC dataset}~\cite{Koch:2019:ABC} contains a set of one million mechanical parts and assemblies that were created with the OnShape~\cite{OnShape} CAD modeling software. In addition to the meshes, the ABC dataset provides the parametric boundary curves and surfaces that were used to author each shape.
The \emph{Fusion 360 Gallery} dataset~\cite{Fusion360Gallery, joinable} is based on $20$k designs created with the Autodesk Fusion 360 CAD modelling software~\cite{Fusion360}. In addition to meshes, several different types of high-level information are provided, including a hierarchical decomposition into sub-assemblies, joints between parts, holes, contact surfaces, `sketch and extrude' construction sequences of some designs, and a segmentation of each shape into the modeling operations used to create each part of the surface.

\paragraph*{Materials and Textures:}
Procedural materials used in practice are typically defined as node graphs, which are visual representations of functional programs. Several recent neurosymbolic methods have experimented with synthesizing these node graphs. Three main data sources are available, only two of which are public. Substance Source~\cite{Substance_Source} is a dataset of roughly $7$k node graphs that were generated by professional material artists. It was used as the training set by MatFormer~\cite{guerrero2022matformer}, but is not publicly available. A public alternative is the Substance 3D Community Assets~\cite{Substance_Community}, which is a website containing publicly available node graphs created by users of Substance 3D Designer~\cite{Substance3D}. However, the graphs on this website have not been assembled into a dataset yet. Nvidia vMaterials~\cite{VMaterials} is a dataset of over $800$ realistic materials that also provide node graphs.

%% file: 05-applications.tex
\definecolor{C1}{RGB}{251,229,214}
\definecolor{C2}{RGB}{226,240,217}
\definecolor{C3}{RGB}{218,227,243}
\definecolor{C4}{RGB}{238,224,248}
\definecolor{C5}{RGB}{255,210,211}
\definecolor{C6}{RGB}{255,242,204}

\newcolumntype{a}{>{\columncolor{C1}}c}
\newcolumntype{b}{>{\columncolor{C2}}c}
\newcolumntype{d}{>{\columncolor{C3}}c}
\newcolumntype{e}{>{\columncolor{C4}}c}
\newcolumntype{f}{>{\columncolor{C5}}c}
\newcolumntype{g}{>{\columncolor{C6}}c}

\section{Application: 2D Shapes}
\label{sec:app_layout}

\begin{table*}[t!]
    \centering    
    \scriptsize
    \begin{tabular}{@{}laabbbddefgg@{}}
        & \multicolumn{2}{c}{\textbf{Task Spec}} & \multicolumn{3}{c}{\textbf{DSL}} & \multicolumn{2}{c}{\textbf{Synthesizer}} & \multicolumn{1}{c}{\textbf{Execution}} & \multicolumn{1}{c}{\textbf{Refinement}} & \multicolumn{2}{c}{\textbf{Learning}} \\
        \cmidrule(lr){2-3} \cmidrule(lr){4-6} \cmidrule(lr){7-8}  \cmidrule(lr){11-12}
        \textbf{Method} & \textit{Input} & \textit{Output} & \textit{Paradigm} & \textit{Primitives} & \textit{Mutability} & \textit{Search} & \textit{Guidance} &  &  & \textit{End-to-end} & \textit{Modular}\\
		\midrule
		Para et al.~\cite{para2021generative} & \visualexamples{} & \progdist{} & \constraint{} & \hardcodedprims{} & \fixed{} & \stochastic{} & \synthguide{} & \solver{} & & & \people{} \\
		PlanIT~\cite{PlanIT} & \visualexamples{} & \progdist{} & \constraint{} & \hardcodedprims{} & \fixed{} & \stochastic{} & \synthguide{} & \solver{} \solverguide{} & & & \people{} \\
		CurveGen~\cite{willis2021engineering} & \visualexamples{} & \progdist{}{} & \imperative{} & \hardcodedprims{} & \fixed{} & \stochastic{} & \synthguide{} & \direct{} & & & \people{} \\
		SketchGen~\cite{para2021sketchgen} & \visualexamples{} & \progdist{}{} & \constraint{} & \hardcodedprims{} & \fixed{} & \stochastic{} & \synthguide{} & \solver{} & & & \people{} \\
		CAD as Language~\cite{Ganin2021ComputerAidedDA} & \visualexamples{} \visualtarget{} & \progdist{} \detprog{}  & \constraint{} & \hardcodedprims{} & \fixed{} & \stochastic{} & \synthguide{} & \solver{} & & & \people{} \\
		Vitruvion~\cite{seff2021vitruvion} & \visualexamples{} \visualtarget{} & \progdist{} \detprog{} & \constraint{} & \hardcodedprims{} & \fixed{} & \stochastic{} & \synthguide{} & \solver{} & & & \people{} \\
		SkexGen~\cite{xu2022skexgen} & \visualexamples{} & \progdist{} & \imperative{} & \hardcodedprims{} & \fixed{} & \stochastic{} & \synthguide{} & \direct{} & & & \people{} \\
        SVG-VAE~\cite{lopes2019learned} & \visualexamples{} & \progdist{} & \imperative{} & \hardcodedprims{} & \fixed{} & \stochastic{} & \synthguide{} & \direct{} & & & \people{} \\
        DeepSVG~\cite{DeepSVG} & \visualexamples{} & \progdist{} & \imperative{} & \hardcodedprims{} & \fixed{} & \stochastic{} & \synthguide{} & \direct{} & & & \people{} \\
        Im2Vec~\cite{reddy2021im2vec} & \visualexamples{} \visualtarget{} & \progdist{} \detprog{} & \imperative{} & \hardcodedprims{} & \fixed{} & \stochastic{} & \synthguide{} & \direct{} & & \smoothrelax{} & \\
        DeepVecFont~\cite{wang2021deepvecfont} & \visualexamples{} & \progdist{} & \imperative{} & \hardcodedprims{} & \fixed{}  & \stochastic{} & \synthguide{} & \direct{} & & \smoothrelax{} & \people{} \\
        Ellis et al.~\cite{ellis2018learning} & \visualtarget{} & \detprog{} & \imperative{} & \hardcodedprims{} & \fixed{} & \stochastic{} \constraintsat{} & \synthguide{} & \direct{} & & & \synthetic{} \\
        Guo et al.~\cite{guo2020inverse} & \visualtarget{} & \detprog{} & \functional{} & \neuralprims{} & \fixed{} & \enumeration{} & & \direct{} & & & \synthetic{} \\
        Ellis et al.~\cite{ellis2019repl} & \visualtarget{} & \detprog{} & \imperative{} & \hardcodedprims{} & \fixed{} & \stochastic{} & \synthguide{} & \direct{} & & \rl{} & \synthetic{} \\
        SPIRAL~\cite{ganin2018synthesizing} & \visualexamples{} \visualtarget{} & \progdist{} \detprog{} & \imperative{} & \hardcodedprims{} & \fixed{} & \stochastic{} & \synthguide{} & \direct{} & & \rl{} & \\
        DreamCoder~\cite{DreamCoder} & \visualtarget{} & \detprog{} & \functional{} & \hardcodedprims{} & \evolving{} & \enumeration{} & \synthguide{} & \direct{} & & & \bootstrap{} \synthetic\\
        \rev{Yang et al.~\cite{SketchConcepts}} & \visualtarget{} & \detprog{} & \constraint{} & \hardcodedprims{} & \evolving{} & \enumeration{} & \synthguide{} & \solver{} & & \smoothrelax{} & \\
        \bottomrule
    \end{tabular}
    \iconlegend
    \caption{A summary of the work on neurosymbolic 2D shape modeling discussed in Section~\ref{sec:app_layout}, where each approach is situated in our design space.}
    \label{tab:2d_methods}
\end{table*}

From freeform sketches of ideas, to detailed technical drawings, 2D shapes are a building block of human communication. Augmenting peoples' ability to communicate and express concepts is at the heart of research in neurosymbolic 2D shape creation. 2D shapes can also play a foundational role in the creation of 3D shapes, as discussed in Section~\ref{sec:app_shapes}. In this section, we review the application areas of layout generation, engineering sketch generation, vector graphics generation, and inverse 2D graphics.
Table~\ref{tab:2d_methods} situations each prior work we discuss within our design space.

\subsection{Layout Generation}

The problem of layout generation arises in a number of domains, including graphic design layout, floor plan synthesis, and furniture layout. Common across these domains is the spatial and topological arrangement of layout primitives (potentially to meet a set of constraints). In floor plan synthesis, for example, an adjacency constraint may dictate that the master bedroom reside next to an en suite bathroom, forming a spatial and topological relationship between the two rooms. A significant amount of research related to layout generation exists; here, we cover only select works that fall under the neurosymbolic modeling umbrella.


\begin{figure}[t!]
    \centering
    \begin{tabular}{cc}
        \includegraphics[width=0.45\linewidth]{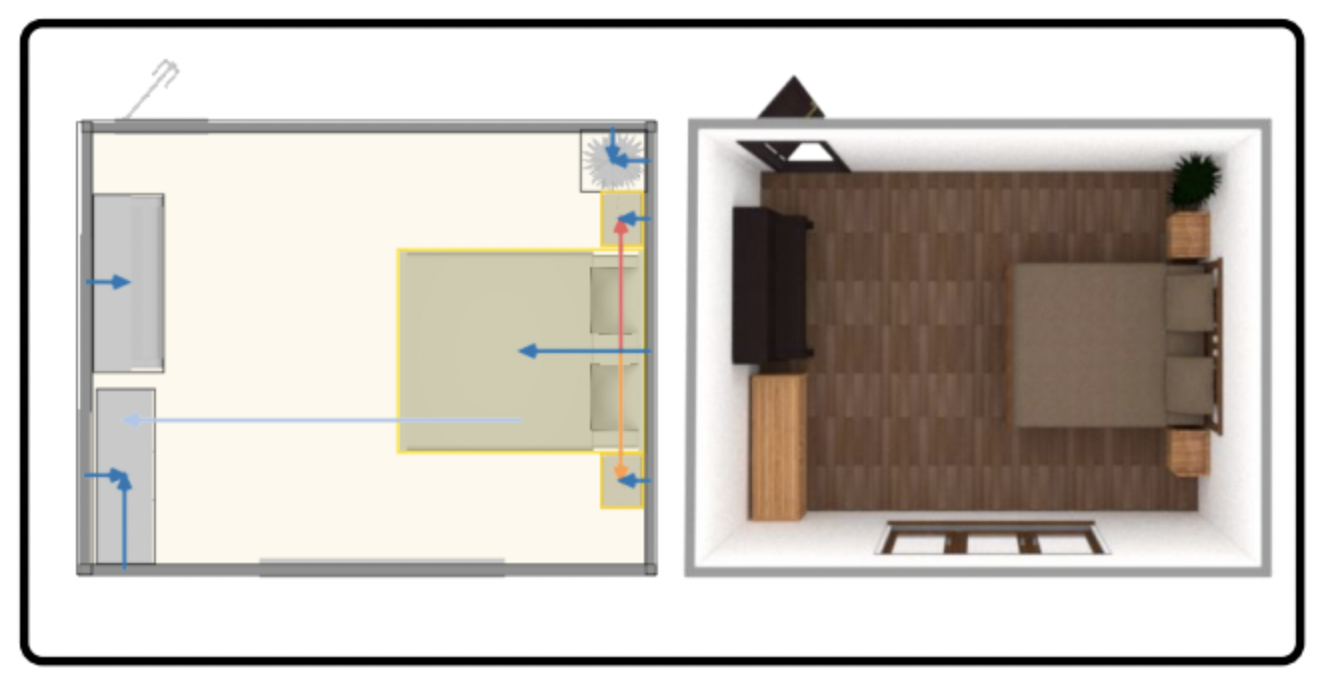} &
        \includegraphics[width=0.45\linewidth]{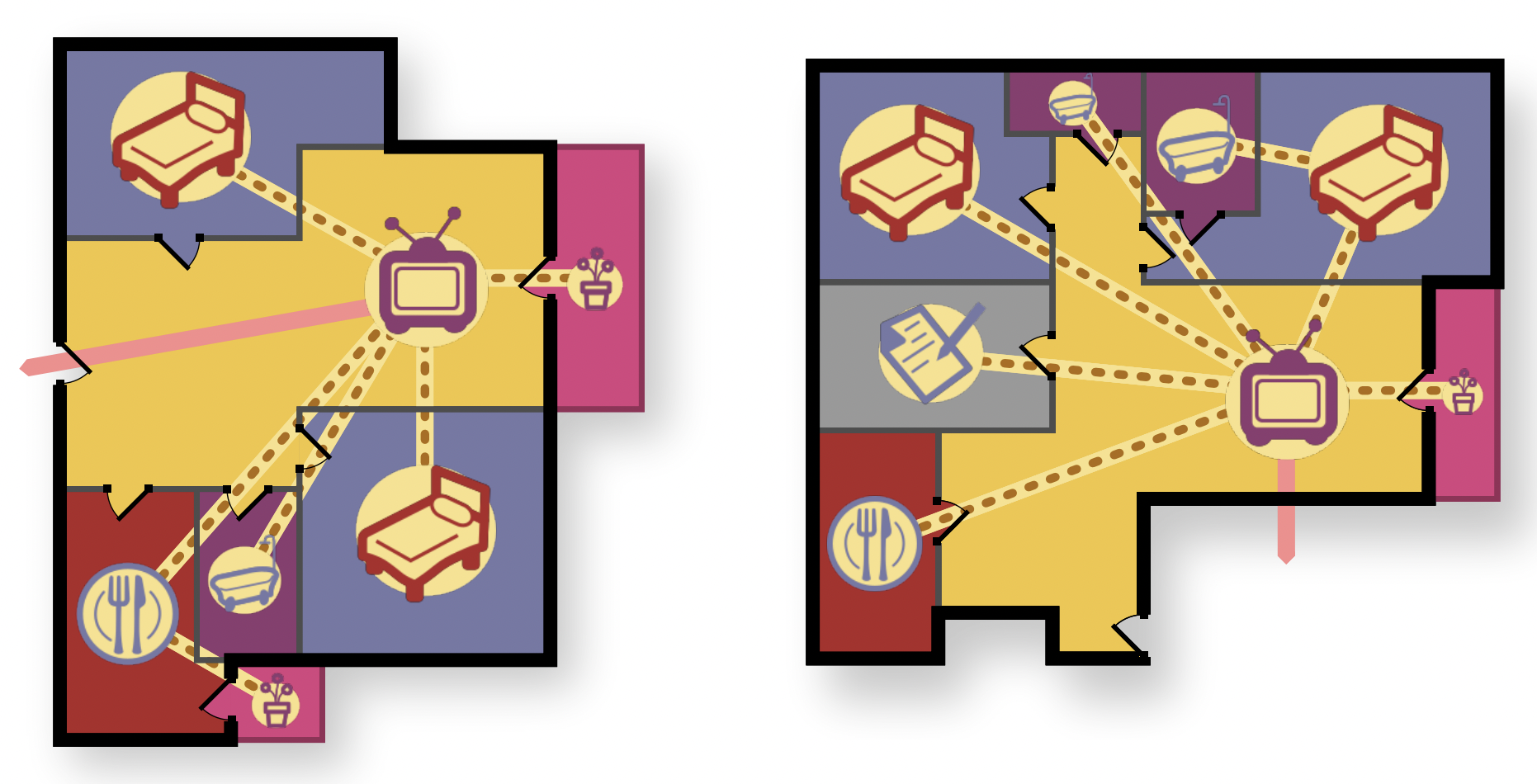}
    \end{tabular}
    \caption{
    Neurosymbolic layout generation via learning to synthesize layout constraint programs.
    \emph{Left:} A graph neural network synthesizes a constraint graph specifying furniture spatial relationships; a neurally-guided search then finds object placements that satisfy those constraints~\cite{PlanIT}.
    \emph{Right:} A transformer + pointer network architecture synthesizes a floor plan layout constraint program, which is then solved to produce a floor plan~\cite{para2021generative}.
    }
    \label{fig:layoutgen}
\end{figure}

Floor plan synthesis involves the generation of realistic floor plans to meet a series of architectural constraints.
Para et al.~\cite{para2021generative} introduce a neurosymbolic approach for this task. First, a transformer network is used to generate constraints on the parameters of layout elements (e.g. room type and width/height ranges); second, relationship constraints (e.g. door or wall edges) are generated between elements using pointer networks.
These constraints together make up a constraint program which is then executed (i.e. solved) to generate a spatial layout consistent with the constraints (Figure~\ref{fig:layoutgen} right).
The goal of furniture layout, or more generally scene generation, is to position objects (e.g. furniture) in a scene (e.g. an empty bedroom with a given shape) in a realistic manner. 
PlanIT~\cite{PlanIT} divides the problem into a planning phase and an instantiation phase. In the planning phase, a deep graph convolutional generative model is used to synthesize relation graphs which form a constraint program: nodes indicate objects to be placed in the scenes, and edges indicate spatial relationship constraints between them.
In the synthesis phase, a backtracking search is used to find placements for the objects implied by the graph nodes.
This search is guided by a convolutional network that operates on top-down views of the partial scene (Figure~\ref{fig:layoutgen} left).

\begin{figure}[t]
    \centering
    \includegraphics[width=\linewidth]{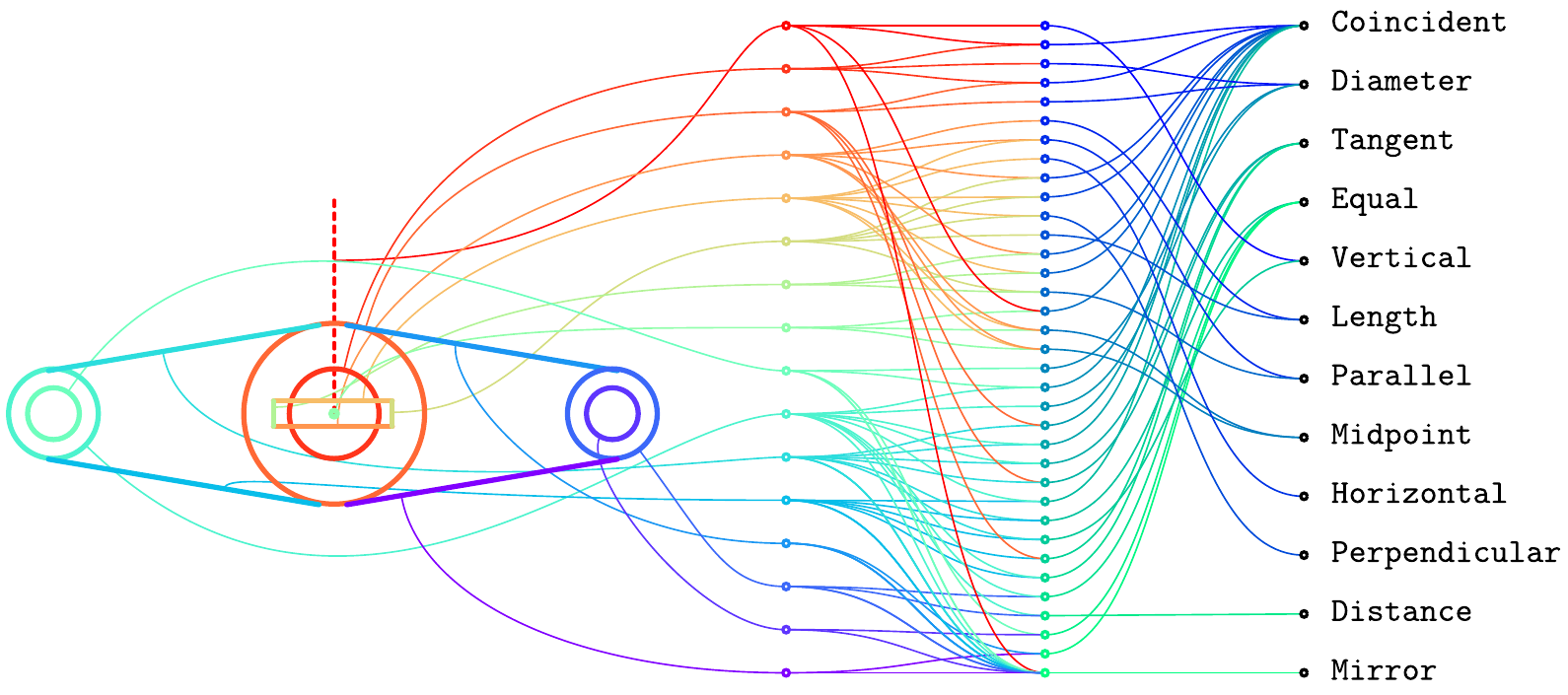}
    \caption{
       Engineering sketches form the 2D basis of parametric CAD. Geometric primitives, such as circles and lines (left), are represented as nodes in the first column. Sketch constraints are applied to the geometric primitives and represented as nodes in the second column. The third column lists the constraint types in order of frequency. The order of nodes (top to bottom) follows the generation order of the learned model from Ganin et al.~\cite{Ganin2021ComputerAidedDA}.
    }
    \label{fig:cad_as_language}
\end{figure}

\subsection{Engineering Sketch Generation}
Software for the creation of 2D engineering sketches dates back to the very first CAD system~\cite{SketchPad}.
Engineering sketches form the 2D basis of parametric CAD, the foremost 3D modeling paradigm used to design manufactured objects from automobile parts, to electronic devices, to furniture. 
Engineering sketches (as shown in Figure~\ref{fig:cad_as_language}) consist of composite curves made up of 2D geometric primitives (e.g. lines, arcs, circles), topological information about how these primitives connect together, and constraints defined using topology (e.g. coincidence, tangency, symmetry).
Essentially, an engineering sketch is a program: given a change to its parameters, the program can be re-executed to produce new 2D geometry (which satisfies any constraints).
The ability to generate high-quality engineering sketches is an enabling technology for the automatic generation of parametric CAD files suitable for manufacturing. The availability of large-scale engineering sketch datasets (see Section~\ref{sec:datasets}) has enabled learning-based approaches to engineering sketch generation. A number of concurrent works~\cite{willis2021engineering, para2021sketchgen, Ganin2021ComputerAidedDA, seff2021vitruvion} approach the task of engineering sketch generation by treating the sketch as a sequential language that can be modeled using Transformers, either with or without constraints. Providing control over the generation of engineering sketches is another outstanding challenge, as designers need ways to influence the generated shape.  
One approach is to condition the network on user-provided images~\cite{Ganin2021ComputerAidedDA} or hand-drawn sketches~\cite{seff2021vitruvion}. Another approach, introduced by SkexGen~\cite{xu2022skexgen}, is the use of codebooks~\cite{razavi2019generating} to separate control of sketch geometry and topology into learned codes that can be selected to guide the generation of topologically or geometrically similar shapes. \rev{Yang et al.~\cite{yang2022} learn modular `concepts' from engineering sketch graphs to capture repetitive design patterns and aid with image-conditioned generation and auto-completion tasks. Transformer-based generation approaches have several limitations. Spatial resolution is currently limited and addressed by using quantized vertex positions on a  8 bit grid or smaller. Higher spatial resolution allows more accurate sketches to be generated but vastly increases the size of the prediction space. Another limitation, due to the well known issue of modeling long sequences with Transformers, is generating complex sketches with higher numbers of curves.}

\begin{figure}[t]
    \centering
    \includegraphics[width=\linewidth]{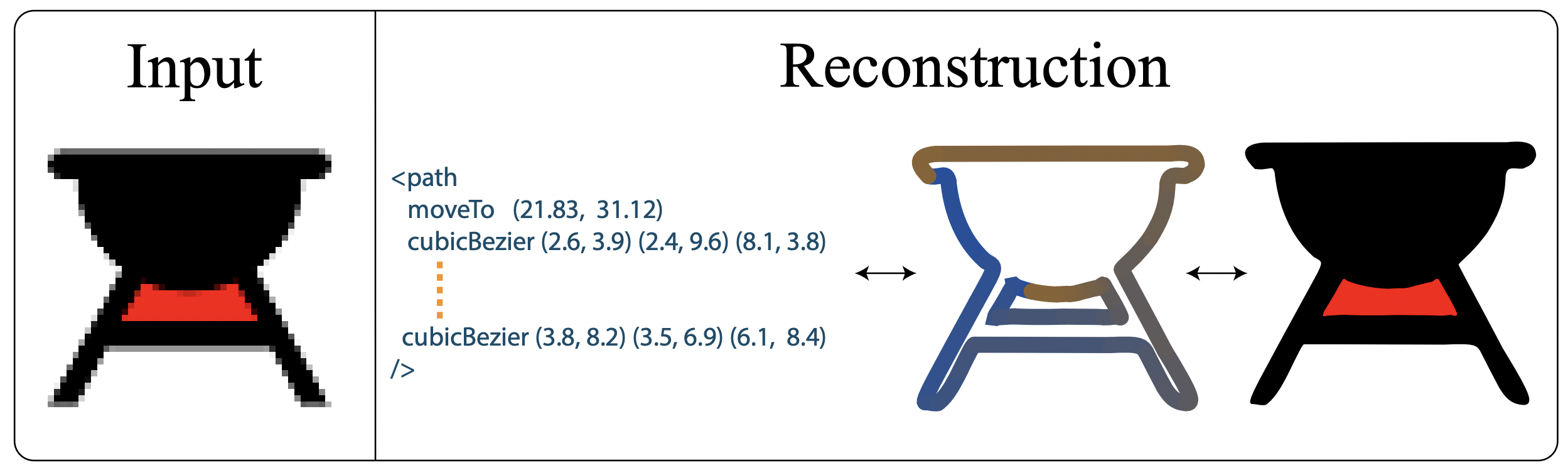}
    \caption{
    The Im2Vec system learns a neural network that can reconstruct, synthesize, and interpolate between vector graphics programs~\cite{reddy2021im2vec}.
    It uses an end-to-end training setup with a differentiable vector graphics rasterizer, avoiding the need for ground-truth SVG programs as training data.
    }
    \label{fig:im2vec}
\end{figure}

\subsection{Vector Graphics Generation}
Adjacent to engineering sketches is the domain of vector graphics, which also makes use of 2D geometric primitives, such as lines and B\'ezier curves, to represent scalable graphical artwork.
An SVG file is essentially a simple program, which uses compositions of parametrized functions (e.g. \texttt{moveTo}, \texttt{lineTo}, \texttt{cubicBezier}) to produce geometry.
 SVG-VAE~\cite{lopes2019learned} was one of the first deep learning-based approaches to generate vector graphics using an LSTM-based VAE trained on font characters. DeepSVG~\cite{DeepSVG} demonstrated both font character and icon generation using a non-autoregressive Transformer-based architecture. Although promising, both works produce results lacking the regularities found in human-designed vector graphics, such as symmetric, concentric, or tangent curves. To address this limitation, recent work has leveraged alternate representations to sequences of vector graphic primitives. DeepVecFont~\cite{wang2021deepvecfont} uses a hybrid raster/vector representation where the raster graphic is used as a supervision signal to improve the vector program output. \rev{This approach combines the benefits of learning human-designed curve topology from vector supervision (i.e. how curves are connected together) and curve geometry from raster supervision (i.e. the location of the curves). As vector graphics do not conform well to a grid structure, like engineering sketches do, higher spatial resolution is required. The geometric loss provided by raster supervision appears to be critical at this time to get good visual results. }

\begin{figure}[t]
    \centering
    \includegraphics[width=\linewidth]{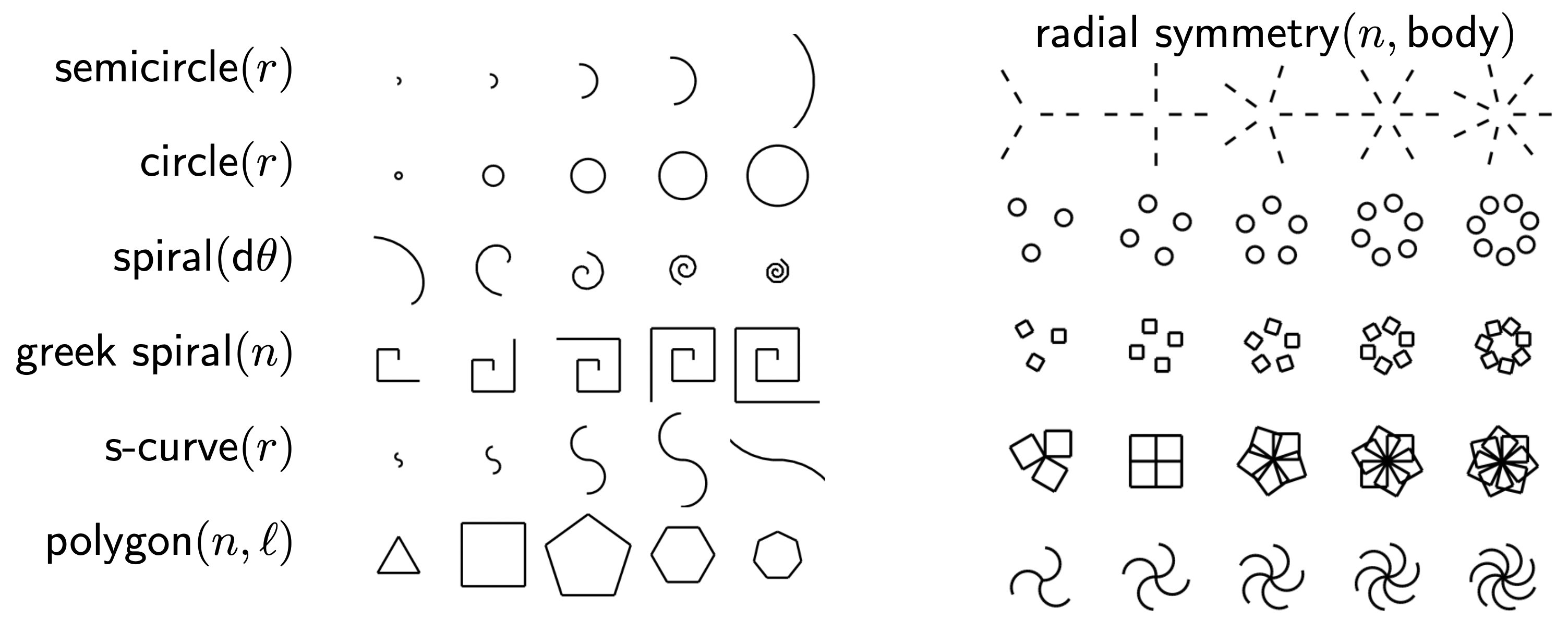}
    \caption{
       Example learned library routines using DreamCoder~\cite{DreamCoder} on a LOGO graphics task. The learned routines include both parametric routines for drawing families of curves (left) as well as higher-order functions that take entire sub-programs as input (right). 
    }
    \label{fig:dreamcoder}
\end{figure}

\subsection{Inverse 2D Graphics}
In addition to the generation of graphics, the inverse problem of recovering a 2D shape program from approximate input, such as an image or hand-drawn sketch, has been studied in recent literature. Ellis et al.~\cite{ellis2018learning} introduce an approach to convert simple hand drawings into a program representation that captures the regularities commonly found in human design, such as symmetry, repetition, and structural reuse.
This approach uses neurally-guided search to extract primitive graphical elements from the input sketch, followed by constraint-based program synthesis to find a program that generates those primitives.
Guo et al.~\cite{guo2020inverse} show how neural guidance can be used to learn an L-system representation of pixel images with branching structures, as found in trees and other natural patterns.
Ellis et al.~\cite{ellis2019repl} use reinforcement learning to synthesize simple 2D CAD programs that reproduce an input shape;
similarly, 
SPIRAL~\cite{ganin2018synthesizing} uses RL and adversarial learning to synthesize simple painting programs that reproduce an input image.
\rev{Im2Vec~\cite{reddy2021im2vec}, shown in Figure~\ref{fig:im2vec}, 
uses a differentiable rasterization pipeline and reconstruction loss between the input image and the rasterized generated vector graphic.}
DreamCoder~\cite{DreamCoder}, \rev{shown in Figure~\ref{fig:dreamcoder}}, can infer a LOGO graphics program for a target shape. It interleaves learning to infer programs via neurally guided search with learning new abstractions (i.e. subroutines) for its drawing language, which then makes it easier to infer programs for more complex targets.
DreamCoder's discovered abstractions capture complex concepts, produceing primitive shapes such as semi-circles and polygons, as well as regularities such as radial symmetry (Figure~\ref{fig:dreamcoder}).



\section{Application: 3D Shapes}
\label{sec:app_shapes}

\begin{table*}[t!]
    \centering    
    \scriptsize
    \begin{tabular}{@{}laabbbddefgg@{}}
        & \multicolumn{2}{c}{\textbf{Task Spec}} & \multicolumn{3}{c}{\textbf{DSL}} & \multicolumn{2}{c}{\textbf{Synthesizer}} & \multicolumn{1}{c}{\textbf{Execution}} & \multicolumn{1}{c}{\textbf{Refinement}} & \multicolumn{2}{c}{\textbf{Learning}} \\
        \cmidrule(lr){2-3} \cmidrule(lr){4-6} \cmidrule(lr){7-8}  \cmidrule(lr){11-12}
        \textbf{Method} & \textit{Input} & \textit{Output} & \textit{Paradigm} & \textit{Primitives} & \textit{Mutability} & \textit{Search} & \textit{Guidance} &  &  & \textit{End-to-end} & \textit{Modular}\\
		\midrule
		Fusion 360 Gallery\cite{Fusion360Gallery}            & \visualtarget   & \detprog  & \imperative & \hardcodedprims & \fixed & \enumeration & \synthguide & \direct & & & \people        \\
		CSGNet\cite{CSGNet}                                & \visualtarget   & \detprog  & \functional & \hardcodedprims & \fixed & \enumeration & \synthguide & \direct & & \rl & \synthetic\\
		Ellis et al.~\cite{ellis2019repl}                           & \visualtarget   & \detprog  & \imperative & \hardcodedprims & \fixed & \stochastic  & \synthguide & \direct & & \rl & \synthetic            \\
	    PLAD\cite{jones2022PLAD}                           & \visualtarget   & \detprog  & \functional \imperative & \hardcodedprims & \fixed & \enumeration & \synthguide & \direct & & & \bootstrap \synthetic \\  
		Shape2Prog\cite{tian2018learning}                            & \visualtarget   & \detprog  & \imperative & \hardcodedprims & \fixed & \enumeration & \synthguide & \proxy  & & \smoothrelax & \synthetic  \\
		ProGRIP\cite{deng2022unsupervised}       & \visualtarget   & \detprog  & \imperative & \neuralprims & \fixed & \enumeration & \synthguide & \proxy  & & \smoothrelax &  \\
		Sketch2CAD\cite{Li:2020:Sketch2CAD}                & \visualtarget   & \detprog  & \imperative & \hardcodedprims & \fixed & \enumeration & \synthguide & \direct & & &  \synthetic\\
		Free2CAD\cite{Li:2022:Free2CAD}                    & \visualtarget   & \detprog  & \imperative & \hardcodedprims & \fixed & \enumeration & \synthguide & \direct & & & \synthetic \\
		Zone Graphs\cite{zoneGraphs}                       & \visualtarget   & \detprog  & \imperative & \hardcodedprims & \fixed & \enumeration & \synthguide & \direct & & & \people    \\
		Lambourne et al. \cite{lambourne2022recon}     & \visualtarget   & \detprog  & \functional & \hardcodedprims & \fixed & \enumeration \retrieval & \synthguide & \direct & & \smoothrelax & \people  \\
		Point2Cyl\cite{uy-point2cyl-cvpr22}                & \visualtarget   & \detprog  & \imperative & \neuralprims & \fixed & \enumeration & \synthguide & \direct & & & \people \\
		CSG-Stump\cite{ren2021csg}                         & \visualtarget   & \detprog  & \functional & \hardcodedprims & \fixed & \enumeration & \synthguide & \direct & & \smoothrelax &  \\
		CAPRI-Net\cite{Yu_2022_CVPR}                       & \visualtarget   & \detprog  & \functional & \hardcodedprims & \fixed & \enumeration & \synthguide & \direct & & \smoothrelax &  \\
		ExtrudeNet\cite{ren2022extrude}                       & \visualtarget   & \detprog  & \functional & \hardcodedprims & \fixed & \enumeration & \synthguide & \direct & & \smoothrelax &  \\
		UCSGNet\cite{kania2020ucsgnet}                     & \visualtarget   & \detprog  & \functional & \hardcodedprims & \fixed & \enumeration & \synthguide & \direct & & \smoothrelax &  \\
		DeepCAD\cite{DeepCAD}                              & \visualexamples & \progdist & \imperative & \hardcodedprims & \fixed & \stochastic  & \synthguide & \direct & &  & \people            \\
		SkexGen\cite{xu2022skexgen}                        & \visualexamples & \progdist & \imperative & \hardcodedprims & \fixed & \stochastic  & \synthguide & \direct & &  & \people            \\
		ShapeAssembly\cite{jones2020SA}                    & \visualexamples & \progdist & \imperative & \hardcodedprims & \fixed & \stochastic  & \synthguide & \direct & \neuralpostproc{} & & \people    \\
		ShapeMOD\cite{jones2021shapeMOD}                   & \visualexamples & \progdist & \imperative & \hardcodedprims & \preprocessed & \stochastic & \synthguide & \direct & & & \people \\
		Ritchie et al. \cite{ExampleBasedProcMod}                         & \visualexamples & \probprog & \functional & \neuralprims{} & \fixed & \stochastic  &             & \direct & & &   *    \\
		
        \bottomrule
    \end{tabular}
    \iconlegend
    \caption{
    A summary of the work on neurosymbolic 3D shape modeling discussed in Section~\ref{sec:app_shapes}, where each approach is situated in our design space.
    * Ritchie et al.~\cite{ExampleBasedProcMod} use supervised training for their neural language primitives.
    }
    \label{tab:3d_methods}
\end{table*}

The demand for 3D models has never been higher. 
Applications from within entertainment \& gaming systems, to augmented \& virtual reality, and even those in vision \& robotics, all desire access to high-quality 3D objects.
Stakeholders in these areas are often not content with unstructured assets: their applications need shapes that are interactive \& editable, yet cover a wide range of outputs while maintaining high fidelity. 
Neurosymbolic methods for 3D shapes have been explored with these criteria in mind.
Table~\ref{tab:3d_methods} situates each prior work we discuss within our design space.

\subsection{Inferring 3D Shape Programs}

Many methods have investigated how to infer the underlying structure of a 3D object.
For instance, if one is able to find a program that is a good representation of an input shape, then the program structure can be used to analyze or manipulate the underlying 3D object. 
This is a sub-problem of program synthesis, where the input specification is a visual representation of a 3D object, which the inferred program's output must geometrically match. This problem is also sometimes referred to as visual program induction.


Within this area, a general approach is to linearize programs into sequences of tokens. 
Then a program inference network can be trained to autoregressively generate program tokens, conditioned on visual data. 
In Fusion 360 Gallery~\cite{Fusion360Gallery}, a model learns to represent 3D shapes as sequences of sketch and extrude CAD operations by training in a supervised fashion on human-written programs.
\rev{While supervised learning is the preferred strategy when program datasets are available, many domain+language combinations lack such information. }
When a program dataset is not available, approaches have investigated how to use reinforcement learning to reconstruct 3D shapes as sequences of commands. 
\begin{figure}[]
    \centering
    \includegraphics[width=\linewidth]{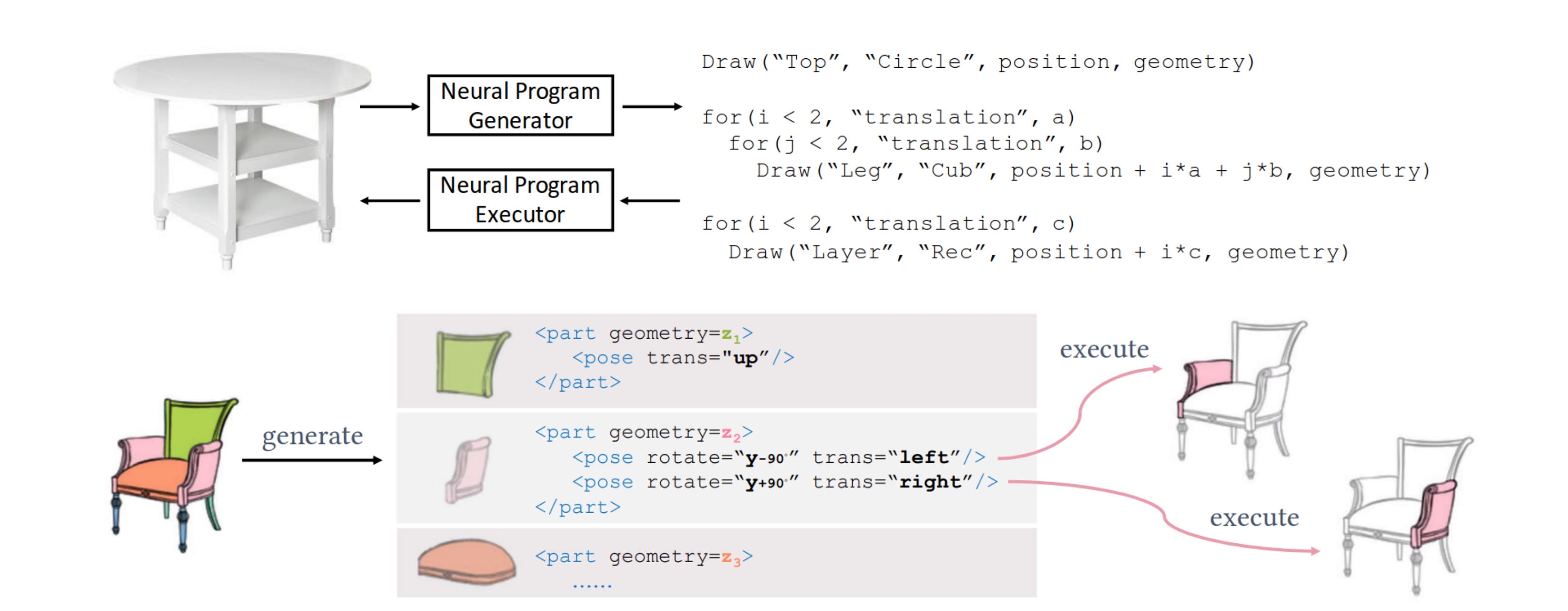}
    \caption{
       (Top) Shape2Prog~\cite{tian2018learning} learns how to infer visual programs that capture the structure of input shapes. The program generator is trained in an end-to-end fashion with a learned proxy executor. (Bottom) ProGRIP~\cite{deng2022unsupervised} improves the reconstruction fidelity of this approach by replacing hard-coded primitives with learned neural implicits.   
    }
    \label{fig:s2p}
\end{figure}

CSGNet~\cite{CSGNet} uses beam search guided by an auto-regressive inference network to find CSG programs that achieve a low-reconstruction error with respect to the input. 
The parameters of these programs can be further improved with a heuristic refinement step.
Ellis et al.~\cite{ellis2019repl} employ a Sequential Monte Carlo search guided by a policy network that synthesizes code and a value network that learns to assess the prospects of partial programs. 
They equip this search with access to a REPL (read-eval-print-loop), framing program search as a Markov Decision Process (MDP) where each network reasons over a \textit{state} represented by the executed output of partially constructed programs.
\rev{While policy gradient reinforcement learning offers a domain-agnostic solution that does not depend on a dataset of ground-truth programs, it is notoriously unstable, due to high variance gradients, which can hurt its convergence speed and performance.  
}

\begin{figure}[]
    \centering
    \includegraphics[width=\linewidth]{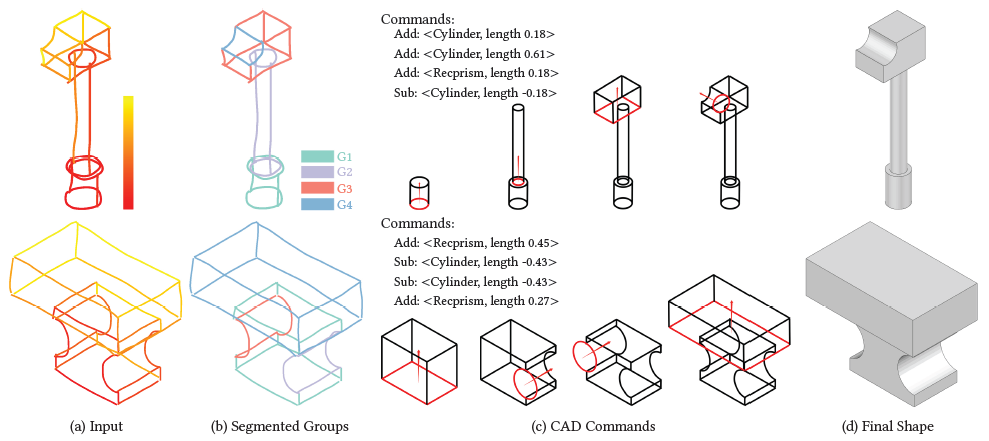}
    \caption{
        Free2CAD \cite{Li:2022:Free2CAD} parses freehand drawings into visual programs. The program is built up through an iterative procedure that alternates between (i) grouping related sketch strokes and (ii) searching for CAD operations that correspond to the segmented group.    
    }
    \label{fig:f2cad}
\end{figure}

\begin{figure*}[t!]
    \centering
    \includegraphics[width=0.8\linewidth]{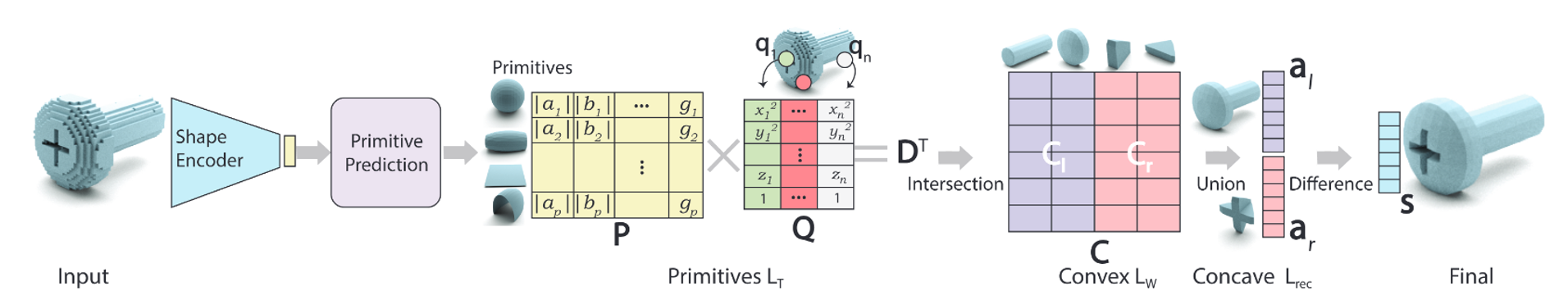}
    \caption{
        CAPRI-Net \cite{Yu_2022_CVPR} infers a CSG program that reconstructs an input shape. Its neural architecture acts as a CSG execution engine, which limits the types of program structures that it can discover but allows the system to train in an end-to-end fashion.
    }
    \label{fig:capri}
\end{figure*}

To deal with more complex DSLs for 3D shapes, PLAD~\cite{jones2022PLAD} introduced a self-supervised learning method that fine-tunes an inference model for a target domain of interest.
\rev{This bootstrapping approach maintains domain-generality, while avoiding the pitfalls of policy-gradient RL by training with maximum likelihood updates on approximately correct (shape, program) pairs.}
\rev{Alternatively, some methods have explored learning a network that acts as  a differentiable relaxation of a program executor: allowing the system to train in an end-to-end fashion with respect to a geometric loss.}
Shape2Prog~\cite{tian2018learning} uses a network that learns from synthetic data to act as a differentiable proxy for their execution engine 
\rev{(Figure \ref{fig:s2p}, Top)}. 
\rev{This approach has been recently extended in ProGRIP~\cite{deng2022unsupervised}
(Figure \ref{fig:s2p}, Bottom)}.
\rev{
ProGRIP replaces the hard-coded primitives used by Shape2Prog with learned neural primitives, which also removes the need to pretrain any proxy network with hand-crafted synthetic data.
}


\rev{Some prior work has developed shape program inference approaches specially tailored for their particular domain}.
When inferring a 3D CAD model conditioned on an input sketch, a network's noisy predictions can be regularized by mapping them to symbolic CAD operations with a fitting procedure.
In the Sketch2CAD framework~\cite{Li:2020:Sketch2CAD}, a network is trained to predict CAD operations that correspond with segmented sketch strokes. 
Free2CAD~\cite{Li:2022:Free2CAD} generalizes this system by additionally learning how to segment a complete sketch into groups that can be mapped to CAD operations 
\rev{(Figure \ref{fig:f2cad})}
. 
As the search space of methods that aim to reverse-engineer CAD models can be expansive, some approaches have investigated heuristics that make the problem more tractable.
Zone Graphs~\cite{zoneGraphs} converts a boundary representation of a 3D Shape into a partition of zones.
While this partitioning permits the use of enumerative search strategies, further guiding this search with a neural module results in improved programs.
\rev{These domain-specific methods perform well for their respective applications, but they rely on heuristics and design principles that do not readily generalize to other domains.}


A central difficulty of neurosymbolic models is that gradients cannot flow through symbolic elements, complicating end-to-end learning.
A workaround for this issue is to train a system where symbolic DSL components are replaced with differentiable proxies.
Then at inference time, these proxies can be replaced by DSL expressions with similar outputs. 
In Lambourne et al.~\cite{lambourne2022recon}, a network learns to reconstruct a target shape with a collection of differentiable extrusions learned from human-written programs. 
At inference time, these extrusions are replaced with observed profiles from an input collection, resulting in a complete sequence of CAD operations that reconstruct the input.
In Point2Cyl~\cite{uy-point2cyl-cvpr22}, a network learns to represent a shape by combining differentiable extrusion proxies with boolean operations. 
These proxies can then be converted into extrusion cylinders through differentiable, closed-form formulations in order to produce an editable CAD model.
\rev{This scheme of relaxation with replacement allows for end-to-end learning with high-fidelity outputs but often requires clever insights for both architecture construction and the replacement scheme that are domain-specific.}

Pushing this trend further, some methods learn how to infer 3D shape programs in a purely end-to-end fashion, by designing neural architectures that act as a smooth relaxation of a DSL executor. 
Some languages are more amenable to this approach than others. 
For instance in CSG, primitives are easily differentiable, and \textit{hard} boolean set operations can be replaced with \textit{soft} versions.
The neural architecture used often places restrictive constraints on the \emph{structures} of programs that can be found with these approaches.
\rev{We depict the architecture of one such method, CAPRI-Net~\cite{Yu_2022_CVPR} in Figure \ref{fig:capri}.}
In \rev{their formulation} primitives are converted into convexes through intersection, \emph{left} and \emph{right} shapes are created by unioning convexes, and the final output is the difference between the \emph{left} shape and \emph{right} shape.
In CSG-Stump~\cite{ren2021csg}, the output CSG program is a union over intersections of either primitives or complements of primitives.
ExtrudeNet~\cite{ren2022extrude} extends upon the CAPRI-Net architecture by replacing quadric primitives with extruded 2D sketches, while maintaining end-to-end differentiability.   
UCSGNet \cite{kania2020ucsgnet} explores a higher diversity of program structures, by evaluating many boolean operations in parallel, but this comes at the cost of more complex programs with worse reconstruction performance.  
\rev{
For domains that permit baking program execution behavior into a neural architecture, such methods typically achieve the best geometric reconstruction performance.
The limitation of these approaches is that each architecture is inherently tied to a specific domain, and even within a domain, the class of programs producible by the neural network is typically constrained.
}




\subsection{3D Shape Generation}

\begin{figure}[]
    \centering
    \includegraphics[width=\linewidth]{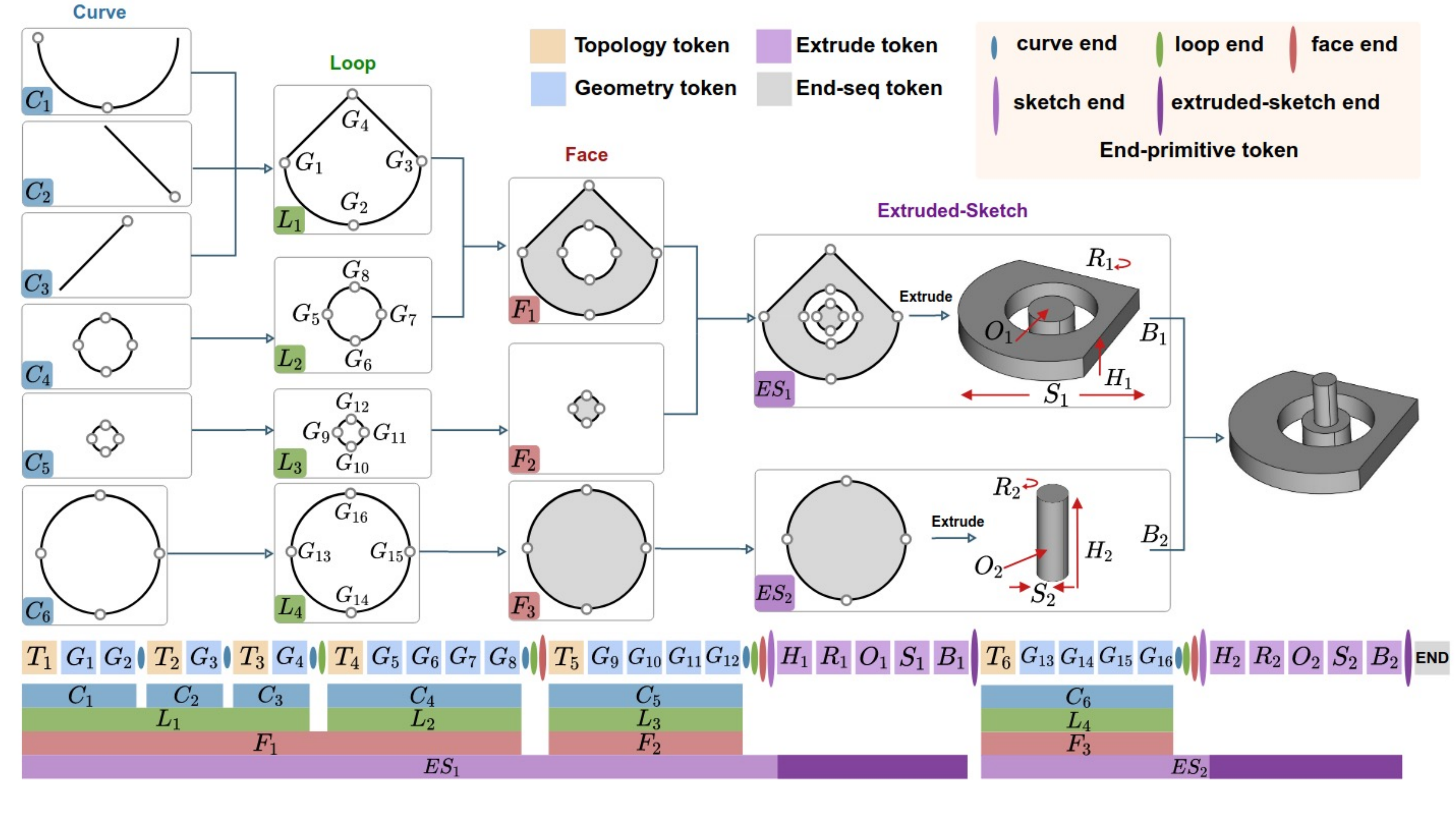}
    \caption{
        SkexGen~\cite{xu2022skexgen} represents 3D shapes as sequences of sketch-and-extrude operations, where the tokens of different operator types are produced by disentangled codebooks. 
        A Transformer learns over a distribution of shapes encoded in this representation, and is able to synthesize novel 3D geometry through auto-regressive sampling.
    }
    \label{fig:skexgen}
\end{figure}

A number of techniques have also explored how neurosymbolic models can be used to generate novel 3D shape instances. 
When a dataset of ground-truth (i.e. human-written) shape programs exists, deep generative modeling techniques can be directly employed.
DeepCAD~\cite{DeepCAD}, trains a transformer-based autoencoder to auto-regressively generate sequences of sketch and extrude commands conditioned on the input vector.
Novel shapes are then synthesized by prompting the decoder with new conditioning vectors output by a latent GAN.
SkexGen \cite{xu2022skexgen} improves upon this paradigm by designing an architecture where sub-networks are specialized for particular types of CAD operations 
\rev{(Figure \ref{fig:skexgen})}
.
Specifically, a "Sketch" branch reasons about 2D sketches (tokens related to their topology and geometry), while an "Extrude" branch predicts how sketches should be lifted into 3D (tokens related to an extrusion direction). 

\begin{figure}[]
    \centering
    \includegraphics[width=\linewidth]{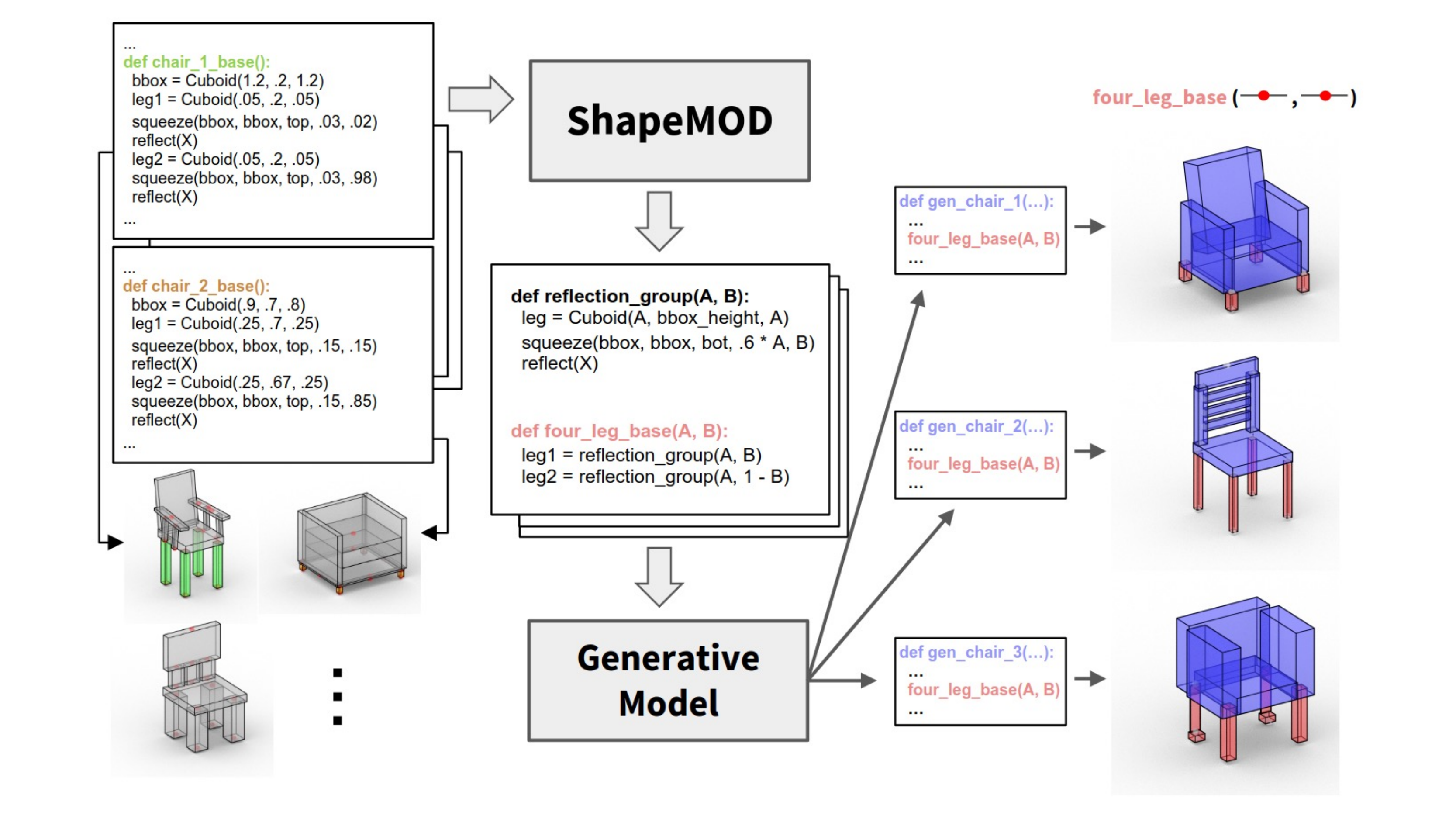}
    \caption{
        ShapeMOD~\cite{jones2021shapeMOD} takes a collection of 3D shape programs as input and makes them more compact by automatically discovering common macros which can be re-used across the collection. Programs rewritten with these macros can benefit downstream applications such as generative modeling. 
    }
    \label{fig:shapemod}
\end{figure}

When human-written programs are not available, an alternative approach is to heuristically \textit{parse} programs from dataset of 3D shapes with structured annotations.
ShapeAssembly~\cite{jones2020SA}, introduced a new DSL designed for specifying the part structure of manufactured 3D objects.
A hierarchical sequence VAE learns to write new ShapeAssembly programs after training over a dataset of programs parsed from PartNet shapes.
One issue with parsing programs directly from shape repositories is that the resulting programs might be overly complex.
The ShapeMOD algorithm~\cite{jones2021shapeMOD}, addresses this issue by automatically discovering macro operations that abstract out common structural and parametric patterns over a collection of shape programs
\rev{(Figure \ref{fig:shapemod})}.
Deep generative models can then benefit from training over program distributions that are rewritten to use macro operations. 

Related to the problem of learning a generative model over a program dataset, Ritchie et al.~\cite{ExampleBasedProcMod} developed a procedure for inferring probabilistic programs that explain a small set of example shapes. 
\rev{As inferring probabilistic programs is very difficult, the method makes a number of simplifying assumptions: it requires that the input shapes have a consistent hierarchical organization of parts and it produces programs that are similar to context-free grammars. }
Under these assumptions, the method demonstrates the power of this representation:
the inferred probabilistic programs capture both hierarchical structure and continuous relationships of parts present in modular inputs, where part transform distributions are parameterized by a learned network.
This allows new objects in the style of the exemplars to be synthesized by sampling from the distributions captured by the probabilistic program.




\section{Application: Materials \& Textures}
\label{sec:app_materials}

\begin{table*}[t!]
    \centering
    \scriptsize
    \begin{tabular}{@{}laabbbddefgg@{}}
        & \multicolumn{2}{c}{\textbf{Task Spec}} & \multicolumn{3}{c}{\textbf{DSL}} & \multicolumn{2}{c}{\textbf{Synthesizer}} & \multicolumn{1}{c}{\textbf{Execution}} & \multicolumn{1}{c}{\textbf{Refinement}} & \multicolumn{2}{c}{\textbf{Learning}} \\
        \cmidrule(lr){2-3} \cmidrule(lr){4-6} \cmidrule(lr){7-8}  \cmidrule(lr){11-12}
        \textbf{Method} & \textit{Input} & \textit{Output} & \textit{Paradigm} & \textit{Primitives} & \textit{Mutability} & \textit{Search} & \textit{Guidance} &  &  & \textit{End-to-end} & \textit{Modular}\\
		\midrule
		Liu et al.~\cite{liu2018perception} & \visualtarget{} & \detprog{} & \functional{} & \hardcodedprims{} & \fixed{} & \retrieval{} & \synthguide{} & \direct{} & & & \people{} \\ 
		Hu et al.~\cite{hu2019inverse} & \visualtarget{} & \detprog{} & \functional{} & \hardcodedprims{} & \fixed{} & \retrieval{} & \synthguide{} & \direct{} & \neuralpostproc{} & & \people{} \\
		Tchapmi et al.~\cite{Tchapmi:2022:Procedural} & \visualtarget{} & \detprog{} & \functional{} & \hardcodedprims{} & \fixed{} & \retrieval{} & \synthguide{} & \direct{} & & & \people{} \synthetic{} \\
		MATch~\cite{Shi2020:MATch} & \visualtarget{} & \detprog{} & \functional{} & \hardcodedprims{} & \fixed{} & \retrieval{} & \synthguide{} & \direct{} & & & \people{} \\
		Differentiable Proxies~\cite{hu2022diff} & \visualtarget{} & \detprog{} & \functional{} & \hardcodedprims{} & \fixed{} & \retrieval{}& \synthguide{} & \direct{} \proxy{} & & \smoothrelax{} & \people{} \\
		Hu et al.~\cite{hu2022inverse} & \visualtarget{} & \detprog{} & \functional{} & \hardcodedprims{} & \fixed{} & \userintheloop{} & & \direct{} & & & \\
		MatFormer~\cite{guerrero2022matformer} & \visualexamples{} & \progdist{} & \functional{} & \hardcodedprims{} & \fixed{} & \stochastic{} & \synthguide{} & \direct{} & & & \people{} \\
        \bottomrule
    \end{tabular}
    \iconlegend
    \caption{A summary of the work on neurosymbolic material \& texture modeling discussed in Section~\ref{sec:app_materials}, where each approach is situated in our design space.}
    \label{tab:material_methods}
\end{table*}


Procedural workflows have gained increased popularity in the material and texture design community over the last few years, \rev{driven by modern authoring tools~\cite{Substance3D, Houdini, Blender}.}
%
Materials are a good fit for procedural workflows, due to the presence of repeated structures and self-similarity, typically with some stochastic variations between the repetitions. Figure~\ref{fig:datasets} bottom-right, shows some examples of procedurally generated materials. Procedural workflows offer several benefits to material artists, such as (i) a non-destructive workflow, where any operation can be changed at any point in the authoring process, (ii) the ability to quickly create variations of a material by adjusting parameters of the procedure, (iii) automatic tileability, since the provided procedural operations typically ensure tileability, (iv) resolution-independence, and (v) the ability to easily scale the material up or down by adjusting the number/frequency of repetitions. Procedural materials offer an increasingly fertile field for research into neurosymbolic models, as the interest in procedural materials increases and more data becomes available.

Early approaches for procedural materials and textures~\cite{Peachey:1985:SolidTexturing, Perlin:1985:ImageSynthesizer, Witkin:1991:reactiondiffusion, Worley:1996:Cellular, Lefebvre:2000:StructuralTextures, Galerne:2012:Gabor} and some more recent approaches~\cite{Liu:2016:wood, Guehl:2020:PPTBF} provide a black-box function to the user and expose parameters that can be modified to control the generated result. This black-box may, for example, implement a parametric noise function, a reaction-diffusion approach~\cite{Witkin:1991:reactiondiffusion}, a physically-inspired simulation~\cite{Liu:2016:wood}, or point process basis functions~\cite{Guehl:2020:PPTBF}.

Procedural materials that are used in \rev{practice~\cite{Substance3D, Houdini, Blender}} also provide control over the content of the black-box using a DSL. To create a material, the program is executed, combining and transforming a set of initial noises and patterns through several image filtering operations. This DSL is exposed to the user as a \emph{node graph} (Figure~\ref{fig:datasets} bottom-right). Nodes correspond to image filtering/processing functions, or to generators for the initial noises/patterns. Each node has a set of inputs and outputs. Inputs typically consist of images or scalar/vector-valued parameters, and outputs are one or multiple processed images. Edges control the data flow between the inputs and outputs of different nodes---in essence, the node graph specifies a functional program.

Neurosymbolic approaches in this domain aim to synthesize such node graph programs: either inferring a program for a given input image, or sampling novel programs whose characteristics match a dataset of examples.
They differ in the approach used to synthesize such programs. Typically these methods have separate approaches to handle the generation of the program structure (node types and edges) and the generation of the program parameters (node parameters).
Table~\ref{tab:material_methods} situates each prior work we discuss within our design space.

\paragraph*{Program and parameter retrieval:} An early method~\cite{liu2018perception} creates both program structure and parameters for a given visual target image by retrieving them from a pre-defined dataset of parameterized programs. A perceptual metric based on psychophysical experiments is introduced as a distance metric for both program and parameter retrieval. Generation of programs through retrieval from a large dataset is limited to the programs available in the dataset and has, therefore, limited coverage over the space of possible visual targets.

\begin{figure}[t]
    \centering
    \includegraphics[width=\linewidth]{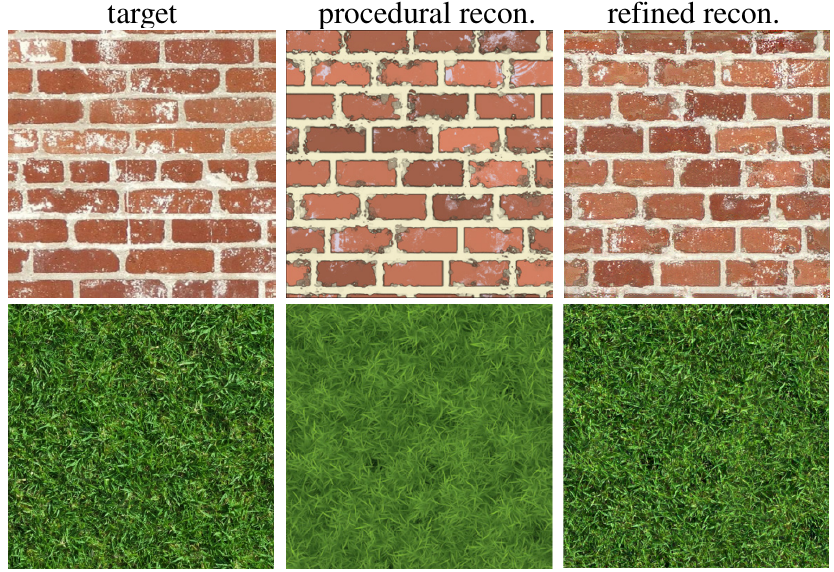}
    \caption{
    Hu et al.~\cite{hu2019inverse} synthesize a procedural material matching a given target image by retrieving a procedural program. Parameters for the program that reproduces the target image are estimated using a trained network. Additionally, Hu et al. use a style transfer network to refine the output of the program to further improve its match with the target.
    }
    \label{fig:hu2019inverse}
\end{figure}

\begin{figure}[t]
    \centering
    \includegraphics[width=\linewidth]{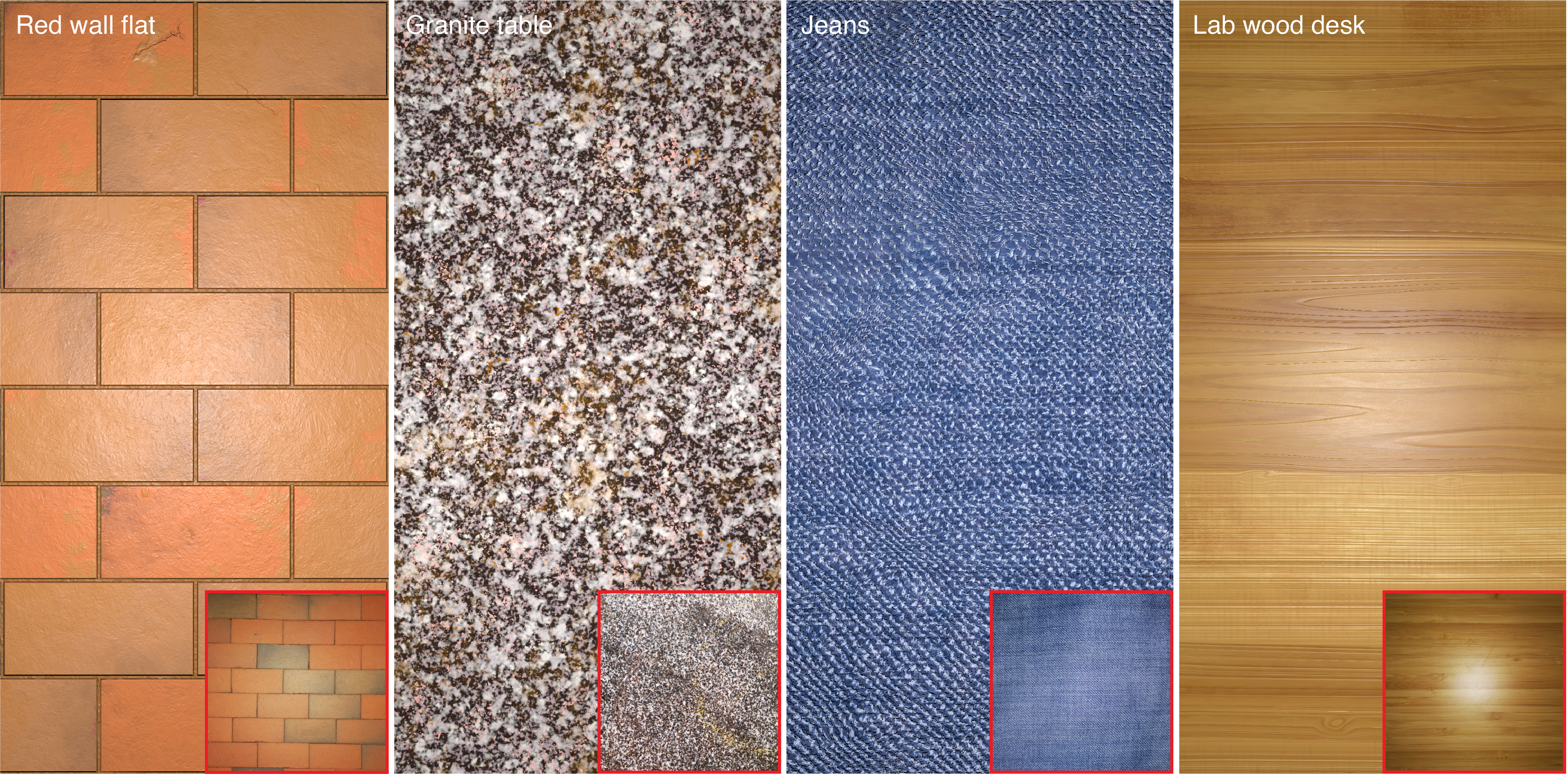}
    \caption{
    MATch~\cite{Shi2020:MATch} synthesizes a procedural material that matches a given target image (lower right corner). A differentiable procedural program is retrieved from a large dataset of material programs, and the parameters of this program are optimized to match the target using gradient descent.
    }
    \label{fig:match}
\end{figure}

\paragraph*{Program retrieval, parameter synthesis:} Two approaches~\cite{hu2019inverse, Tchapmi:2022:Procedural} improve upon this work by using a neural network to estimate parameters given a retrieved program structure and a visual target, rather than retrieving parameters (Figure~\ref{fig:hu2019inverse}). \rev{The earlier approach~\cite{hu2019inverse} requires training a parameter estimator for each program, while the more recent approach~\cite{Tchapmi:2022:Procedural} improves scalability by training a single estimator for multiple programs.} MATch~\cite{Shi2020:MATch} goes a different route by directly optimizing the parameters of a retrieved program to match the program's output to the given visual target, rather than using a network to amortize the optimization. This improves accuracy over amortized optimization. The authors introduce differentiable versions of most operators in the program to enable optimization with gradient descent. A few examples of optimized materials are shown in Figure~\ref{fig:match}. Since not all operators can be made differentiable, only a subset of the program parameters can be optimized. Differentiable Proxies~\cite{hu2022diff} tackles the problem of non-differentiable operators by training small neural networks to approximate these operators. The networks act as proxies that are differentiable and have parameters that can be optimized with gradient descent.

\begin{figure}[t]
    \centering
    \includegraphics[width=\linewidth]{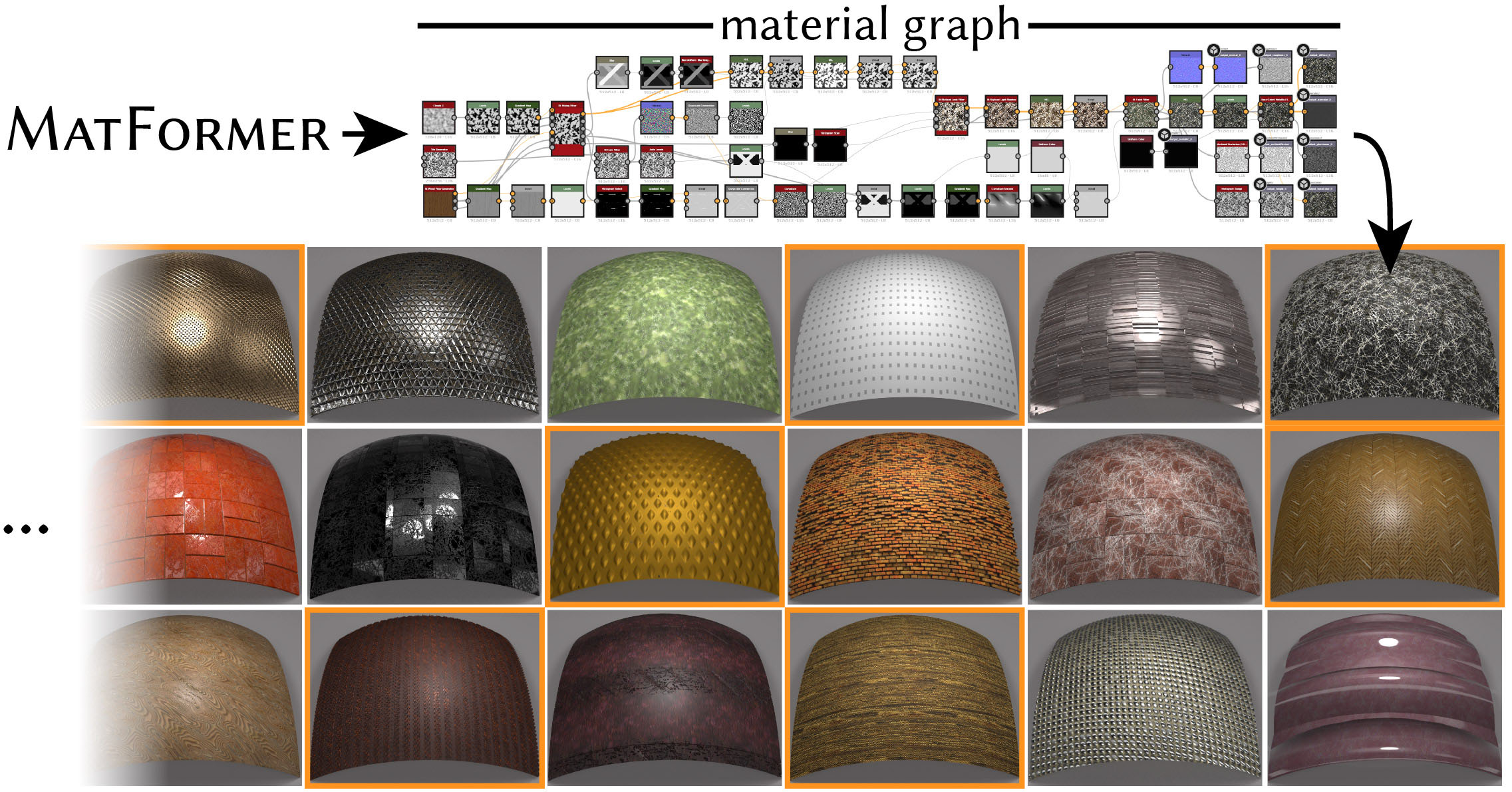}
    \caption{
    MatFormer~\cite{guerrero2022matformer} is an unconditional generator for procedural materials represented as node graphs. It creates a node graph with three Transformers that are trained to generate nodes, node parameters, and edges, respectively.
    }
    \label{fig:matformer}
\end{figure}

\paragraph*{Program and parameter synthesis:}
Recently, methods have started to synthesize the program structure in addition to the parameters, rather than retrieving it from a dataset. Hu et al.~\cite{hu2022inverse} propose to \rev{synthesize programs that reconstruct given target images. The programs have a special constrained structure}, where procedural masks based on point process basis functions~\cite{Guehl:2020:PPTBF} are first used to define a tree of sub-regions in a material that contains increasingly uniform texture, and noise generators are then applied to the leaves to generate a texture. The procedural masks are found using interactive segmentation of a visual target, and noise generators are fit to the content of a region based on local spectra. \rev{The required user interaction for segmentation and the and the reduced generality of the programs due to their constrained structure are limitations of the method.
MatFormer~\cite{guerrero2022matformer} proposes an unconditional generator for more general material programs. It is based} on three Transformer models that are trained to synthesize nodes, edges, and node parameters of a node graph, respectively. Since Transformers work on linear sequences, various strategies to linearize a node graph are proposed. Figure~\ref{fig:matformer} shows an example of a generated node graph and several examples of materials that were created with generated node graphs. \rev{However, MatFormer does not support conditional generation, for example, to reproduce a given target image.}



%% file: 06-conclusion.tex
\section{Conclusion and Future Work}
\label{sec:conclusion}

In this report, we reviewed \emph{neurosymbolic} models: techniques that combine the best features of procedural models and machine learning to generate visual data for computer graphics applications.
We defined a formal design space for such models, providing a framework for organizing different possible instantiations of neurosymbolic models.
We then surveyed recent work on neurosymbolic modeling in 2D shape modeling, 3D shape modeling, and material \& texture modeling, placing each prior work into our design space.

In addition to organizing past work, our design space provides
another benefit: identifying areas of the space which are sparsely populated by prior work (or empty), as such areas may warrant further exploration.
With this in mind, the following are some open problems and opportunities for future work in neurosymbolic modeling, in the context of Computer Graphics. 

\paragraph*{New application domains:}
Perhaps the most immediate opportunity for future work is to apply some of the techniques described in this report to modeling data in other graphics domains.
Any type of visual data is fair game, but it is especially worth looking at domains where programmatic representations have already proven useful.
One such example is shader programming, a domain for which large repositories of data exist~\cite{Shadertoy}.
Character animation \& behavior modeling could also warrant investigation, as some crowd simulation software already uses hand-authored animation procedures~\cite{Massive}.
This report covered some recent work in fabrication-aware 3D shape modeling; other fabrication-aware design domains are worth exploring.
For example, sewing patterns for clothing exhibit recurring structural patterns which a neurosymbolic model could learn to predict as programs~\cite{ParsingSewingPatterns}.
Finally, there also exist opportunities in non-visual domains which are relevant to computer graphics.
For instance, games can use procedural representations of music to dynamically alter their soundtrack in response to player actions~\cite{IntelligentMusic}; neurosymbolic modeling techniques might bring new capabilities to this application or could produce such representations more automatically.

\paragraph*{More complex programs:}
Learning to produce generative programs with complex structure is a challenging problem; as a consequence, much of the early work in neurosymbolic modeling, which we have covered in this report, uses simple languages (e.g. CSG for 3D shape modeling).
To move beyond the research lab and be useful for real-world design applications, neurosymbolic models must be able to produce complex programs in rich, fully-featured domain-specific languages: for example, full CAD languages such as OpenSCAD~\cite{OpenSCAD}, or shading languages such as GLSL.
The MatFormer model, which produces programs in the form of detailed Substance material graphs, is a step in this direction~\cite{guerrero2022matformer}.
However, this model was trained on a large dataset of human-authored programs, which brings us to our next open problem:

\paragraph*{Learning without direct supervision:}
The most straightforward way for a neurosymbolic model to learn to produce programs is to train on programs authored by \textbf{people} \people{}.
However, such data is rarely available at scale for most languages of interest.
Learning without such supervision, especially as the target language becomes more complex, remains a challenge.
Our design space identifies three learning approaches for this setting: \smoothrelax{} \textbf{Smooth Relaxation}, \rl{} \textbf{Reinforcement Learning}, and  \bootstrap{} \textbf{Bootstrapping}.
Each of these approaches warrants further investigation:
can one design more general-purpose smooth relaxations for new languages?
Are there new techniques from the RL literature which work well for synthesizing generative programs?
Can bootstrapping approaches such as PLAD~\cite{jones2022PLAD} be extended to handle more complex languages,  possibly involving hierarchical structure?

\paragraph*{Discovering new languages:}
Thus far, work on neurosymbolic modeling has targeted existing domain-specific languages, potentially modifying them by adding new abstractions.
Might it be possible for entirely new languages to be \textbf{invented} \invented{}, to best fit the domain of data being modeled?
Neurosymbolic models make it possible to consider this question: one could imagine learning a set of \textbf{neural primitives} \neuralprims{} which represent the visual ``atoms'' of the data domain being modeled (e.g. objects, parts) and then learning a set of primitive functions which combine these atoms to create more complex visual structures.
The language discovered could adapt to the particular data distribution on which it is trained, lending a new definition to the abbreviation ``DSL'': \emph{distribution}-specific languages.


\paragraph*{Capturing user intent:}
The usefulness of a neurosymbolic model is largely determined by how well its \emph{task specification} captures a user's intent.
On the input side of this specification, supporting more input types can help make neurosymbolic models more useful for a wider population of users.
While defining our design space (Section~\ref{sec:taskspec}), we have already mentioned several new types of input that may warrant future exploration.
For instance: we are not aware of any prior neurosymbolic models which take a \textbf{text prompt} \textprompt{}  as input, though this seems like an appealing way to specify user intent (if the recent success of text-to-image generators is any indicator~\cite{DALLE2,Imagen,StableDiffusion}).
Such natural language input could specify desired attributes of the visual data to be generated, of the programs which will generate that data, or both (including connections between the two, e.g. that a particular program construct should be used to generate a particular visual feature). 

On the output side of the task specification, neurosymbolic models must produce programs that are usable by people.
How do we assess how usable a program is, and how can we encourage neurosymbolic models to produce such programs?
As mentioned in Section~\ref{sec:progsynth}, the only prior on program structure that most previous work considers is a preference for shorter programs.
This is justified with appeals to information theory or Occam's razor, but not to usability.
In fact, excessively short programs can become \emph{un}usable (see obfuscated code contests or business card raytracers~\cite{BusinessCardRaytracer}).
Furthermore, different users may prefer different program structures.
Are there better priors than program length for the usability of a program?
What are effective mechanisms for users to specify their preferences about program structure?

\paragraph*{Human interpretation and interaction:} While program representations have the key advantage of exposing an interpretable structure, reasoning about programs and manipulating them is often quite challenging and requires domain and coding expertise. By combining program representations with learning, we can advance symbolic reasoning in three fundamental directions: interpretation, manipulation, and composition.    

Code \textbf{\emph{interpretation}} refers to the ability to analyze code: exposing its structure, capabilities, and potential bugs. 
The programming language community has made significant advances in this space by using theorem provers or SMT solvers to detect and expose bugs~\cite{SMTBugFinding}. 
Such methods, however, are still limited in scale and domain. How can we leverage advances in machine learning to help detect code errors and expose them? Can such systems go a step beyond detection and also propose solutions?  

For code that generates a visual output, \textbf{\emph{manipulation}} is extremely important to enable customization and iterative design.
While recent work describes techniques to optimize code based on the direct manipulation of the visual output~\cite{hempel2019sketch,cascaval2022differentiable,AutoDiffProcMod}, these methods are limited in the types of variations they enable, only allowing the program parameters to change, but not its structure. Furthermore, existing techniques fundamentally struggle to infer the user intent, since direct visual manipulation is a \emph{partial} (and therefore ambiguous) specification. How can neurosymbolic models address these challenges to enable a wider class of variations that are not limited to parameter changes but also require program rewrites? Can we leverage existing datasets or extract information from user interactions to learn to disambiguate the partial specifications?  

Typical programs that generate visual outputs are designed with hierarchical and compositional structures. 
This design decision is not only critical for human-understandable editing and interaction, it also allows analysis methods to reason over different semantic parts of a visual model.
Creating new visual content by mixing and matching existing components is a fundamental design strategy in computer graphics~\cite{funkhouser2004modeling}. However, code \textbf{\emph{composition}} is challenging, as it requires that certain properties hold at the interface of program components. Generating code that can be easily decomposed and re-used is a tedious, error-prone task. Further, understanding how to constrain or manipulate parts to enable seamless compositions is challenging. 
Can neurosymbolic reasoning enable design through composition with symbolic representations?

%% file: 07-bios.tex
\section{Author Bios}
\label{sec:bios}

\paragraph*{Daniel Ritchie}
is the Eliot Horowitz Assistant Professor of Computer Science at Brown University. He received his PhD from Stanford University. His research sits at the intersection of computer graphics and artificial intelligence, where he is particularly interested in data-driven methods for designing, synthesizing, and manipulating visual content.
In the area of neurosymbolic modeling, he has worked on applying probabilistic programming to procedural modeling, on learning procedural models from a small number of examples, and on deep learning models for generating/inferring procedural representations of 3D shapes.
His most recent work focuses on learning procedural representations in the absence of ground-truth programs.

\paragraph*{Paul Guerrero} 
is a research scientist at Adobe. He received his PhD from the Institute for Computer Graphics and Algorithms, Vienna University of Technology, and at the Visual Computing Center in KAUST.
He is working on the analysis of irregular and compositional structures, such as graphs, meshes, or vector graphics, by combining methods from machine learning, optimization, and computational geometry
[\href{https://dl.acm.org/doi/10.1145/2897824.2925950}{1}, 
\href{https://dl.acm.org/doi/10.1145/3355089.3356527}{2}, 
\href{https://openaccess.thecvf.com/content_CVPR_2020/papers/Mo_StructEdit_Learning_Structural_Shape_Variations_CVPR_2020_paper.pdf}{3}, 
\href{https://dl.acm.org/doi/abs/10.1145/3414685.3417812}{4}, 
\href{https://dl.acm.org/doi/abs/10.1145/3414685.3417812}{5}, 
\href{https://proceedings.neurips.cc/paper/2021/file/28891cb4ab421830acc36b1f5fd6c91e-Paper.pdf}{6}]. 

\paragraph*{R. Kenny Jones}
is a PhD student at Brown University where he is supported by a Brown University Presidential Fellowship and advised by Daniel Ritchie. His research explores how machine learning and artificial intelligence techniques can be leveraged to better understand and represent visual data. Relevant to this report, his recent publications have investigated using neurosymbolic models for 3D shape generation, automatic macro discovery, and inferring visual programs.

\paragraph*{Niloy J. Mitra} leads the Smart Geometry Processing group in the Department of Computer Science at University College London and the Adobe Research London Lab. He received his PhD from Stanford University under the guidance of Leonidas Guibas. His current research focuses on developing machine learning frameworks towards generative models for high-quality geometric and appearance content for CG applications. He received the 2019 Eurographics Outstanding Technical Contributions Award, the 2015 British Computer Society Roger Needham Award, and the 2013 ACM SIGGRAPH Significant New Researcher Award. He was elected as a fellow of Eurographics in 2021 and was the SIGGRAPH Technical Papers Chair in 2022. 

\paragraph*{Adriana Schulz} is an Assistant Professor at the Paul G. Allen School of Computer Science and Engineering at the University of Washington. She received her PhD from the Massachusetts Institute of Technology. Her research is in the area of computational design and fabrication.  Relevant to this report, her recent work uses programable abstractions to represent and optimize both 3D designs and their fabrication plans. She also uses machine learning and program synthesis techniques to support the design and manipulation of CAD models. 

\paragraph*{Karl D.D. Willis} is a Senior Research Manager at Autodesk Research focused on data-driven design software for manufacturing. He holds a PhD in Computational Design from Carnegie Mellon University and has presented his research internationally at conferences such as ACM SIGGRAPH, IEEE/CVF CVPR, and ICML. His work at Autodesk has won numerous awards including Fast Company Innovation By Design Honoree and Core77 Design Awards Research and Strategy Honoree. 

\paragraph*{Jiajun Wu} is an Assistant Professor of Computer Science at Stanford University. He received his PhD from Massachusetts Institute of Technology. His research is in the area of computer vision, artificial intelligence, graphics, and robotics. In the area of neurosymbolic modeling, he has worked on building and learning structured representations for visual data of various modalities (images, shapes, videos) by integrating domain knowledge with data-driven methods.

%% file: 08-acks.tex
\section*{Acknowledgments}

Daniel Ritchie was supported by NSF awards \#1907547 and \#1941808.
He is also an advisor to Geopipe and owns equity in the company. Geopipe is a start-up that is developing 3D technology to build immersive virtual copies of the real world with applications in various fields, including games and architecture. Jiajun Wu was supported by NSF awards \#2120095 and \#2211258, Autodesk, IBM, and Salesforce. Adriana Schulz was supported by NSF awards \#2219864 and \#2017927, Adobe, Intel, Meta, and Amazon. 

%% file: main.bbl
\newcommand{\etalchar}[1]{$^{#1}$}
\begin{thebibliography}{\uppercase{LZCvdP20}}

\bibitem[ABJ{\etalchar{*}}13]{alur2013syntax}
\textsc{Alur R., Bodik R., Juniwal G., Martin M.~M., Raghothaman M., Seshia
  S.~A., Singh R., Solar-Lezama A., Torlak E., Udupa A.}:
\newblock Syntax-guided synthesis.
\newblock In \emph{Formal Methods in Computer--Aided Design (FMCAD)} (2013).

\bibitem[{Ado}a]{SubstanceDesigner}
\textsc{{Adobe}}:
\newblock {Substance Designer}.
\newblock \url{https://www.adobe.com/products/substance3d-designer.html}.
\newblock Accessed: 2022-09-26.

\bibitem[{Ado}b]{Substance3D}
\textsc{{Adobe Inc.}}:
\newblock {Substance 3D}.
\newblock URL: \url{https://www.substance3d.com/}.

\bibitem[Ado21a]{Substance_Community}
\textsc{Adobe}:
\newblock Substance {3D} community assets, 2021.
\newblock \url{https://substance3d.adobe.com/community-assets}.

\bibitem[Ado21b]{Substance_Source}
\textsc{Adobe}:
\newblock Substance source, 2021.
\newblock \url{https://substance3d.adobe.com/assets}.

\bibitem[AFDJ03]{MCMCIntro}
\textsc{Andrieu C., Freitas N., Doucet A., Jordan M.}:
\newblock An introduction to mcmc for machine learning.
\newblock \emph{Machine Learning 50} (2003), 5--43.

\bibitem[{Aut}a]{Fusion360}
\textsc{{Autodesk}}:
\newblock {Fusion 360}.
\newblock \url{https://www.autodesk.com/products/fusion-360/}.
\newblock Accessed: 2022-10-16.

\bibitem[{Aut}b]{MayaHypershade}
\textsc{{Autodesk Maya Wiki}}:
\newblock {Hypershade}.
\newblock \url{https://autodeskmaya.fandom.com/wiki/Hypershade}.
\newblock Accessed: 2022-10-16.

\bibitem[BBB{\etalchar{*}}57]{backus1957fortran}
\textsc{Backus J.~W., Beeber R.~J., Best S., Goldberg R., Haibt L.~M., Herrick
  H.~L., Nelson R.~A., Sayre D., Sheridan P.~B., Stern H., et~al.}:
\newblock The {FORTRAN} automatic coding system.
\newblock In \emph{Western Joint Computer Conference: Techniques for
  Reliability} (1957), pp.~188--198.

\bibitem[BGK{\etalchar{*}}13]{ParsingSewingPatterns}
\textsc{Berthouzoz F., Garg A., Kaufman D.~M., Grinspun E., Agrawala M.}:
\newblock Parsing sewing patterns into {3D} garments.
\newblock \emph{ACM Transactions on Graphics (TOG) 32}, 4 (2013).

\bibitem[BHB{\etalchar{*}}18]{RelationalInductiveBiases}
\textsc{Battaglia P.~W., Hamrick J.~B., Bapst V., Sanchez-Gonzalez A., Zambaldi
  V., Malinowski M., Tacchetti A., Raposo D., Santoro A., Faulkner R., et~al.}:
\newblock Relational inductive biases, deep learning, and graph networks.
\newblock \emph{arXiv preprint arXiv:1806.01261} (2018).

\bibitem[{Ble}]{Blender}
\textsc{{Blender Online Community}}:
\newblock Blender - a {3D} modelling and rendering package.
\newblock URL: \url{http://www.blender.org}.

\bibitem[BLW{\etalchar{*}}20]{bau2020rewriting}
\textsc{Bau D., Liu S., Wang T., Zhu J.-Y., Torralba A.}:
\newblock Rewriting a deep generative model.
\newblock In \emph{European Conference on Computer Vision (ECCV)} (2020).

\bibitem[BMR{\etalchar{*}}20]{GPT3Paper}
\textsc{Brown T., Mann B., Ryder N., Subbiah M., Kaplan J.~D., Dhariwal P.,
  Neelakantan A., Shyam P., Sastry G., Askell A., Agarwal S., Herbert-Voss A.,
  Krueger G., Henighan T., Child R., Ramesh A., Ziegler D., Wu J., Winter C.,
  Hesse C., Chen M., Sigler E., Litwin M., Gray S., Chess B., Clark J., Berner
  C., McCandlish S., Radford A., Sutskever I., Amodei D.}:
\newblock Language models are few-shot learners.
\newblock In \emph{Advances in Neural Information Processing Systems (NeurIPS)}
  (2020), vol.~33, pp.~1877--1901.

\bibitem[BT18]{SMT}
\textsc{Barrett C., Tinelli C.}:
\newblock Satisfiability modulo theories.
\newblock In \emph{Handbook of model checking}. Springer, 2018, pp.~305--343.

\bibitem[BTLLW22]{deep_gen_survey}
\textsc{Bond-Taylor S., Leach A., Long Y., Willcocks C.~G.}:
\newblock Deep generative modelling: A comparative review of {VAEs}, {GANs},
  normalizing flows, energy-based and autoregressive models.
\newblock \emph{IEEE Transactions on Pattern Analysis and Machine intelligence
  (TPAMI) 44}, 11 (2022), 7327--7347.

\bibitem[BZS{\etalchar{*}}20]{GanDissection}
\textsc{Bau D., Zhu J.-Y., Strobelt H., Lapedriza A., Zhou B., Torralba A.}:
\newblock Understanding the role of individual units in a deep neural network.
\newblock \emph{Proceedings of the National Academy of Sciences} (2020).

\bibitem[Cat74]{CatmulThesis}
\textsc{Catmull E.~E.}:
\newblock \emph{A Subdivision Algorithm for Computer Display of Curved
  Surfaces.}
\newblock PhD thesis, The University of Utah, 1974.

\bibitem[CDAT20]{DeepSVG}
\textsc{Carlier A., Danelljan M., Alahi A., Timofte R.}:
\newblock {DeepSVG}: A hierarchical generative network for vector graphics
  animation.
\newblock In \emph{Advances in Neural Information Processing Systems (NeurIPS)}
  (2020), vol.~33, pp.~16351--16361.

\bibitem[CEP{\etalchar{*}}21]{chaudhuri2021neurosymbolic}
\textsc{Chaudhuri S., Ellis K., Polozov O., Singh R., Solar-Lezama A., Yue Y.,
  et~al.}:
\newblock Neurosymbolic programming.
\newblock \emph{Foundations and Trends{\textregistered} in Programming
  Languages 7}, 3 (2021), 158--243.

\bibitem[CFG{\etalchar{*}}15]{shapenet2015}
\textsc{Chang A.~X., Funkhouser T., Guibas L., Hanrahan P., Huang Q., Li Z.,
  Savarese S., Savva M., Song S., Su H., Xiao J., Yi L., Yu F.}:
\newblock {ShapeNet}: An information-rich {3D} model repository.
\newblock \emph{arXiv preprint arXiv:1512.03012} (2015).

\bibitem[CHIS22]{croitoru2022diffusion}
\textsc{Croitoru F.-A., Hondru V., Ionescu R.~T., Shah M.}:
\newblock Diffusion models in vision: A survey.
\newblock \emph{arXiv preprint arXiv:2209.04747} (2022).

\bibitem[Cla76]{JimClarkLOD}
\textsc{Clark J.~H.}:
\newblock Hierarchical geometric models for visible surface algorithms.
\newblock \emph{Communications of the ACM 19}, 10 (1976), 547–554.

\bibitem[Cla03]{Clausen2003BranchAB}
\textsc{Clausen J.}:
\newblock \emph{Branch and Bound Algorithms-Principles and Examples}.
\newblock Tech. rep., University of Copenhagen, 2003.

\bibitem[Coo84]{ShadeTrees}
\textsc{Cook R.~L.}:
\newblock Shade trees.
\newblock In \emph{Annual Conference on Computer Graphics and Interactive
  Techniques (SIGGRAPH)} (1984), p.~223–231.

\bibitem[CRW{\etalchar{*}}20]{3d_struct_star}
\textsc{Chaudhuri S., Ritchie D., Wu J., Xu K., Zhang H.}:
\newblock Learning generative models of {3D} structures.
\newblock \emph{Computer Graphics Forum (CGF) 39}, 2 (2020), 643--666.

\bibitem[CSQ{\etalchar{*}}22]{cascaval2022differentiable}
\textsc{Cascaval D., Shalah M., Quinn P., Bodik R., Agrawala M., Schulz A.}:
\newblock Differentiable {3D} {CAD} programs for bidirectional editing.
\newblock \emph{Computer Graphics Forum (CGF) 41}, 2 (2022), 309--323.

\bibitem[CZ19]{IMNet}
\textsc{Chen Z., Zhang H.}:
\newblock Learning implicit fields for generative shape modeling.
\newblock In \emph{IEEE/CVF Conference on Computer Vision and Pattern
  Recognition (CVPR)} (2019).

\bibitem[DAD{\etalchar{*}}18]{SingleImageSVBRDF}
\textsc{Deschaintre V., Aittala M., Durand F., Drettakis G., Bousseau A.}:
\newblock Single-image {SVBRDF} capture with a rendering-aware deep network.
\newblock \emph{ACM Transactions on Graphics (TOG) 37}, 128 (2018), 15.

\bibitem[{Das}]{Solidworks}
\textsc{{Dassault Systemes}}:
\newblock {SOLIDWORKS}.
\newblock \url{https://www.solidworks.com/}.
\newblock Accessed: 2022-10-16.

\bibitem[dB78]{SplineBook}
\textsc{de~Boor C.}:
\newblock A practical guide to splines.
\newblock In \emph{Applied Mathematical Sciences} (1978).

\bibitem[DDFG01]{SMCBook}
\textsc{Doucet A., De~Freitas N., Gordon N.} (Eds.):
\newblock \emph{Sequential Monte Carlo Methods in Practice}.
\newblock Springer, 2001.

\bibitem[DIP{\etalchar{*}}18]{du2018inversecsg}
\textsc{Du T., Inala J.~P., Pu Y., Spielberg A., Schulz A., Rus D.,
  Solar-Lezama A., Matusik W.}:
\newblock {InverseCSG}: Automatic conversion of {3D} models to {CSG} trees.
\newblock \emph{ACM Transactions on Graphics (TOG) 37}, 6 (2018), 1--16.

\bibitem[DKD{\etalchar{*}}22]{deng2022unsupervised}
\textsc{Deng B., Kulal S., Dong Z., Deng C., Tian Y., Wu J.}:
\newblock Unsupervised learning of shape programs with repeatable implicit
  parts.
\newblock In \emph{Advances in Neural Information Processing Systems (NeurIPS)}
  (2022).

\bibitem[DNJ20]{overfit_implicits}
\textsc{Davies T., Nowrouzezahrai D., Jacobson A.}:
\newblock On the effectiveness of weight-encoded neural implicit {3D} shapes.
\newblock \emph{arXiv preprint arXiv:2009.09808} (2020).

\bibitem[Edw63]{SketchPad}
\textsc{Edward S.~I.}:
\newblock \emph{{SketchPad}: A man-machine graphical communication system}.
\newblock PhD thesis, Massachusetts Institute of Technology, 1963.

\bibitem[ENP{\etalchar{*}}19]{ellis2019repl}
\textsc{Ellis K., Nye M., Pu Y., Sosa F., Tenenbaum J.~B., Solar-Lezama A.}:
\newblock Write, execute, assess: Program synthesis with a repl.
\newblock In \emph{Advances in Neural Information Processing Systems (NeurIPS)}
  (2019).

\bibitem[ERSLT18]{ellis2018learning}
\textsc{Ellis K., Ritchie D., Solar-Lezama A., Tenenbaum J.}:
\newblock Learning to infer graphics programs from hand-drawn images.
\newblock In \emph{Advances in Neural Information Processing Systems (NeurIPS)}
  (2018), vol.~31.

\bibitem[{Esr}]{CityEngine}
\textsc{{Esri}}:
\newblock {ArcGIS CityEngine}.
\newblock
  \url{https://www.esri.com/en-us/arcgis/products/arcgis-cityengine/overview}.
\newblock Accessed: 2022-09-26.

\bibitem[EWN{\etalchar{*}}21]{DreamCoder}
\textsc{Ellis K., Wong C., Nye M., Sabl{\'e}-Meyer M., Morales L., Hewitt L.,
  Cary L., Solar-Lezama A., Tenenbaum J.~B.}:
\newblock {DreamCoder}: Bootstrapping inductive program synthesis with
  wake-sleep library learning.
\newblock In \emph{ACM SIGPLAN International Symposium on New Ideas, New
  Paradigms, and Reflections on Programming and Software (SIGPLAN)} (2021),
  pp.~835--850.

\bibitem[FAW19]{TileGAN}
\textsc{Fr\"{u}hst\"{u}ck A., Alhashim I., Wonka P.}:
\newblock {TileGAN}: Synthesis of large-scale non-homogeneous textures.
\newblock \emph{ACM Transactions on Graphics (TOG) 38}, 4 (2019).

\bibitem[FBA22]{UrbanTreeGenerator}
\textsc{Firoze A., Benes B., Aliaga D.}:
\newblock Urban tree generator: spatio-temporal and generative deep learning
  for urban tree localization and modeling.
\newblock \emph{The Visual Computer 38} (06 2022), 1--13.
\newblock \href {https://doi.org/10.1007/s00371-022-02526-x}
  {\path{doi:10.1007/s00371-022-02526-x}}.

\bibitem[FKS{\etalchar{*}}04]{funkhouser2004modeling}
\textsc{Funkhouser T., Kazhdan M., Shilane P., Min P., Kiefer W., Tal A.,
  Rusinkiewicz S., Dobkin D.}:
\newblock Modeling by example.
\newblock \emph{ACM Transactions on Graphics (TOG) 23}, 3 (2004), 652--663.

\bibitem[Fow10]{fowler2010domain}
\textsc{Fowler M.}:
\newblock \emph{Domain-specific languages}.
\newblock Pearson Education, 2010.

\bibitem[GAD{\etalchar{*}}20]{Guehl:2020:PPTBF}
\textsc{Guehl P., Allegre R., Dischler J.-M., Benes B., Galin E.}:
\newblock Semi-procedural textures using point process texture basis functions.
\newblock \emph{Computer Graphics Forum (CGF) 39}, 4 (2020), 159--171.

\bibitem[GBL{\etalchar{*}}21]{Ganin2021ComputerAidedDA}
\textsc{Ganin Y., Bartunov S., Li Y., Keller E., Saliceti S.}:
\newblock Computer-aided design as language.
\newblock In \emph{Advances in Neural Information Processing Systems (NeurIPS)}
  (2021).

\bibitem[GEB15]{GatysStyleTransfer}
\textsc{Gatys L.~A., Ecker A.~S., Bethge M.}:
\newblock A neural algorithm of artistic style.
\newblock \emph{arXiv preprint arXiv:1508.06576} (2015).

\bibitem[Gha08]{CSGBookChapter}
\textsc{Ghali S.}:
\newblock \emph{Constructive solid geometry}.
\newblock Springer, 2008, pp.~277--283.

\bibitem[GHS{\etalchar{*}}22]{guerrero2022matformer}
\textsc{Guerrero P., Hasan M., Sunkavalli K., Mech R., Boubekeur T., Mitra N.}:
\newblock {MatFormer}: A generative model for procedural materials.
\newblock \emph{ACM Transactions on Graphics (TOG) 41}, 4 (2022).

\bibitem[GJB{\etalchar{*}}20]{guo2020inverse}
\textsc{Guo J., Jiang H., Benes B., Deussen O., Zhang X., Lischinski D., Huang
  H.}:
\newblock Inverse procedural modeling of branching structures by inferring
  l-systems.
\newblock \emph{ACM Transactions on Graphics (TOG) 39}, 5 (2020), 1--13.

\bibitem[GKB{\etalchar{*}}18]{ganin2018synthesizing}
\textsc{Ganin Y., Kulkarni T., Babuschkin I., Eslami S.~A., Vinyals O.}:
\newblock Synthesizing programs for images using reinforced adversarial
  learning.
\newblock In \emph{International Conference on Machine Learning (ICML)} (2018),
  PMLR, pp.~1666--1675.

\bibitem[GKG{\etalchar{*}}22]{AutoDiffProcMod}
\textsc{Gaillard M., Krs V., Gori G., Měch R., Benes B.}:
\newblock Automatic differentiable procedural modeling.
\newblock \emph{Computer Graphics Forum 41}, 2 (2022), 289--307.
\newblock URL: \url{https://onlinelibrary.wiley.com/doi/abs/10.1111/cgf.14475},
  \href
  {http://arxiv.org/abs/https://onlinelibrary.wiley.com/doi/pdf/10.1111/cgf.14475}
  {\path{arXiv:https://onlinelibrary.wiley.com/doi/pdf/10.1111/cgf.14475}},
  \href {https://doi.org/https://doi.org/10.1111/cgf.14475}
  {\path{doi:https://doi.org/10.1111/cgf.14475}}.

\bibitem[GLLD12]{Galerne:2012:Gabor}
\textsc{Galerne B., Lagae A., Lefebvre S., Drettakis G.}:
\newblock Gabor noise by example.
\newblock \emph{ACM Transactions on Graphics (TOG) 31}, 4 (2012).

\bibitem[GLP{\etalchar{*}}22]{GuoComplexGen2022}
\textsc{Guo H., Liu S., Pan H., Liu Y., Tong X., Guo B.}:
\newblock Complexgen: Cad reconstruction by b-rep chain complex generation.
\newblock \emph{ACM Trans. Graph. (SIGGRAPH) 41}, 4 (July 2022).
\newblock URL: \url{https://doi.org/10.1145/3528223.3530078}, \href
  {https://doi.org/10.1145/3528223.3530078}
  {\path{doi:10.1145/3528223.3530078}}.

\bibitem[GPS{\etalchar{*}}17]{gulwani2017program}
\textsc{Gulwani S., Polozov O., Singh R., et~al.}:
\newblock Program synthesis.
\newblock \emph{Foundations and Trends{\textregistered} in Programming
  Languages 4}, 1-2 (2017), 1--119.

\bibitem[GSH{\etalchar{*}}20]{Guo:2020:MaterialGAN}
\textsc{Guo Y., Smith C., Ha\v{s}an M., Sunkavalli K., Zhao S.}:
\newblock {MaterialGAN}: Reflectance capture using a generative {SVBRDF} model.
\newblock \emph{ACM Transactions on Graphics (TOG) 39}, 6 (2020),
  254:1--254:13.

\bibitem[GSLT{\etalchar{*}}18]{gottschlich2018three}
\textsc{Gottschlich J., Solar-Lezama A., Tatbul N., Carbin M., Rinard M.,
  Barzilay R., Amarasinghe S., Tenenbaum J.~B., Mattson T.}:
\newblock The three pillars of machine programming.
\newblock In \emph{ACM SIGPLAN International Symposium on New Ideas, New
  Paradigms, and Reflections on Programming and Software (SIGPLAN)} (2018),
  pp.~69--80.

\bibitem[GSR{\etalchar{*}}17]{NeuralMessagePassing}
\textsc{Gilmer J., Schoenholz S.~S., Riley P.~F., Vinyals O., Dahl G.~E.}:
\newblock Neural message passing for quantum chemistry.
\newblock In \emph{International Conference on Machine Learning (ICML)} (2017),
  p.~1263–1272.

\bibitem[Gul11]{gulwani2011automating}
\textsc{Gulwani S.}:
\newblock Automating string processing in spreadsheets using input-output
  examples.
\newblock In \emph{ACM SIGPLAN International Symposium on New Ideas, New
  Paradigms, and Reflections on Programming and Software (SIGPLAN)} (2011).

\bibitem[HDMR21]{henzler2021neuralmaterial}
\textsc{Henzler P., Deschaintre V., Mitra N.~J., Ritschel T.}:
\newblock Generative modelling of {BRDF} textures from flash images.
\newblock \emph{ACM Transactions on Graphics (TOG) 40}, 6 (2021).

\bibitem[HDR19]{hu2019inverse}
\textsc{Hu Y., Dorsey J., Rushmeier H.}:
\newblock A novel framework for inverse procedural texture modeling.
\newblock \emph{ACM Transactions on Graphics (TOG) 38}, 6 (2019).

\bibitem[HE17]{SketchRNN}
\textsc{Ha D., Eck D.}:
\newblock A neural representation of sketch drawings.
\newblock \emph{arXiv preprint arXiv:1704.03477} (2017).

\bibitem[HGH{\etalchar{*}}22]{hu2022diff}
\textsc{Hu Y., Guerrero P., Hasan M., Rushmeier H., Deschaintre V.}:
\newblock Node graph optimization using differentiable proxies.
\newblock In \emph{Annual Conference on Computer Graphics and Interactive
  Techniques (SIGGRAPH)} (2022).

\bibitem[HHD{\etalchar{*}}22]{hu2022inverse}
\textsc{Hu Y., He C., Deschaintre V., Dorsey J., Rushmeier H.}:
\newblock An inverse procedural modeling pipeline for {SVBRDF} maps.
\newblock \emph{ACM Transactions on Graphics (TOG) 41}, 2 (2022).

\bibitem[HLB19]{han2019image}
\textsc{Han X.-F., Laga H., Bennamoun M.}:
\newblock Image-based {3D} object reconstruction: State-of-the-art and trends
  in the deep learning era.
\newblock \emph{IEEE Transactions on Pattern Analysis and Machine intelligence
  (TPAMI) 43}, 5 (2019), 1578--1604.

\bibitem[HLC19]{hempel2019sketch}
\textsc{Hempel B., Lubin J., Chugh R.}:
\newblock {Sketch-n-Sketch}: Output-directed programming for {SVG}.
\newblock In \emph{ACM Symposium on User Interface Software and Technology
  (UIST)} (2019), pp.~281--292.

\bibitem[HLHF22]{3DWaveletDiffusion}
\textsc{Hui K.-H., Li R., Hu J., Fu C.-W.}:
\newblock Neural wavelet-domain diffusion for {3D} shape generation.
\newblock In \emph{Annual Conference on Computer Graphics and Interactive
  Techniques Asia (SIGGRAPH Asia)} (2022), pp.~1--9.

\bibitem[Hof89]{hoffmann1989geometric}
\textsc{Hoffmann C.~M.}:
\newblock \emph{Geometric and solid modeling}.
\newblock CUMINCAD, 1989.

\bibitem[Ico]{Icons8}
{Icons8}.
\newblock \url{https://icons8.com/}.
\newblock Accessed: 2022-20-20.

\bibitem[{IDV}]{SpeedTree}
\textsc{{IDV, Inc.}}:
\newblock {SpeedTree} -- {3D} vegetation modeling and middleware.
\newblock \url{https://store.speedtree.com/}.
\newblock Accessed: 2022-09-26.

\bibitem[IK16]{SMTBugFinding}
\textsc{Ishii Y., Kutsuna T.}:
\newblock Effective fault localization using dynamic slicing and an smt solver.
\newblock In \emph{IEEE International Conference on Software Testing,
  Verification and Validation Workshops (ICSTW)} (2016), pp.~180--188.

\bibitem[{Ini}]{Shadertoy}
\textsc{{Inigo Quilez and Pol Jeremias}}:
\newblock {Shadertoy}.
\newblock \url{https://www.shadertoy.com/}.
\newblock Accessed: 2022-10-16.

\bibitem[IZZE17]{pix2pix2017}
\textsc{Isola P., Zhu J.-Y., Zhou T., Efros A.~A.}:
\newblock Image-to-image translation with conditional adversarial networks.
\newblock In \emph{IEEE/CVF Conference on Computer Vision and Pattern
  Recognition (CVPR)} (2017).

\bibitem[JBX{\etalchar{*}}20]{jones2020SA}
\textsc{Jones R.~K., Barton T., Xu X., Wang K., Jiang E., Guerrero P., Mitra
  N.~J., Ritchie D.}:
\newblock {ShapeAssembly}: Learning to generate programs for {3D} shape
  structure synthesis.
\newblock \emph{ACM Transactions on Graphics (TOG) 39}, 6 (2020).

\bibitem[JCG{\etalchar{*}}21]{jones2021shapeMOD}
\textsc{Jones R.~K., Charatan D., Guerrero P., Mitra N.~J., Ritchie D.}:
\newblock {ShapeMOD}: Macro operation discovery for {3D} shape programs.
\newblock \emph{ACM Transactions on Graphics (TOG) 40}, 4 (2021).

\bibitem[JWR22]{jones2022PLAD}
\textsc{Jones R.~K., Walke H., Ritchie D.}:
\newblock {PLAD}: Learning to infer shape programs with pseudo-labels and
  approximate distributions.
\newblock In \emph{IEEE/CVF Conference on Computer Vision and Pattern
  Recognition (CVPR)} (2022).

\bibitem[Kel21]{CityEngineBookChapter}
\textsc{Kelly T.}:
\newblock {CityEngine}: An introduction to rule-based modeling.
\newblock In \emph{Urban Informatics}. Springer, 2021, pp.~637--662.

\bibitem[KLA21]{StyleGAN}
\textsc{Karras T., Laine S., Aila T.}:
\newblock A style-based generator architecture for generative adversarial
  networks.
\newblock \emph{IEEE Transactions on Pattern Analysis and Machine intelligence
  (TPAMI) 43}, 12 (2021), 4217--4228.

\bibitem[KMJ{\etalchar{*}}19]{Koch:2019:ABC}
\textsc{Koch S., Matveev A., Jiang Z., Williams F., Artemov A., Burnaev E.,
  Alexa M., Zorin D., Panozzo D.}:
\newblock {ABC}: A big {CAD} model dataset for geometric deep learning.
\newblock In \emph{IEEE/CVF Conference on Computer Vision and Pattern
  Recognition (CVPR)} (2019).

\bibitem[Koz92]{GeneticProgrammingBook}
\textsc{Koza J.~R.}:
\newblock \emph{Genetic Programming: On the Programming of Computers by Means
  of Natural Selection}.
\newblock MIT Press, Cambridge, MA, USA, 1992.

\bibitem[KZK20]{kania2020ucsgnet}
\textsc{Kania K., Zieba M., Kajdanowicz T.}:
\newblock {UCSG-NET} - unsupervised discovering of constructive solid geometry
  tree.
\newblock In \emph{Advances in Neural Information Processing Systems (NeurIPS)}
  (2020), vol.~33, pp.~8776--8786.

\bibitem[LBH15]{lecun2015deep}
\textsc{LeCun Y., Bengio Y., Hinton G.}:
\newblock Deep learning.
\newblock \emph{Nature 521}, 7553 (2015), 436--444.

\bibitem[LCC{\etalchar{*}}22]{li2022competition}
\textsc{Li Y., Choi D., Chung J., Kushman N., Schrittwieser J., Leblond R.,
  Eccles T., Keeling J., Gimeno F., Lago A.~D., et~al.}:
\newblock Competition-level code generation with {AlphaCode}.
\newblock \emph{arXiv preprint arXiv:2203.07814} (2022).

\bibitem[LDHM16]{Liu:2016:wood}
\textsc{Liu A.~J., Dong Z., Ha\v{s}an M., Marschner S.}:
\newblock Simulating the structure and texture of solid wood.
\newblock \emph{ACM Transactions on Graphics (TOG) 35}, 6 (2016).

\bibitem[LGB{\etalchar{*}}21]{TreePartNet21}
\textsc{Liu Y., Guo J., Benes B., Deussen O., Zhang X., Huang H.}:
\newblock Treepartnet: Neural decomposition of point clouds for 3d tree
  reconstruction.
\newblock \emph{ACM Transactions on Graphics (Proceedings of SIGGRAPH ASIA)
  40}, 6 (2021), 232:1--232:16.

\bibitem[LGD{\etalchar{*}}18]{liu2018perception}
\textsc{Liu J., Gan Y., Dong J., Qi L., Sun X., Jian M., Wang L., Yu H.}:
\newblock Perception-driven procedural texture generation from examples.
\newblock \emph{Neurocomputing 291} (2018), 21--34.

\bibitem[LHES19]{lopes2019learned}
\textsc{Lopes R.~G., Ha D., Eck D., Shlens J.}:
\newblock A learned representation for scalable vector graphics.
\newblock In \emph{IEEE/CVF International Conference on Computer Vision (ICCV)}
  (2019).

\bibitem[LKK{\etalchar{*}}21]{SingleImageTrees}
\textsc{Li B., Ka\l{}u\.{z}ny J., Klein J., Michels D.~L., Pa\l{}ubicki W.,
  Benes B., Pirk S.}:
\newblock Learning to reconstruct botanical trees from single images.
\newblock \emph{ACM Trans. Graph. 40}, 6 (dec 2021).
\newblock URL: \url{https://doi.org/10.1145/3478513.3480525}, \href
  {https://doi.org/10.1145/3478513.3480525}
  {\path{doi:10.1145/3478513.3480525}}.

\bibitem[LLHF21]{li2021spgan}
\textsc{Li R., Li X., Hui K.-H., Fu C.-W.}:
\newblock {SP-GAN}: Sphere-guided {3D} shape generation and manipulation.
\newblock \emph{ACM Transactions on Graphics (TOG) 40}, 4 (2021).

\bibitem[LP00]{Lefebvre:2000:StructuralTextures}
\textsc{Lefebvre L., Poulin P.}:
\newblock Analysis and synthesis of structural textures.
\newblock In \emph{Graphics Interface} (2000), vol.~2000, pp.~77--86.

\bibitem[LPBM20]{Li:2020:Sketch2CAD}
\textsc{Li C., Pan H., Bousseau A., Mitra N.~J.}:
\newblock {Sketch2CAD}: Sequential {CAD} modeling by sketching in context.
\newblock \emph{ACM Transactions on Graphics (TOG) 39}, 6 (2020),
  164:1--164:14.

\bibitem[LPBM22]{Li:2022:Free2CAD}
\textsc{Li C., Pan H., Bousseau A., Mitra N.~J.}:
\newblock {Free2CAD}: Parsing freehand drawings into {CAD} commands.
\newblock \emph{ACM Transactions on Graphics (TOG) 41}, 4 (2022), 93:1--93:16.

\bibitem[LST15]{Brenden:2015:Omniglot}
\textsc{Lake B.~M., Salakhutdinov R., Tenenbaum J.~B.}:
\newblock Human-level concept learning through probabilistic program induction.
\newblock \emph{Science 350}, 6266 (2015), 1332--1338.

\bibitem[LST19]{lake2019omniglot}
\textsc{Lake B.~M., Salakhutdinov R., Tenenbaum J.~B.}:
\newblock The {Omniglot} challenge: a 3-year progress report.
\newblock \emph{Current Opinion in Behavioral Sciences 29} (2019), 97--104.

\bibitem[LWJ{\etalchar{*}}22]{lambourne2022recon}
\textsc{Lambourne J.~G., Willis K. D.~D., Jayaraman P.~K., Zhang L., Sanghi A.,
  Malekshan K.~R.}:
\newblock Reconstructing editable prismatic {CAD} from rounded voxel models.
\newblock \emph{arXiv preprint arXiv:2209.01161} (2022).

\bibitem[LXC{\etalchar{*}}17]{GRASS}
\textsc{Li J., Xu K., Chaudhuri S., Yumer E., Zhang H., Guibas L.}:
\newblock {GRASS}: Generative recursive autoencoders for shape structures.
\newblock \emph{ACM Transactions on Graphics (TOG) 36}, 4 (2017), 1--14.

\bibitem[LZCvdP20]{ling2020character}
\textsc{Ling H.~Y., Zinno F., Cheng G., van~de Panne M.}:
\newblock Character controllers using motion {VAEs}.
\newblock \emph{ACM Transactions on Graphics (TOG) 39}, 4 (2020).

\bibitem[{Mas}]{Massive}
\textsc{{Massive Software}}:
\newblock {Massive Software}.
\newblock \url{https://www.massivesoftware.com/}.
\newblock Accessed: 2022-09-26.

\bibitem[MBBO22]{Mezghanni_2022_CVPR}
\textsc{Mezghanni M., Bodrito T., Boulkenafed M., Ovsjanikov M.}:
\newblock Physical simulation layer for accurate 3d modeling.
\newblock In \emph{Proceedings of the IEEE/CVF Conference on Computer Vision
  and Pattern Recognition (CVPR)} (June 2022), pp.~13514--13523.

\bibitem[MGA{\etalchar{*}}22]{ma2022searching}
\textsc{Ma K., Gharbi M., Adams A., Kamil S., Li T.-M., Barnes C., Ragan-Kelley
  J.}:
\newblock Searching for fast demosaicking algorithms.
\newblock \emph{ACM Transactions on Graphics (TOG)} (2022).

\bibitem[MGY{\etalchar{*}}19]{StructureNet}
\textsc{Mo K., Guerrero P., Yi L., Su H., Wonka P., Mitra N., Guibas L.}:
\newblock {StructureNet}: Hierarchical graph networks for {3D} shape
  generation.
\newblock \emph{ACM Transactions on Graphics (TOG) 38}, 6 (2019).

\bibitem[Mit77]{Mitchell1977VersionSA}
\textsc{Mitchell T.~M.}:
\newblock Version spaces: A candidate elimination approach to rule learning.
\newblock In \emph{International Joint Conference on Artificial Intelligence
  (IJCAI)} (1977).

\bibitem[MKG{\etalchar{*}}18]{creativeAI}
\textsc{Mitra N.~J., Kokkinos I., Guerrero P., Thuerey N., Ritschel T.}:
\newblock {CreativeAI}: Deep learning for graphics.
\newblock In \emph{SIGGRAPH Asia 2018 Courses} (2018).

\bibitem[MST{\etalchar{*}}20]{mildenhall2020nerf}
\textsc{Mildenhall B., Srinivasan P.~P., Tancik M., Barron J.~T., Ramamoorthi
  R., Ng R.}:
\newblock {NeRF}: Representing scenes as neural radiance fields for view
  synthesis.
\newblock In \emph{European Conference on Computer Vision (ECCV)} (2020).

\bibitem[MVG13]{FacadeParsing}
\textsc{Martinovic A., Van~Gool L.}:
\newblock Bayesian grammar learning for inverse procedural modeling.
\newblock In \emph{IEEE/CVF Conference on Computer Vision and Pattern
  Recognition (CVPR)} (2013), pp.~201--208.

\bibitem[MWH{\etalchar{*}}06]{CGAShape}
\textsc{M{\"u}ller P., Wonka P., Haegler S., Ulmer A., Van~Gool L.}:
\newblock Procedural modeling of buildings.
\newblock In \emph{Annual Conference on Computer Graphics and Interactive
  Techniques (SIGGRAPH)} (2006), pp.~614--623.

\bibitem[MZC{\etalchar{*}}19]{Mo_2019_CVPR}
\textsc{Mo K., Zhu S., Chang A.~X., Yi L., Tripathi S., Guibas L.~J., Su H.}:
\newblock {PartNet}: A large-scale benchmark for fine-grained and hierarchical
  part-level {3D} object understanding.
\newblock In \emph{IEEE/CVF Conference on Computer Vision and Pattern
  Recognition (CVPR)} (2019).

\bibitem[NGEB20]{polygen}
\textsc{Nash C., Ganin Y., Eslami S. M.~A., Battaglia P.~W.}:
\newblock {PolyGen}: An autoregressive generative model of {3D} meshes.
\newblock In \emph{International Conference on Machine Learning (ICML)} (2020).

\bibitem[NO80]{nelson1980fast}
\textsc{Nelson G., Oppen D.~C.}:
\newblock Fast decision procedures based on congruence closure.
\newblock \emph{Journal of the ACM (JACM) 27}, 2 (1980), 356--364.

\bibitem[{Nvi}22]{VMaterials}
\textsc{{Nvidia}}:
\newblock {VMaterials}, 2022.
\newblock \url{https://developer.nvidia.com/vmaterials}.

\bibitem[NWA{\etalchar{*}}20]{nandi2020synthesizing}
\textsc{Nandi C., Willsey M., Anderson A., Wilcox J.~R., Darulova E., Grossman
  D., Tatlock Z.}:
\newblock Synthesizing structured {CAD} models with equality saturation and
  inverse transformations.
\newblock In \emph{ACM SIGPLAN International Symposium on New Ideas, New
  Paradigms, and Reflections on Programming and Software (SIGPLAN)} (2020),
  pp.~31--44.

\bibitem[NWP{\etalchar{*}}18]{nandi2018functional}
\textsc{Nandi C., Wilcox J.~R., Panchekha P., Blau T., Grossman D., Tatlock
  Z.}:
\newblock Functional programming for compiling and decompiling computer-aided
  design.
\newblock In \emph{ACM SIGPLAN International Conference on Functional
  Programming (ICFP)} (2018).

\bibitem[NZIS13]{niessner2013hashing}
\textsc{Nie{\ss}ner M., Zollh\"ofer M., Izadi S., Stamminger M.}:
\newblock Real-time {3D} reconstruction at scale using voxel hashing.
\newblock \emph{ACM Transactions on Graphics (TOG)} (2013).

\bibitem[{Ope}]{OpenSCAD}
\textsc{{OpenSCAD}}:
\newblock {OpenSCAD} - the programmers solid {3D} modeler.
\newblock \url{https://openscad.org/}.
\newblock Accessed: 2022-10-21.

\bibitem[PBG{\etalchar{*}}21]{para2021sketchgen}
\textsc{Para W.~R., Bhat S.~F., Guerrero P., Kelly T., Mitra N., Guibas L.,
  Wonka P.}:
\newblock {SketchGen}: Generating constrained {CAD} sketches.
\newblock In \emph{Advances in Neural Information Processing Systems (NeurIPS)}
  (2021).

\bibitem[Pea85]{Peachey:1985:SolidTexturing}
\textsc{Peachey D.~R.}:
\newblock Solid texturing of complex surfaces.
\newblock \emph{ACM Transactions on Graphics (TOG) 19}, 3 (1985), 279–286.

\bibitem[Per85]{Perlin:1985:ImageSynthesizer}
\textsc{Perlin K.}:
\newblock An image synthesizer.
\newblock In \emph{Annual Conference on Computer Graphics and Interactive
  Techniques (SIGGRAPH)} (1985).

\bibitem[PGK{\etalchar{*}}21]{para2021generative}
\textsc{Para W., Guerrero P., Kelly T., Guibas L., Wonka P.}:
\newblock Generative layout modeling using constraint graphs.
\newblock In \emph{IEEE/CVF International Conference on Computer Vision (ICCV)}
  (2021), pp.~6670--6680.

\bibitem[PHHM96]{LSystems}
\textsc{Prusinkiewicz P., Hammel M., Hanan J., M\v{e}ch R.}:
\newblock L-systems: From the theory to visual models of plants.
\newblock In \emph{CSIRO Symposium on Computational Challenges in Life
  Sciences} (1996).

\bibitem[PJM94]{SyntheticTopiary}
\textsc{Prusinkiewicz P., James M., M{\v{e}}ch R.}:
\newblock Synthetic topiary.
\newblock In \emph{Annual Conference on Computer Graphics and Interactive
  Techniques (SIGGRAPH)} (1994), pp.~351--358.

\bibitem[PKGF21]{Paschalidou2021CVPR}
\textsc{Paschalidou D., Katharopoulos A., Geiger A., Fidler S.}:
\newblock Neural parts: Learning expressive {3D} shape abstractions with
  invertible neural networks.
\newblock In \emph{IEEE/CVF Conference on Computer Vision and Pattern
  Recognition (CVPR)} (2021).

\bibitem[PL96]{ABOP}
\textsc{Prusinkiewicz P., Lindenmayer A.}:
\newblock \emph{The Algorithmic Beauty of Plants}.
\newblock Springer-Verlag, Berlin, Heidelberg, 1996.

\bibitem[{PTC}]{OnShape}
\textsc{{PTC Inc.}}:
\newblock {OnShape}.
\newblock URL: \url{https://www.onshape.com/}.

\bibitem[Qui]{QuickDraw}
{The Quick, Draw! Dataset}.
\newblock \url{https://quickdraw.withgoogle.com/data}.
\newblock Accessed: 2022-20-20.

\bibitem[RBCP20]{Sketchformer2020}
\textsc{Ribeiro L. S.~F., Bui T., Collomosse J., Ponti M.}:
\newblock Sketchformer: Transformer-based representation for sketched
  structure.
\newblock In \emph{IEEE/CVF Conference on Computer Vision and Pattern
  Recognition (CVPR)} (2020).

\bibitem[RBL{\etalchar{*}}22]{StableDiffusion}
\textsc{Rombach R., Blattmann A., Lorenz D., Esser P., Ommer B.}:
\newblock High-resolution image synthesis with latent diffusion models.
\newblock In \emph{IEEE/CVF Conference on Computer Vision and Pattern
  Recognition (CVPR)} (2022).

\bibitem[RDN{\etalchar{*}}22]{DALLE2}
\textsc{Ramesh A., Dhariwal P., Nichol A., Chu C., Chen M.}:
\newblock Hierarchical text-conditional image generation with {CLIP} latents.
\newblock \emph{arXiv preprint arXiv:2204.06125} (2022).

\bibitem[Red76]{BeamSearch}
\textsc{Reddy D.}:
\newblock Speech understanding systems: summary of results of the five-year
  research effort.
\newblock \emph{Computer Science, Carnegie-Mellon University, Pittsburgh, PA}
  (1976).

\bibitem[RGLM21]{reddy2021im2vec}
\textsc{Reddy P., Gharbi M., Lukac M., Mitra N.~J.}:
\newblock {Im2Vec}: Synthesizing vector graphics without vector supervision.
\newblock In \emph{IEEE/CVF Conference on Computer Vision and Pattern
  Recognition (CVPR)} (2021), pp.~7342--7351.

\bibitem[RJT18]{ExampleBasedProcMod}
\textsc{Ritchie D., Jobalia S., Thomas A.}:
\newblock Example-based authoring of procedural modeling programs with
  structural and continuous variability.
\newblock \emph{Computer Graphics Forum (CGF) 37}, 2 (2018), 401--413.

\bibitem[RVdOV19]{razavi2019generating}
\textsc{Razavi A., Van~den Oord A., Vinyals O.}:
\newblock Generating diverse high-fidelity images with {VQ-VAE-2}.
\newblock In \emph{Advances in Neural Information Processing Systems (NeurIPS)}
  (2019), vol.~32.

\bibitem[RZC{\etalchar{*}}21]{ren2021csg}
\textsc{Ren D., Zheng J., Cai J., Li J., Jiang H., Cai Z., Zhang J., Pan L.,
  Zhang M., Zhao H., et~al.}:
\newblock {CSG-Stump}: A learning friendly {CSG}-like representation for
  interpretable shape parsing.
\newblock In \emph{IEEE/CVF International Conference on Computer Vision (ICCV)}
  (2021), pp.~12478--12487.

\bibitem[RZC{\etalchar{*}}22]{ren2022extrude}
\textsc{Ren D., Zheng J., Cai J., Li J., Zhang J.}:
\newblock {ExtrudeNet}: Unsupervised inverse sketch-and-extrude for shape
  parsing.
\newblock In \emph{European Conference on Computer Vision (ECCV)} (2022).

\bibitem[San]{BusinessCardRaytracer}
\textsc{Sanglard F.}:
\newblock Decyphering the business card raytracer.
\newblock \url{https://fabiensanglard.net/rayTracing_back_of_business_card/}.
\newblock Accessed: 2022-10-19.

\bibitem[SCS{\etalchar{*}}22]{Imagen}
\textsc{Saharia C., Chan W., Saxena S., Li L., Whang J., Denton E., Ghasemipour
  S. K.~S., Ayan B.~K., Mahdavi S.~S., Lopes R.~G., et~al.}:
\newblock Photorealistic text-to-image diffusion models with deep language
  understanding.
\newblock \emph{arXiv preprint arXiv:2205.11487} (2022).

\bibitem[SG71]{ShapeGrammars}
\textsc{Stiny G., Gips J.}:
\newblock Shape grammars and the generative specification of painting and
  sculpture.
\newblock \emph{Information Processing} (1971).

\bibitem[SGL{\etalchar{*}}18]{CSGNet}
\textsc{Sharma G., Goyal R., Liu D., Kalogerakis E., Maji S.}:
\newblock {CSGNet}: Neural shape parser for constructive solid geometry.
\newblock In \emph{IEEE/CVF Conference on Computer Vision and Pattern
  Recognition (CVPR)} (2018).

\bibitem[{Sid}]{Houdini}
\textsc{{SideFX}}:
\newblock {Houdini}.
\newblock \url{https://www.sidefx.com/products/houdini/}.
\newblock Accessed: 2022-09-26.

\bibitem[SLH{\etalchar{*}}20]{Shi2020:MATch}
\textsc{Shi L., Li B., Ha{\v s}an M., Sunkavalli K., Boubekeur T., Mech R.,
  Matusik W.}:
\newblock {MATch}: Differentiable material graphs for procedural material
  capture.
\newblock \emph{ACM Transactions on Graphics (TOG) 39}, 6 (2020), 1--15.

\bibitem[SLM{\etalchar{*}}20]{sharmaParseNet}
\textsc{Sharma G., Liu D., Maji S., Kalogerakis E., Chaudhuri S., Mech R.}:
\newblock Parsenet: {A} parametric surface fitting network for 3d point clouds.
\newblock In \emph{Computer Vision - {ECCV} 2020 - 16th European Conference,
  Glasgow, UK, August 23-28, 2020, Proceedings, Part {VII}} (2020), Vedaldi A.,
  Bischof H., Brox T., Frahm J., (Eds.), vol.~12352 of \emph{Lecture Notes in
  Computer Science}, Springer, pp.~261--276.
\newblock URL: \url{https://doi.org/10.1007/978-3-030-58571-6\_16}, \href
  {https://doi.org/10.1007/978-3-030-58571-6\_16}
  {\path{doi:10.1007/978-3-030-58571-6\_16}}.

\bibitem[SLTB{\etalchar{*}}06]{solar2006combinatorial}
\textsc{Solar-Lezama A., Tancau L., Bodik R., Seshia S., Saraswat V.}:
\newblock Combinatorial sketching for finite programs.
\newblock In \emph{International Conference on Architectural Support for
  Programming Languages and Operating Systems (ASPLOS)} (2006), pp.~404--415.

\bibitem[SOZA20]{Ari2020}
\textsc{Seff A., Ovadia Y., Zhou W., Adams R.~P.}:
\newblock {SketchGraphs}: A large-scale dataset for modeling relational
  geometry in computer-aided design.
\newblock In \emph{International Conference on Machine Learning Workshops (ICML
  Workshop)} (2020).

\bibitem[Str06]{BrepBook}
\textsc{Stroud I.}:
\newblock \emph{Boundary representation modelling techniques}.
\newblock Springer Science \& Business Media, 2006.

\bibitem[SWB21]{UNIT-DDPM}
\textsc{Sasaki H., Willcocks C.~G., Breckon T.~P.}:
\newblock {UNIT-DDPM}: Unpaired image translation with denoising diffusion
  probabilistic models.
\newblock \emph{arXiv preprint arXiv:2104.05358} (2021).

\bibitem[SWD{\etalchar{*}}17]{PPO}
\textsc{Schulman J., Wolski F., Dhariwal P., Radford A., Klimov O.}:
\newblock Proximal policy optimization algorithms.
\newblock \emph{arXiv preprint arXiv:1707.06347} (2017).

\bibitem[Sys]{IntelligentMusic}
\textsc{Systems I.~M.}:
\newblock Intelligent music systems.
\newblock \url{https://www.intelligentmusicsystems.com/}.
\newblock Accessed: 2022-10-19.

\bibitem[SZRA22]{seff2021vitruvion}
\textsc{Seff A., Zhou W., Richardson N., Adams R.~P.}:
\newblock Vitruvion: A generative model of parametric {CAD} sketches.
\newblock In \emph{International Conference on Learning Representations (ICLR)}
  (2022).

\bibitem[TB13]{torlak2013growing}
\textsc{Torlak E., Bodik R.}:
\newblock Growing solver-aided languages with rosette.
\newblock In \emph{ACM SIGPLAN International Symposium on New Ideas, New
  Paradigms, and Reflections on Programming and Software (SIGPLAN)} (2013),
  pp.~135--152.

\bibitem[TLS{\etalchar{*}}19]{tian2018learning}
\textsc{Tian Y., Luo A., Sun X., Ellis K., Freeman W.~T., Tenenbaum J.~B., Wu
  J.}:
\newblock Learning to infer and execute {3D} shape programs.
\newblock In \emph{International Conference on Learning Representations (ICLR)}
  (2019).

\bibitem[TRT{\etalchar{*}}22]{Tchapmi:2022:Procedural}
\textsc{Tchapmi L.~P., Ray T., Tchapmi M., Shen B., Martin-Martin R., Savarese
  S.}:
\newblock Generating procedural {3D} materials from images using neural
  networks.
\newblock In \emph{International Conference on Image, Video and Signal
  Processing (IVSP)} (2022), p.~32–40.

\bibitem[TSG{\etalchar{*}}17]{abstractionTulsiani17}
\textsc{Tulsiani S., Su H., Guibas L.~J., Efros A.~A., Malik J.}:
\newblock Learning shape abstractions by assembling volumetric primitives.
\newblock In \emph{IEEE/CVF Conference on Computer Vision and Pattern
  Recognition (CVPR)} (2017).

\bibitem[TSTL09]{eqsat-lmcs}
\textsc{Tate R., Stepp M., Tatlock Z., Lerner S.}:
\newblock Equality saturation: a new approach to optimization.
\newblock In \emph{ACM SIGPLAN International Symposium on New Ideas, New
  Paradigms, and Reflections on Programming and Software (SIGPLAN)} (2009),
  pp.~264--276.

\bibitem[UyCS{\etalchar{*}}22]{uy-point2cyl-cvpr22}
\textsc{Uy M.~A., yu~Chang Y., Sung M., Goel P., Lambourne J., Birdal T.,
  Guibas L.}:
\newblock {Point2Cyl}: Reverse engineering {3D} objects from point clouds to
  extrusion cylinders.
\newblock In \emph{IEEE/CVF Conference on Computer Vision and Pattern
  Recognition (CVPR)} (2022).

\bibitem[VFJ15]{PointerNetworks}
\textsc{Vinyals O., Fortunato M., Jaitly N.}:
\newblock Pointer networks.
\newblock In \emph{Advances in Neural Information Processing Systems (NeurIPS)}
  (2015), vol.~28.

\bibitem[vLA87]{Laarhoven1987SimulatedAT}
\textsc{van Laarhoven P. J.~M., Aarts E. H.~L.}:
\newblock Simulated annealing: Theory and applications.
\newblock In \emph{Mathematics and Its Applications} (1987).

\bibitem[VPB{\etalchar{*}}22]{vinker2022clipasso}
\textsc{Vinker Y., Pajouheshgar E., Bo J.~Y., Bachmann R.~C., Bermano A.~H.,
  Cohen-Or D., Zamir A., Shamir A.}:
\newblock {CLIPasso}: Semantically-aware object sketching.
\newblock In \emph{Annual Conference on Computer Graphics and Interactive
  Techniques (SIGGRAPH)} (2022).

\bibitem[VWM15]{SAT}
\textsc{Vizel Y., Weissenbacher G., Malik S.}:
\newblock Boolean satisfiability solvers and their applications in model
  checking.
\newblock \emph{Proceedings of the IEEE 103} (2015), 2021--2035.

\bibitem[Wil92]{REINFORCE}
\textsc{Williams R.~J.}:
\newblock Simple statistical gradient-following algorithms for connectionist
  reinforcement learning.
\newblock \emph{Machine Learning 8} (1992).

\bibitem[WJC{\etalchar{*}}22]{joinable}
\textsc{Willis K.~D., Jayaraman P.~K., Chu H., Tian Y., Li Y., Grandi D.,
  Sanghi A., Tran L., Lambourne J.~G., Solar-Lezama A., Matusik W.}:
\newblock {JoinABLe}: Learning bottom-up assembly of parametric {CAD} joints.
\newblock In \emph{IEEE/CVF Conference on Computer Vision and Pattern
  Recognition (CVPR)} (2022), pp.~15849--15860.

\bibitem[WJL{\etalchar{*}}21]{willis2021engineering}
\textsc{Willis K. D.~D., Jayaraman P.~K., Lambourne J.~G., Chu H., Pu Y.}:
\newblock Engineering sketch generation for computer-aided design.
\newblock In \emph{IEEE/CVF Conference on Computer Vision and Pattern
  Recognition Workshops (CVPR Workshop)} (2021).

\bibitem[WK91]{Witkin:1991:reactiondiffusion}
\textsc{Witkin A., Kass M.}:
\newblock Reaction-diffusion textures.
\newblock \emph{ACM Transactions on Graphics (TOG) 25}, 4 (1991), 299–308.

\bibitem[WL21]{wang2021deepvecfont}
\textsc{Wang Y., Lian Z.}:
\newblock {DeepVecFont}: Synthesizing high-quality vector fonts via
  dual-modality learning.
\newblock \emph{ACM Transactions on Graphics (TOG) 40}, 6 (2021).

\bibitem[WLW{\etalchar{*}}19]{PlanIT}
\textsc{Wang K., Lin Y.-A., Weissmann B., Savva M., Chang A.~X., Ritchie D.}:
\newblock {PlanIT}: Planning and instantiating indoor scenes with relation
  graph and spatial prior networks.
\newblock \emph{ACM Transactions on Graphics (TOG) 38}, 4 (2019).

\bibitem[WMG{\etalchar{*}}22]{wong2022identifying}
\textsc{Wong C., McCarthy W.~P., Grand G., Friedman Y., Tenenbaum J.~B.,
  Andreas J., Hawkins R.~D., Fan J.~E.}:
\newblock Identifying concept libraries from language about object structure.
\newblock In \emph{Annual Meeting of the Cognitive Science Society (CogSci)}
  (2022).

\bibitem[WNW{\etalchar{*}}21]{2021-egg}
\textsc{Willsey M., Nandi C., Wang Y.~R., Flatt O., Tatlock Z., Panchekha P.}:
\newblock egg: Fast and extensible equality saturation.
\newblock \emph{Proc. ACM Program. Lang. 5}, POPL (Jan. 2021).
\newblock URL: \url{https://doi.org/10.1145/3434304}, \href
  {https://doi.org/10.1145/3434304} {\path{doi:10.1145/3434304}}.

\bibitem[Wor96]{Worley:1996:Cellular}
\textsc{Worley S.}:
\newblock A cellular texture basis function.
\newblock In \emph{Annual Conference on Computer Graphics and Interactive
  Techniques (SIGGRAPH)} (1996), p.~291–294.

\bibitem[WPL{\etalchar{*}}21]{Fusion360Gallery}
\textsc{Willis K. D.~D., Pu Y., Luo J., Chu H., Du T., Lambourne J.~G.,
  Solar-Lezama A., Matusik W.}:
\newblock {Fusion 360 Gallery}: A dataset and environment for programmatic
  {CAD} construction from human design sequences.
\newblock \emph{ACM Transactions on Graphics (TOG) 40}, 4 (2021).

\bibitem[WSCR18]{deep_synth}
\textsc{Wang K., Savva M., Chang A.~X., Ritchie D.}:
\newblock Deep convolutional priors for indoor scene synthesis.
\newblock \emph{ACM Transactions on Graphics (TOG) 37}, 4 (2018).

\bibitem[WXZ21]{DeepCAD}
\textsc{Wu R., Xiao C., Zheng C.}:
\newblock {DeepCAD}: A deep generative network for computer-aided design
  models.
\newblock In \emph{IEEE/CVF International Conference on Computer Vision (ICCV)}
  (2021), pp.~6772--6782.

\bibitem[WZN{\etalchar{*}}19]{wu2019carpentry}
\textsc{Wu C., Zhao H., Nandi C., Lipton J.~I., Tatlock Z., Schulz A.}:
\newblock Carpentry compiler.
\newblock \emph{ACM Transactions on Graphics (TOG) 38}, 6 (2019), 1--14.

\bibitem[WZX{\etalchar{*}}16]{3DGAN}
\textsc{Wu J., Zhang C., Xue T., Freeman W.~T., Tenenbaum J.~B.}:
\newblock Learning a probabilistic latent space of object shapes via {3D}
  generative-adversarial modeling.
\newblock In \emph{Advances in Neural Information Processing Systems (NeurIPS)}
  (2016), pp.~82--90.

\bibitem[XPC{\etalchar{*}}21]{zoneGraphs}
\textsc{Xu X., Peng W., Cheng C.-Y., Willis K. D.~D., Ritchie D.}:
\newblock Inferring {CAD} modeling sequences using zone graphs.
\newblock In \emph{IEEE/CVF Conference on Computer Vision and Pattern
  Recognition (CVPR)} (2021).

\bibitem[XTS{\etalchar{*}}22]{neural_fields}
\textsc{Xie Y., Takikawa T., Saito S., Litany O., Yan S., Khan N., Tombari F.,
  Tompkin J., Sitzmann V., Sridhar S.}:
\newblock Neural fields in visual computing and beyond.
\newblock \emph{Computer Graphics Forum (CGF)} (2022).

\bibitem[XWL{\etalchar{*}}22]{xu2022skexgen}
\textsc{Xu X., Willis K.~D., Lambourne J.~G., Cheng C.-Y., Jayaraman P.~K.,
  Furukawa Y.}:
\newblock {SkexGen}: Autoregressive generation of {CAD} construction sequences
  with disentangled codebooks.
\newblock In \emph{International Conference on Machine Learning (ICML)} (2022).

\bibitem[YCL{\etalchar{*}}22]{Yu_2022_CVPR}
\textsc{Yu F., Chen Z., Li M., Sanghi A., Shayani H., Mahdavi-Amiri A., Zhang
  H.}:
\newblock {CAPRI-Net}: Learning compact {CAD} shapes with adaptive primitive
  assembly.
\newblock In \emph{IEEE/CVF Conference on Computer Vision and Pattern
  Recognition (CVPR)} (2022), pp.~11768--11778.

\bibitem[YLM{\etalchar{*}}22]{yan2022shapeformer}
\textsc{Yan X., Lin L., Mitra N.~J., Lischinski D., Cohen-Or D., Huang H.}:
\newblock {ShapeFormer}: Transformer-based shape completion via sparse
  representation.
\newblock In \emph{IEEE/CVF Conference on Computer Vision and Pattern
  Recognition (CVPR)} (2022), pp.~6239--6249.

\bibitem[YP22a]{SketchConcepts}
\textsc{Yang Y., Pan H.}:
\newblock Discovering design concepts for cad sketches.
\newblock In \emph{Advances in Neural Information Processing Systems (NeurIPS)}
  (2022).

\bibitem[YP22b]{yang2022}
\textsc{Yang Y., Pan H.}:
\newblock Discovering design concepts for cad sketches.
\newblock In \emph{Advances in Neural Information Processing Systems (NeurIPS)}
  (2022).

\bibitem[ZPIE17]{CycleGAN2017}
\textsc{Zhu J.-Y., Park T., Isola P., Efros A.~A.}:
\newblock Unpaired image-to-image translation using cycle-consistent
  adversarial networks.
\newblock In \emph{IEEE/CVF International Conference on Computer Vision (ICCV)}
  (2017).

\bibitem[ZVW{\etalchar{*}}22]{zeng2022lion}
\textsc{Zeng X., Vahdat A., Williams F., Gojcic Z., Litany O., Fidler S., Kreis
  K.}:
\newblock {LION}: Latent point diffusion models for {3D} shape generation.
\newblock In \emph{Advances in Neural Information Processing Systems (NeurIPS)}
  (2022).

\bibitem[ZWZ{\etalchar{*}}21]{zhao2021co}
\textsc{Zhao H., Willsey M., Zhu A., Nandi C., Tatlock Z., Solomon J., Schulz
  A.}:
\newblock Co-optimization of design and fabrication plans for carpentry.
\newblock \emph{arXiv preprint arXiv:2107.12265} (2021).

\end{thebibliography}
